\documentclass[referee,onecolumn]{raa}         
\usepackage{graphicx,times}             
\usepackage{natbib}
\usepackage{amssymb,amsmath}
\usepackage{amsfonts}
\usepackage{multirow}
\usepackage{url}
\usepackage{caption}
\usepackage{soul}
\usepackage{ulem}
\usepackage{threeparttable}
\usepackage{setspace}
\usepackage{xcolor}
\bibpunct{(}{)}{;}{a}{}{,}

\usepackage{booktabs}
\usepackage[pagebackref=true]{hyperref}
\hypersetup{
    colorlinks,
    linkcolor={red!50!black},
    citecolor={blue!50!black},
    urlcolor={blue!80!black}
}

\begin{document}

  \title{\emph{\texttt{Mesiri}}: Mephisto Early Supernovae Ia Rapid Identifier}

   \volnopage{Vol.0 (20xx) No.0, 000--000}     
   \setcounter{page}{1}          
   \author{Lun-Wei Zhang 
      \inst{1}
   \and Zhen-Yu Wang
      \inst{2}
   \and De-Zi Liu
      \inst{1}
   \and Yuan Fang
      \inst{1}
   \and Brajesh Kumar
      \inst{1}
   \and Bing-Qiu Chen
      \inst{1}
   \and Xin-Zhong Er
      \inst{1}
   \and Xiao-Wei Liu
      \inst{1}
   }
   \institute{South-Western Institute for Astronomy Research, Yunnan University, Kunming, Yunnan 650500, China; {\it xer@ynu.edu.cn}; {\it x.liu@ynu.edu.cn}\\
   \and
   Yunnan Observatories, Chinese Academy of Sciences, Kunming 650216, China; \\ 
\vs\no
   {\small Received 2024 month day; accepted 20xx month day}
   }

\abstract{
The early time observations of Type Ia supernovae (SNe Ia) play a crucial role in investigating and resolving longstanding questions about progenitor stars and the explosion mechanisms of these events. Colors of supernovae (SNe) in the initial days after the explosion can help differentiate between different types of SNe. However, the use of true color information to identify SNe Ia at the early-time explosion is still in its infancy. The Multi-channel Photometric Survey Telescope (Mephisto) is a photometric survey telescope equipped with three CCD cameras, capable of simultaneously imaging the same patch of sky in three bands (\emph{u, g, i} or \emph{v, r, z}), yielding real-time colors of astronomical objects. In this paper, we introduce a new time-series classification tool named Mephisto Early Supernovae Ia Rapid Identifier (\emph{\texttt{Mesiri}}), which for the first time, utilizes real-time color information to distinguish early-time SNe Ia from core-collapse supernovae (CCSNe). \emph{\texttt{Mesiri}} is based on the deep learning approach and can achieve an accuracy of $96.75\pm0.79$\%, and AUC of $98.87\pm0.53$\% in case of single epoch random observation before the peak brightness.
These values reach towards perfectness if additional data points on several night observations are considered. The classification with real-time color significantly outperforms that with pseudo-color, especially at the early time, i.e., with only a few points of observations. The architecture of BiLSTM shows the best performance than the others that have been tested in this work.
}
   \authorrunning{Zhang et al.}
   \titlerunning{Mephisto Early Supernovae Ia Rapid Identifier}  
   \maketitle

\section{Introduction}           
\label{sect:intro}

The investigation of supernovae (SNe) is of paramount significance. As an illustration, SNe Ia, as standardized candles, provide a reliable means of measuring cosmic distance. They have led to the discovery of the universe's accelerated expansion \citep{Riess1998, Perlmutter1999} and also provide a valuable means of constraining the Hubble constant \citep[e.g.][]{2001ApJ...553...47F,2021A&ARv..29....9S}. The importance of obtaining early signals from SNe Ia explosions is crucial for various reasons. The radiative signals in the early time can offer import constraints on the explosion mechanism, progenitor system, and their physical origins \citep{2010ApJ...708.1025K, 2011Natur.480..344N, 2016ApJ...820...92M, 2017ApJ...845L..11H, 2018ApJ...865..149J,  2019ApJ...870L...1D, 2020ApJ...902...48B, 2021ApJ...919..142B, 2022MNRAS.514.3541S, 2022ApJ...932L...2A}.
 Moreover, variations in luminosity during this phase can reveal several aspects such as interactions with a companion star in a binary system, the surrounding envelope of the progenitor, or the circumstellar material ejected by the progenitor before the explosion \citep{1996MNRAS.281..263M, 2010ApJ...708.1025K, 2015Natur.521..328C, 2019AAS...23322905D, 2019ApJ...870...13S, 2021ApJ...923L...8J}. Following that, spectroscopic or photometric follow-up at early stages can also serve as a valuable benchmark for further observations at later epochs. The earlier an object can be classified, the more opportunities there are for the community to perform follow-up observation, the more likely to bring about entirely new discoveries
\citep{2016ApJ...826...96P, 2018ApJ...865..149J, 2018ApJ...864L..35S, 2021ApJ...908...51F, 2024arXiv240715768L}. 
 Last but not least, in the initial days following an explosion, changes in color can reveal an asymmetric distribution of elements created through nucleosynthesis \citep{2022NatAs...6..568N} and the helium burning on the surface of a white dwarf \citep{2017Natur.550...80J}. The color evolution at early times can also be used to probe the location within the ejecta of $^{56}$Ni and other radioactive isotopes \citep{2014MNRAS.441..532D}. The studies of early-time color curves of SNe Ia indicate that two branches may exist, i.e., the red and blue branches \citep{2018ApJ...864L..35S, 2020ApJ...902...48B}. This implies that using them as standard candles for cosmic distance measurements can introduce systematic errors, thus affecting the accuracy of cosmic distance measurements. Additionally, there exist similar limitations in current SNe explosion models and the Hubble constant \citep[][and references therein]{2021A&ARv..29....9S}. Thus, a precise and detailed identification of early-time SNe Ia is crucial and indispensable.  

To address these challenges and further unravel the mysteries of the universe, an increasing number of large-scale survey telescopes are currently in operation or planned, including the Vera C. Rubin Observatory Large Synoptic Survey Telescope \citep[LSST;][]{2019ApJ...873..111I}, the Panoramic Survey Telescope and Rapid Response System (Pan-STARRS; \citealt{2004SPIE.5489...11K,2020ApJS..251....7F}), the Catalina Real-Time Transient Survey (CRTS; \citealt{2009ApJ...696..870D}), the Dark Energy Survey (DES; \citealt{2018ApJS..239...18A}), the Asteroid Terrestrial-impact Last Alert System (ATLAS; \citealt{2018PASP..130f4505T}), the Zwicky Transient Facility (ZTF; \citealt{2019PASP..131a8003M,2019PASP..131f8003B}), the 2.5-m Wide Field Survey Telescope (WFST; \citealt{2016SPIE10154E..2AL,2018AcASn..59...22S,2020SPIE11445E..4AL,2022MNRAS.513.2422L,LeiWFST,2022Univ....9....7H}), and so forth. These on-going and future surveys can generate enormous amounts of data, and trigger millions of real-time alerts each night. This presents both unprecedented opportunities for studying transients and new challenges in efficiently processing and analyzing vast datasets.

The observation of SNe can be accomplished through the utilization of both spectroscopic and photometric techniques. Spectroscopic observations provide accurate constraints but require longer observation time and cannot be applied in SNe searching from large-scale surveys. On the other hand, due to the high efficiency of photometric observations, more effects have focused on the early and rapid classification by photometric surveys. Various methods have been employed to analyze the observational data, including visual inspection, template matching, spectroscopic analysis, and machine learning, particularly deep learning-based methods \citep{2023RSOS...1021454S, 2023PASA...40....1H, 2023RASTI...2..441H}. The color of celestial objects contains rich information and can be used for measurement of stellar atmospheric parameters and accurate flux calibration \citep{2016A&A...595A.129A}. \citet{2002PASP..114..833P} pointed out that based on $V-R$ and $R-I$ colors, SNe Ia with redshift less than $0.1$ can be distinguished from the other SNe. Thus, the color and the color evolution can provide valuable information on celestial objects.

In the investigation of early-time photometric classification of SNe, \emph{\texttt{YSE}} \citep{2021ApJ...908..143J} aims to obtain well-sampled Pan-STARRS \emph{g,r,i,z} light curves of thousands of transient events, capable of discovering young transient events with a luminosity of about $21.5$ magnitude; \citet{2017ApJ...837L..28C} achieves high accuracy in classifying the SNe using deep recurrent neural networks with the data including redshift before the night of the sixth observation with S/N $>$ $4$. \emph{\texttt{RAPID}} \citep{2019PASP..131k8002M} uses deep recurrent neural networks with gated recurrent units (GRUs) to automatically identify transient phenomena from the initial alert to the entire light curve by using PLAsTiCC; \emph{\texttt{SuperNNova}} \citep{2020MNRAS.491.4277M} use Bayesian neural networks and can incorporate additional information to improve early time classification of SNe Ia and CCSNe, but require between $2.4 \pm 1.2$ and $3.3 \pm 1.4$ photometric epochs in average to start accurately classifying SNe. \emph{\texttt{ALeRCE}} \citep{2021AJ....161..242F, 2021AJ....162..231C, 2021AJ....161..141S}, is a broker system light curve classifier for processing the alert stream from the ZTF by employing a balanced random forest method based on feature extraction; \citet{2022AJ....163...57Q} proposed a photometric classifier,  \emph{\texttt{SCONE}}, based on convolutional neural network (CNN) using wavelength-time heatmaps, and achieving good classification results for early-time SNe light curves. \emph{\textbf{Fink}} \citep{2022A&A...663A..13L} utilizes feature extraction and active learning to identify early-stage SNe Ia.
Recurrent neural network and its variants, including Long Short-Term Memory (LSTM) \citep{hochreiter1997_LSTM}, Gated Recurrent Unit (GRU) \citep{2014arXiv1412.3555C,cho-etal-2014-learning},   Bidirectional Long Short-Term Memory (BiLSTM) network, have been shown to be particularly powerful where sequential data are accompanied by a set of discrete labels \citep{2017ApJ...837L..28C, 2020arXiv200906109C, 2020ApJ...898...46D, 2020arXiv200612333C, 2021A&C....3600488C, 2021PhRvD.104f4046C, 2022ApJS..260...16A}. However, until now, the overwhelming majority of the previous work has focused on the single-band light curve for the classification of SNe. As illustrated here, only a few studies aim to promptly identify early-time SNe Ia after the explosions. Multi-band simultaneous observational real-color information has not yet been explored for early-time classification.

This paper is organized as follows. 
A brief overview of Mephisto will be provided in Section \ref{sect:telescope}. 
We elaborate on our utilization of the SNCosmo \citep{Barbady2015sncosmo, Barbady2016sncosmo} to simulate the data in section \ref{sect:data}. The deep learning architectures based on RNN, LSTM, GRU, and BiLSTM, as well as the model assessment metrics, are outlined in section \ref{sect:Methods}. Our findings are presented in section \ref{sect:results}. Additional discussion is provided in Section \ref{sect:discussion}. In section \ref{sect:summary}, we encapsulate the culmination of our research findings. We implement a flat $\Lambda$CDM standard cosmology with $\Omega_{\Lambda}$ = $0.7$, $\Omega_{M}$ = $0.3$, $H_{0} = 70.0 \, km \, s^{-1} \, Mpc^{-1}$ in our simulations, and assume the comoving volumetric rate of SNe is $10^{-4}\, yr^{-1}\, Mpc^{-3}$~\citep{2012ApJ...745...31B}. 

\section{Multi-channel Photometric Survey Telescope}
\label{sect:telescope}
\begin{figure}
\centering
\includegraphics[width=0.8\textwidth, angle=0]{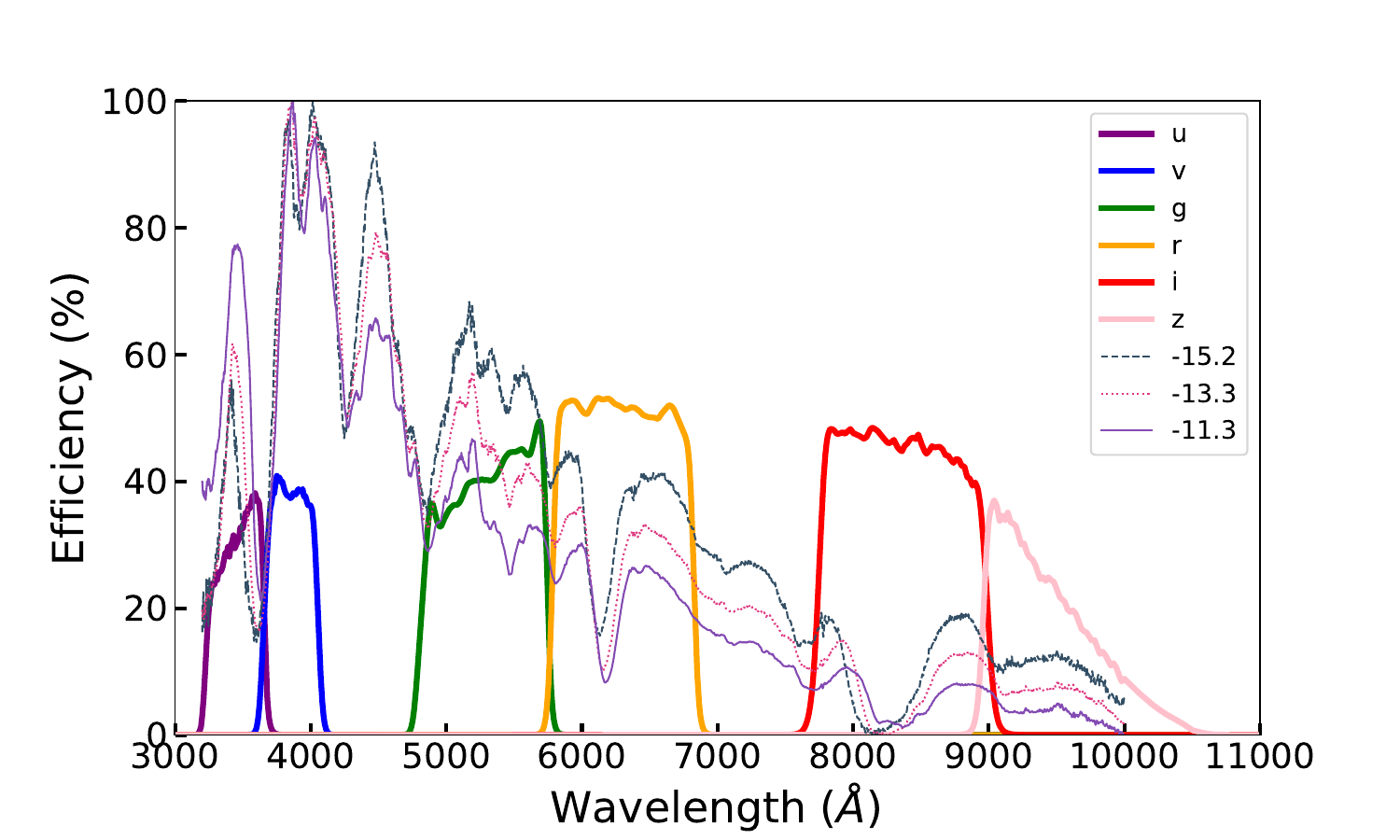}
\caption{\label{fig:mephisto}{Mephisto transmission curve of~\emph{u, v, g, r, i, z} filters combined with early spectra of SN~2011fe~\citep{2013A&A...554A..27P}. The legend indicates SN~2011fe spectra corresponding to $-15.2$ day, $-13.3$ day, and $-11.3$ day before reaching its peak luminosity.}}
\end{figure}

The Multi-channel Photometric Survey Telescope (Mephisto\footnote{\url{http://www.mephisto.ynu.edu.cn/}}; \citealt{Yuan+etal+2020})  is a wide-field of view telescope with a $1.6$-meter primary mirror located at the Lijiang Observatory in Lijiang City, Yunnan Province, China. It is equipped with three CCD cameras, each with a field of view of $2$ square degrees, capable of simultaneously imaging the same patch of sky region in three different bands (\emph{u, g, i} or~\emph{v, r, z}). Fig.~\ref{fig:mephisto} presents the transmission curves of the Mephisto \emph{u, v, g, r, i, z} filters. For comparison, we also over-plot the early spectra of one classical well-observed Type Ia SN~2011fe \citep{2013A&A...554A..27P}. For the purpose of facilitating visual comparison, we have re-scaled the spectra of the SN.

Mephisto survey mode, with sampling intervals ranging from days (D) to minutes (M), allows for systematic searches and studies of various types of explosive events in the universe \citep[e.g.][]{2024RAA....24d5017W}. 
In this work, we investigate the role of real-time color information in SNe classification with an ideal case of 1-day cadence.

\section{Simulation}
\label{sect:data}

With the objective of preparing the training sample, we simulate the photometric light curve of SNe for \emph{u, v, g, r, i, z} filters of Mephisto using SNCosmo\footnote{\url{https://sncosmo.readthedocs.io/en/stable/index.html}}, a python package for SN cosmology analysis. 
Different SNe parameters in our simulation are adopted from Open Supernova Catalog \citep[OSC,][]{2017ApJ...835...64G}. The corresponding number of SNe in different class (in column $N^{b}$), average peak magnitude (in column $M_{peak}^{b}$) and standard deviation of the peak magnitude (in column $\sigma_{M}^{b}$) are derived from all bands in OSC mentioned in Table \ref{tab:para_samples}. $t_{0}$ column denotes the average rise time~\footnote{All the rising time in this work refers to that in the rest frame}, we employ a Gaussian distribution across distinct categories. The $models$ column indicates the number of built-in SN models in SNCosmo. The last column represents the overall number of samples for training, validation, and testing, which is obtained from simulations and the subsequent filtering. The selection criteria are described later. For a typical SN Ia, the average rising time of SNe Ia is about $20$ days after the explosion. However, in \citet{2019ApJ...886..152Y}, among a sample of $127$ SNe, $50$ are detected at least $14$ days prior to the peak of the light curve, and with a subset of $9$ events being detected more than $17$ days before the \emph{g-ZTF} band peak. In \citet{2020ApJ...902...47M}, the mean rise time of SNe Ia is estimated to be $\sim$\,$18.9$ days. Other rise times of SNe Ia have also been presented in other work \citep{2015MNRAS.446.3895F, branchSupernovaExplosions2017}, which is between $17$ and $20$ days. Since our work focuses on the early time classification of SNe Ia, we only select samples with a minimum rise time of $17$ days in the light curves after the explosion for model training and testing. 

\begin{table}
\begin{center}
\begin{threeparttable}
\caption[]{Build-in model used parameters and samples for our model}
\label{tab:para_samples}
\setlength{\tabcolsep}{3.3mm}
\begin{tabular}{l|c|c|c|c|c|c|c}
  \hline\noalign{\smallskip}
    Class & supernovae &  $N^{b}$  & $\rm M^{b}_{peak}$ & $\sigma^{b}_M$ & $\rm t_{0}$ & models&samples\\
  \hline\noalign{\smallskip}
      TNSNe$^{a}$ &Ia & 6995 & -19.03 & 0.86 & $\mathcal{N}(20,1)$ & 15 & 2141\\
      \hline
      \multirow{9}{*}{CCSNe} &Ib & 110 & -17.44 & 1.128 & $\mathcal{N}(20,1)$& 34 & \multirow{9}{*}{6420}\\
      &Ic & 139 & -17.70 & 1.727 & $\mathcal{N}(20,5)$& 31& \\
      &Ib/c & 56 & -17.71 & 1.189 & $\mathcal{N}(20,5)$& 2& \\
      &Ic BL & 21 & -18.57 & 3.616 & $\mathcal{N}(20,1)$& 12& \\
      &II & 1513 & -17.75 & 1.780 & $\mathcal{N}(20,1)$& 46& \\
      &IIb & 69 & -17.72 & 1.189 & $\mathcal{N}(20,1)$& 24& \\
      &IIn & 188 & -18.84 & 1.60 & $\mathcal{N}(20,1)$& 15& \\
      &IIP & 147 & -17.02 & 1.250 & $\mathcal{N}(15,1)$& 28& \\
      &IIL & 3 & -18.60 & 0.086 & $\mathcal{N}(13,1)$& 1& \\
    \noalign{\smallskip}\hline
\end{tabular}
    \begin{tablenotes}    
    \footnotesize  
    \item[a] TNSNe denotes thermonuclear supernovae.
    \item[b] The data derived from all bands in the OSC statistics.	
    \end{tablenotes}           
  \end{threeparttable} 
\end{center}
\end{table}

\begin{figure}
\centering
\includegraphics[width=0.475\textwidth]{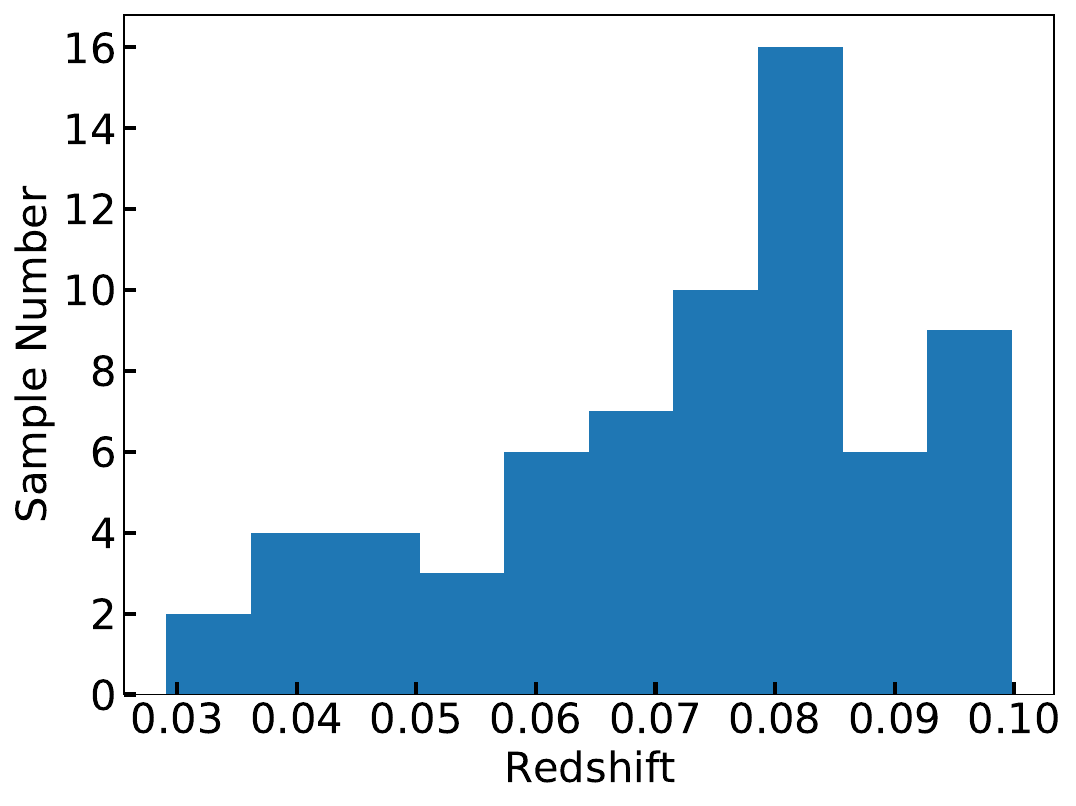}
\includegraphics[width=0.49\textwidth]{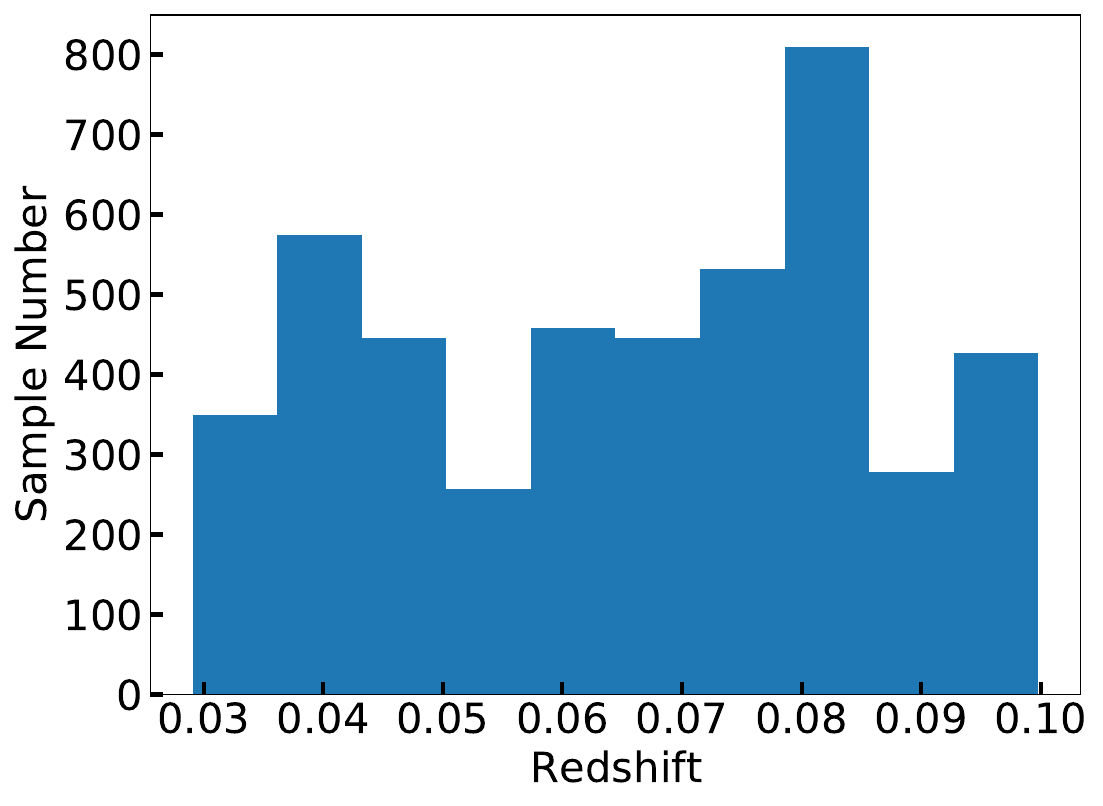}
\caption{Left: the initial redshift distribution of SNe generated by SNCosmo \emph{zdist} function, given the input volumetric SN rate, the cosmology, and the observed area and time. Right: the redshift distribution of the train sample which has a three-day observation before the peak in the light curve.}
\label{fig:redshift} 
\end{figure}

\begin{figure}[h]
\centering
  \begin{minipage}[t]{0.415\linewidth}
  \centering
   \includegraphics[width=\textwidth]{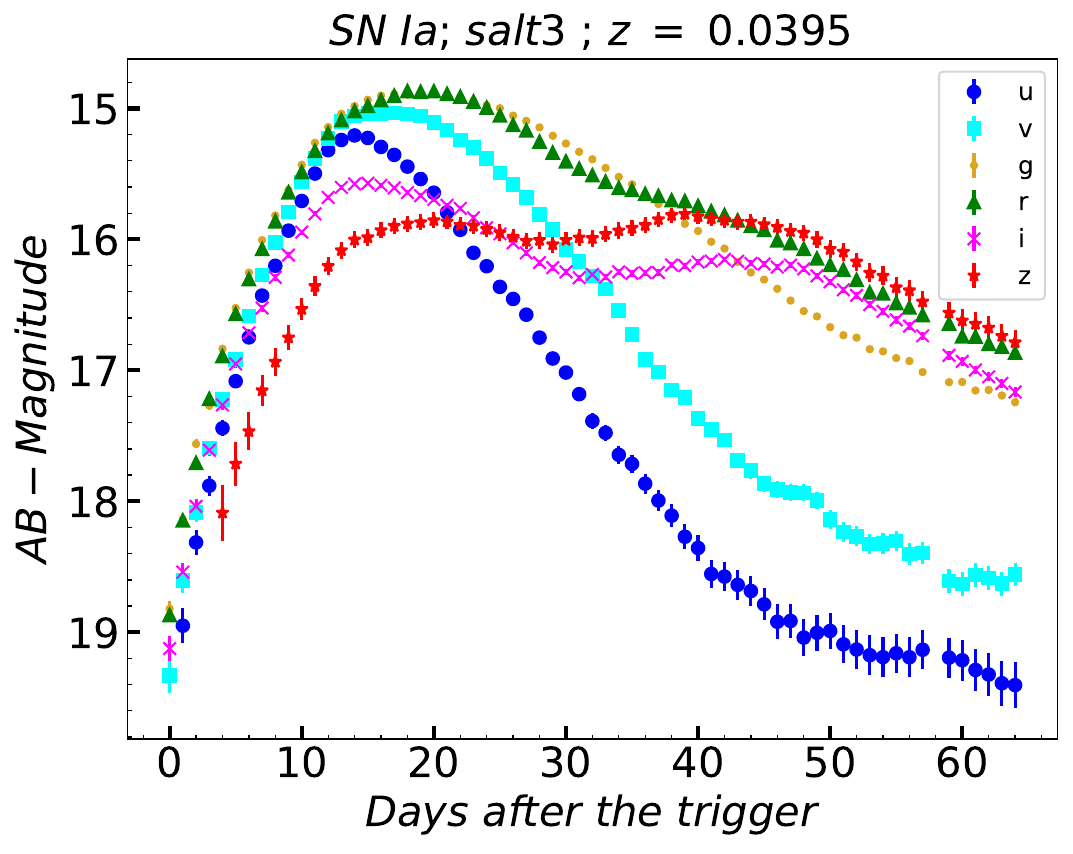}
  \end{minipage}
  \begin{minipage}[t]{0.445\textwidth}
  \centering
   \includegraphics[width=\textwidth]{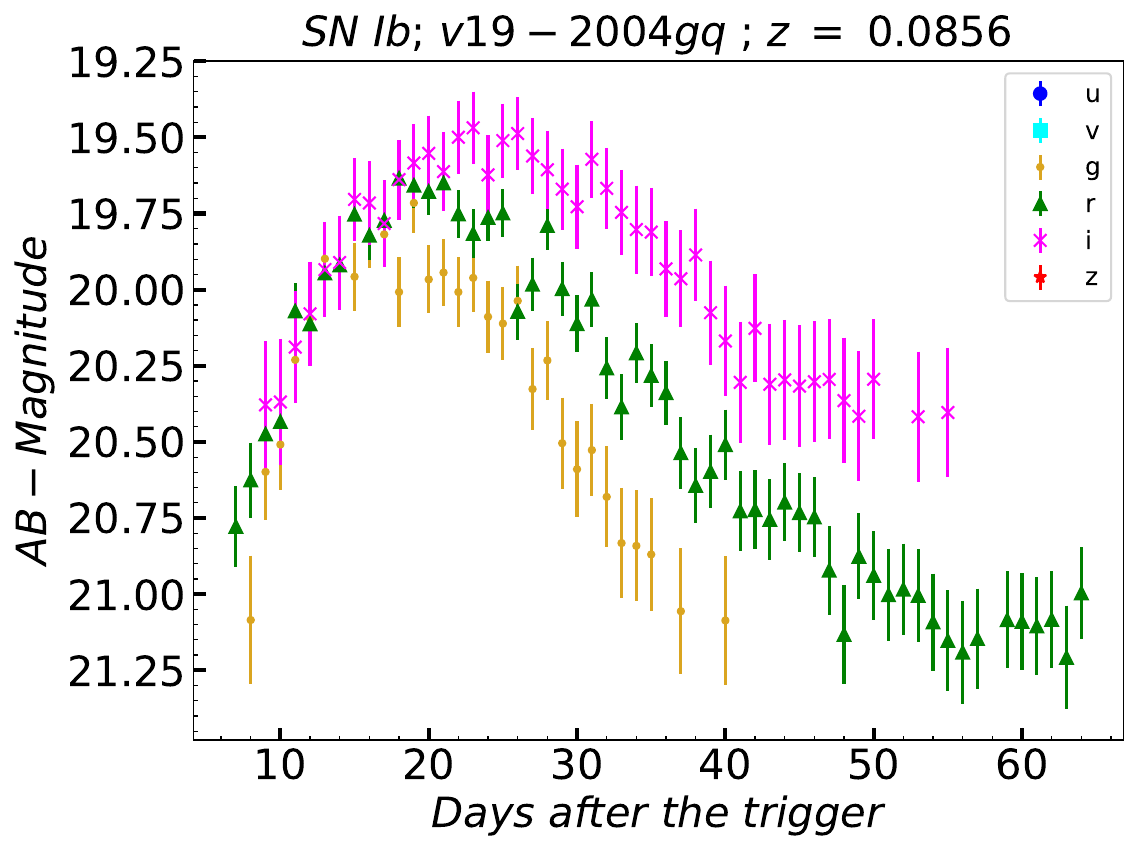}
  \end{minipage}
 \quad
  \begin{minipage}[t]{0.435\textwidth}
  \centering
   \includegraphics[width=\textwidth]{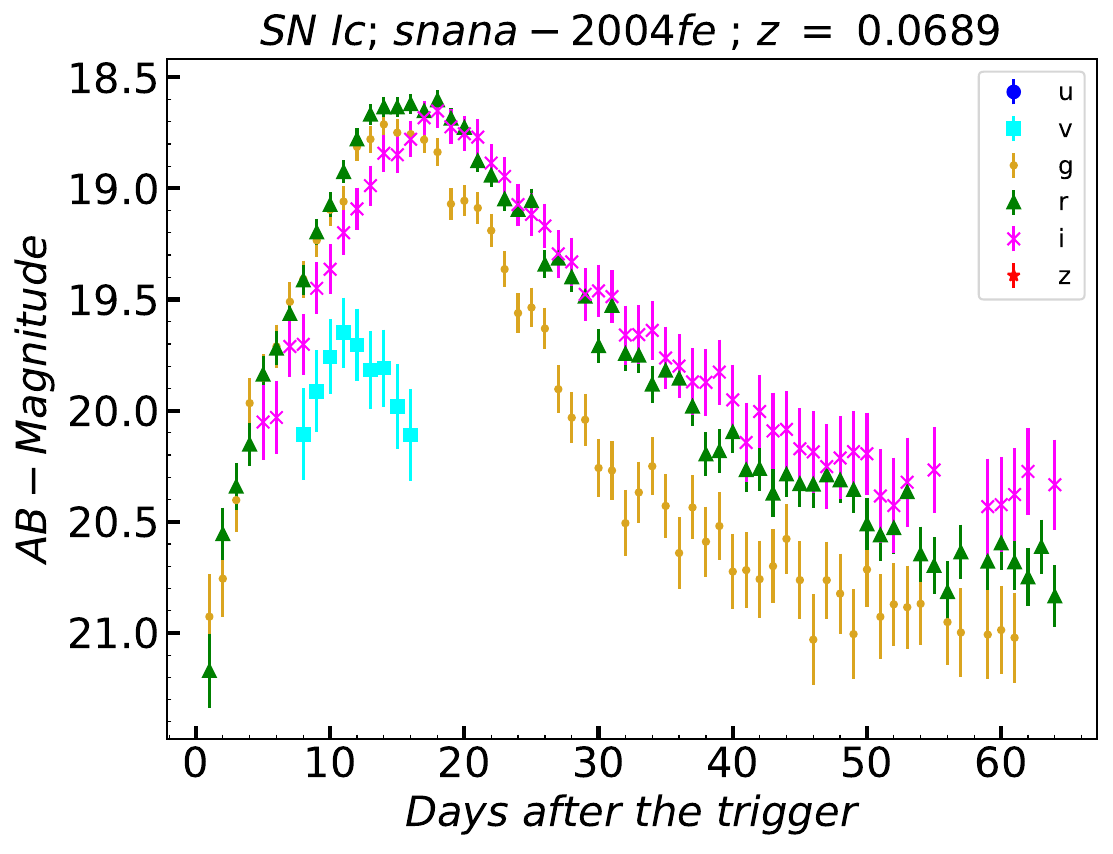}
  \end{minipage}
  \begin{minipage}[t]{0.435\textwidth}
  \centering
   \includegraphics[width=\textwidth]{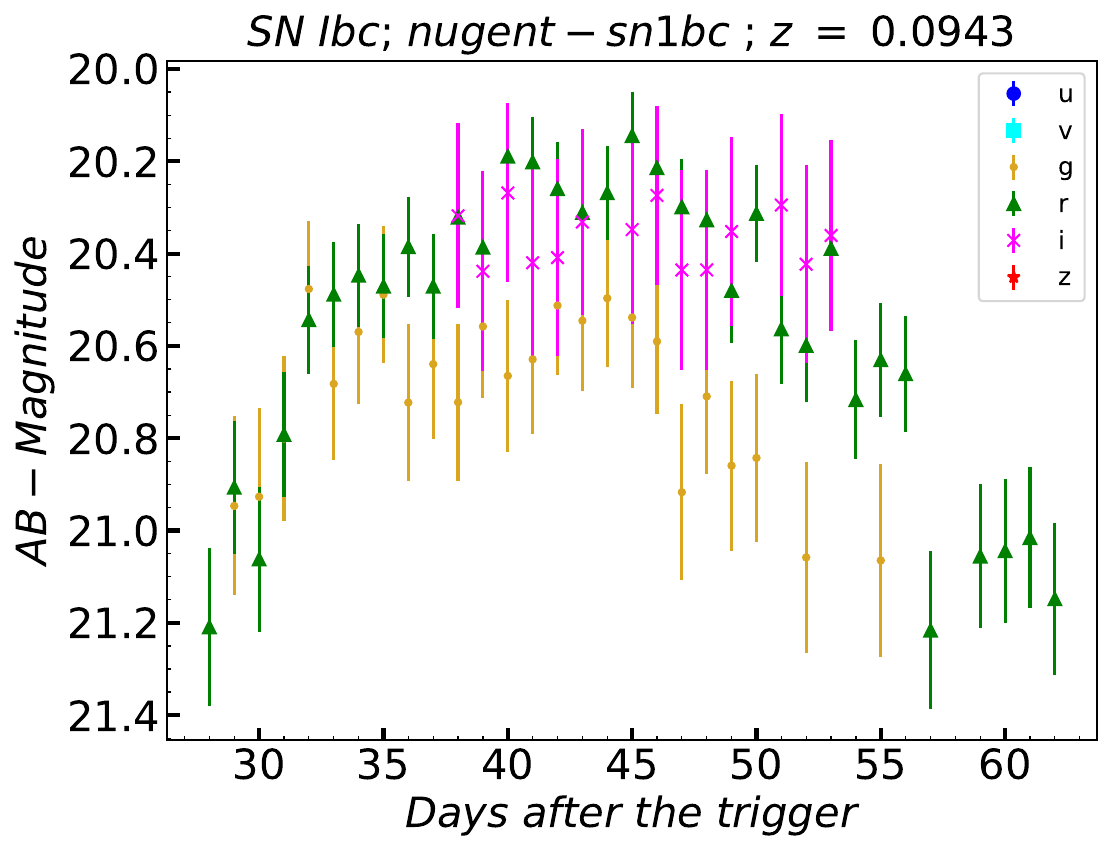}
  \end{minipage}
 \quad
  \begin{minipage}[t]{0.425\textwidth}
  \centering
   \includegraphics[width=\textwidth]{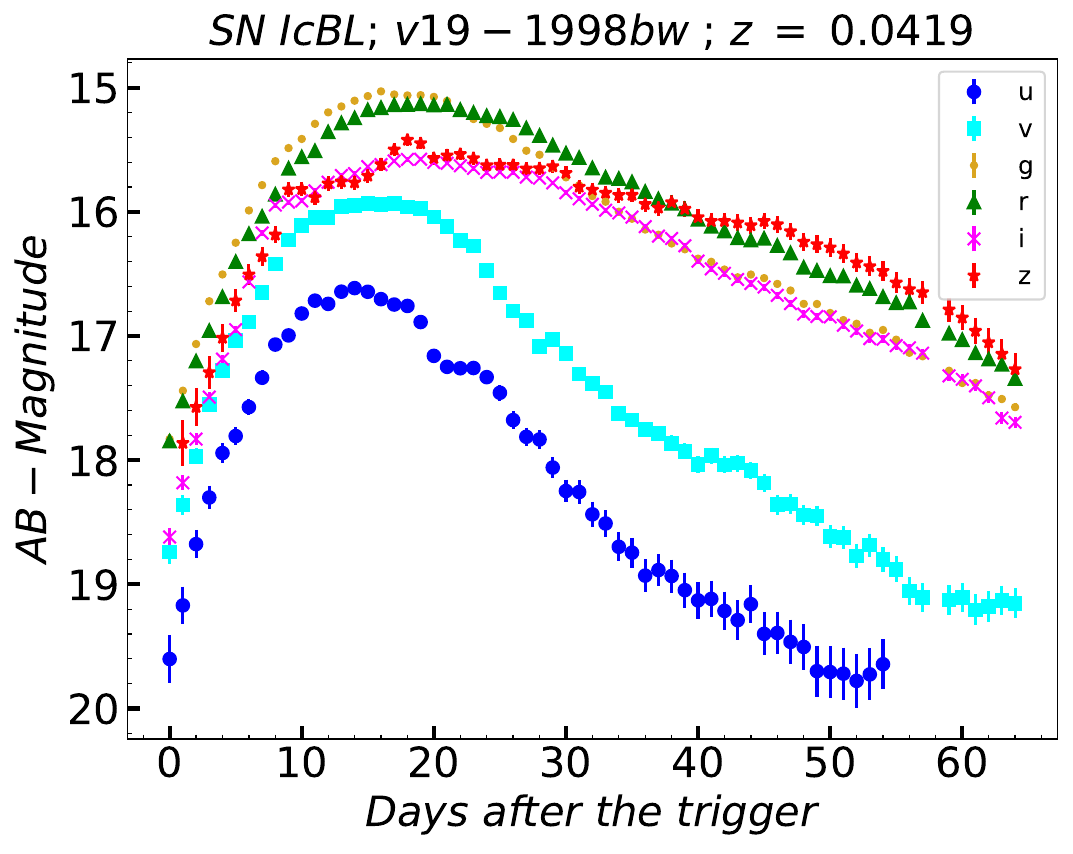}
  \end{minipage}
  \begin{minipage}[t]{0.445\textwidth}
  \centering
   \includegraphics[width=\textwidth]{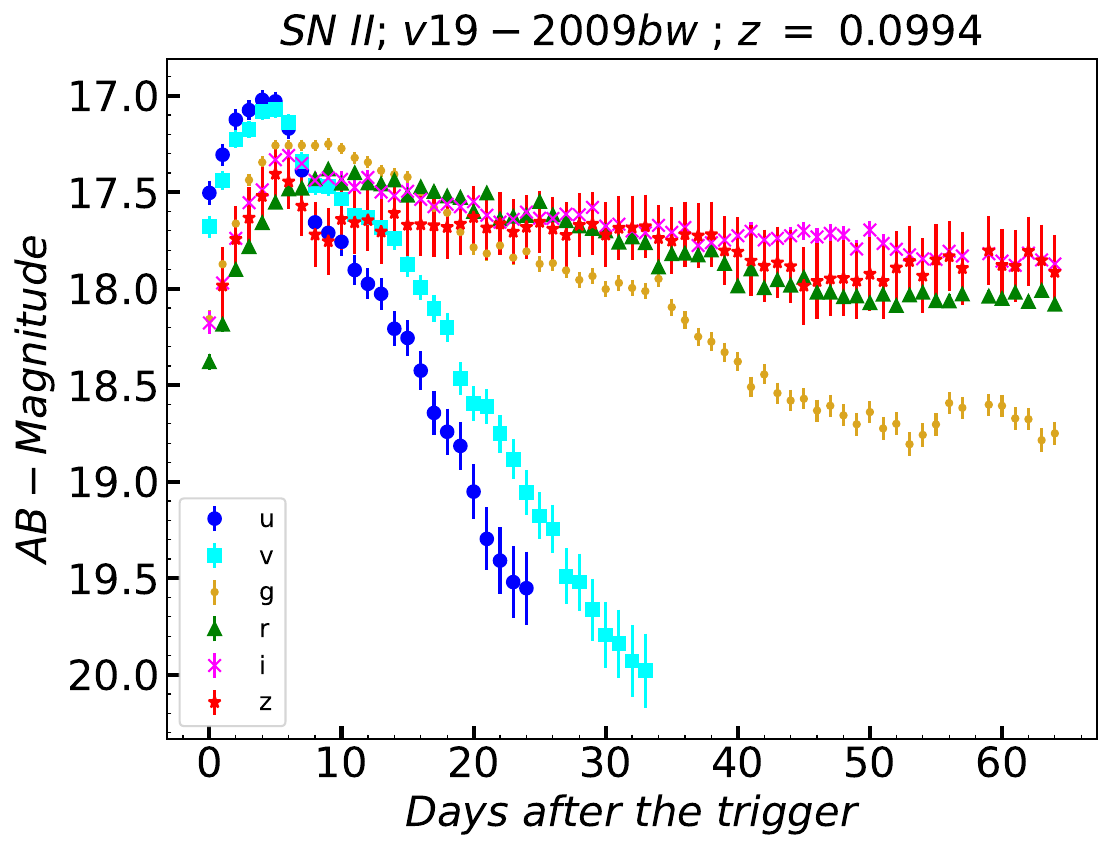}
  \end{minipage}
  \quad
 \begin{minipage}[t]{0.435\textwidth}
  \centering
   \includegraphics[width=\textwidth]{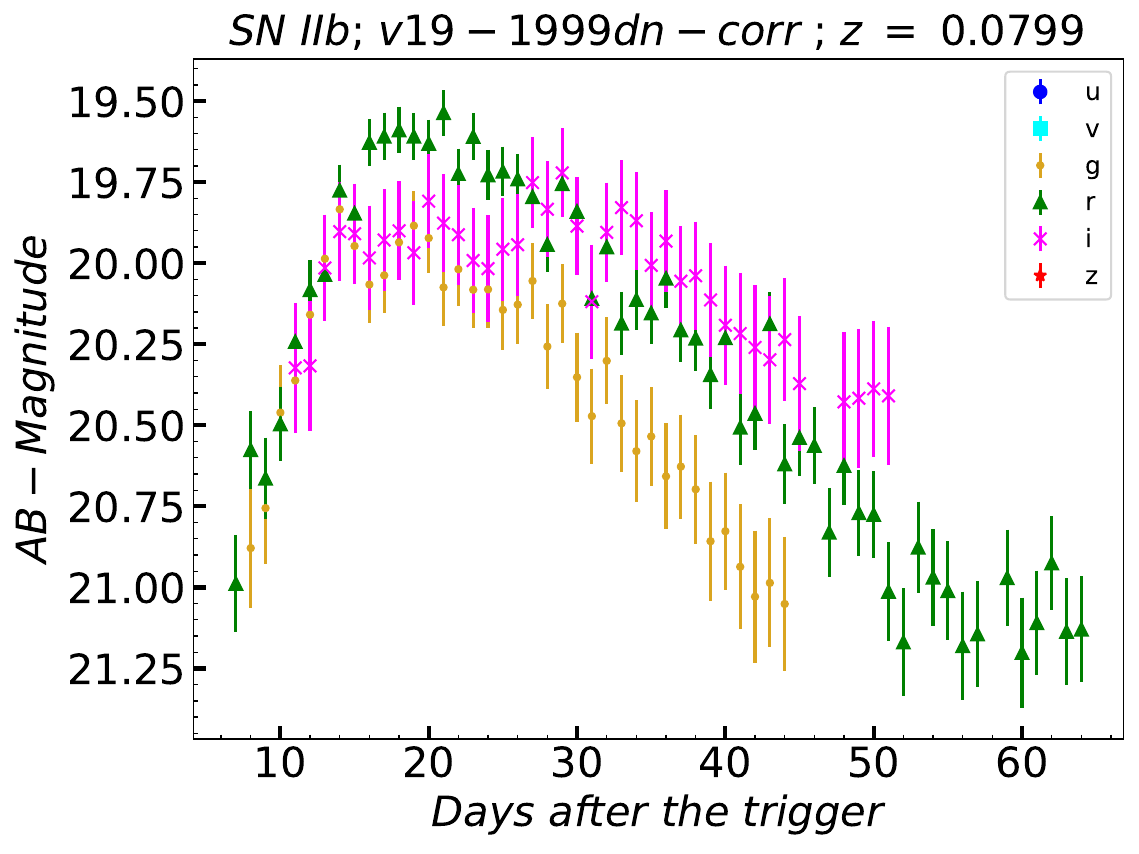}
  \end{minipage}
    \begin{minipage}[t]{0.435\textwidth}
  \centering
   \includegraphics[width=\textwidth]{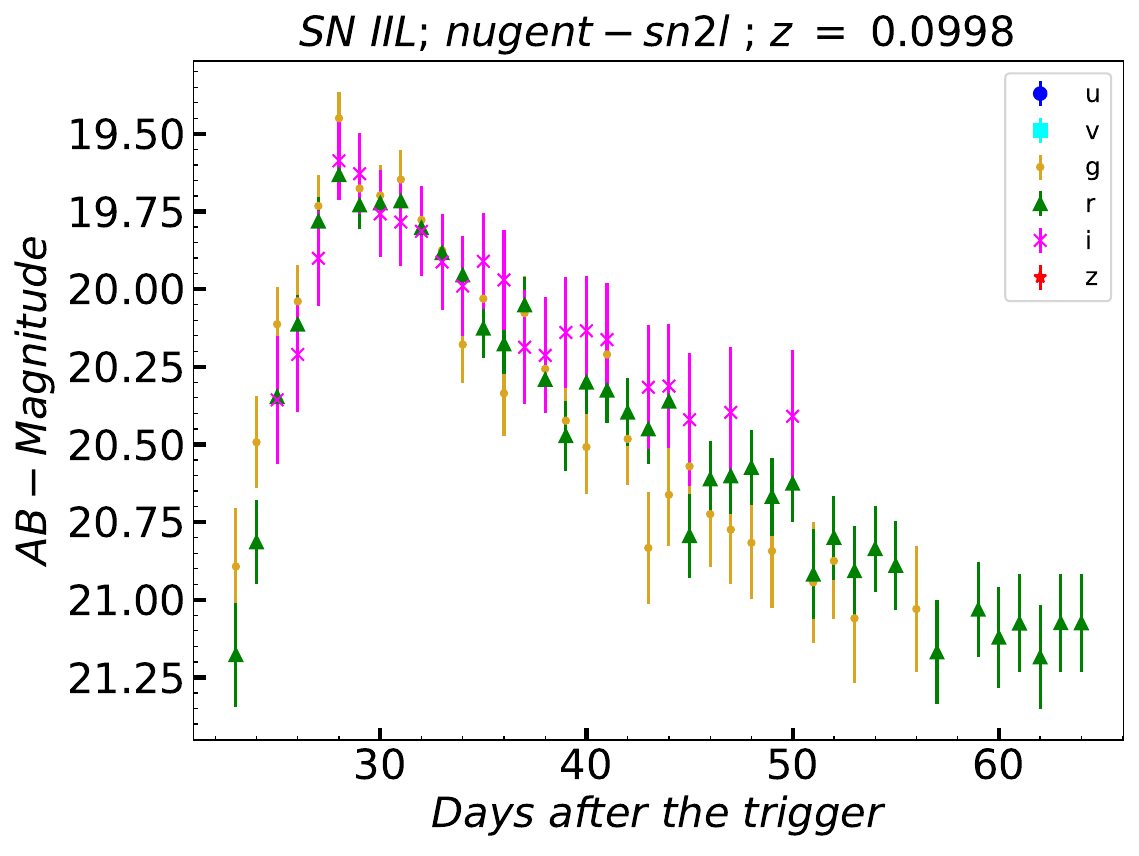}
  \end{minipage}
\caption{A sample of the SNe light curves from Mephisto is plotted with different symbols. The error bars are also indicated. The observation points with low SNR ($<5$) are excluded (see Section~\ref{sect:data}).}
\label{fig:meph_lc}
\end{figure}

\begin{figure}[h]
\centering
  \begin{minipage}[t]{0.415\linewidth}
  \centering
   \includegraphics[width=\textwidth]{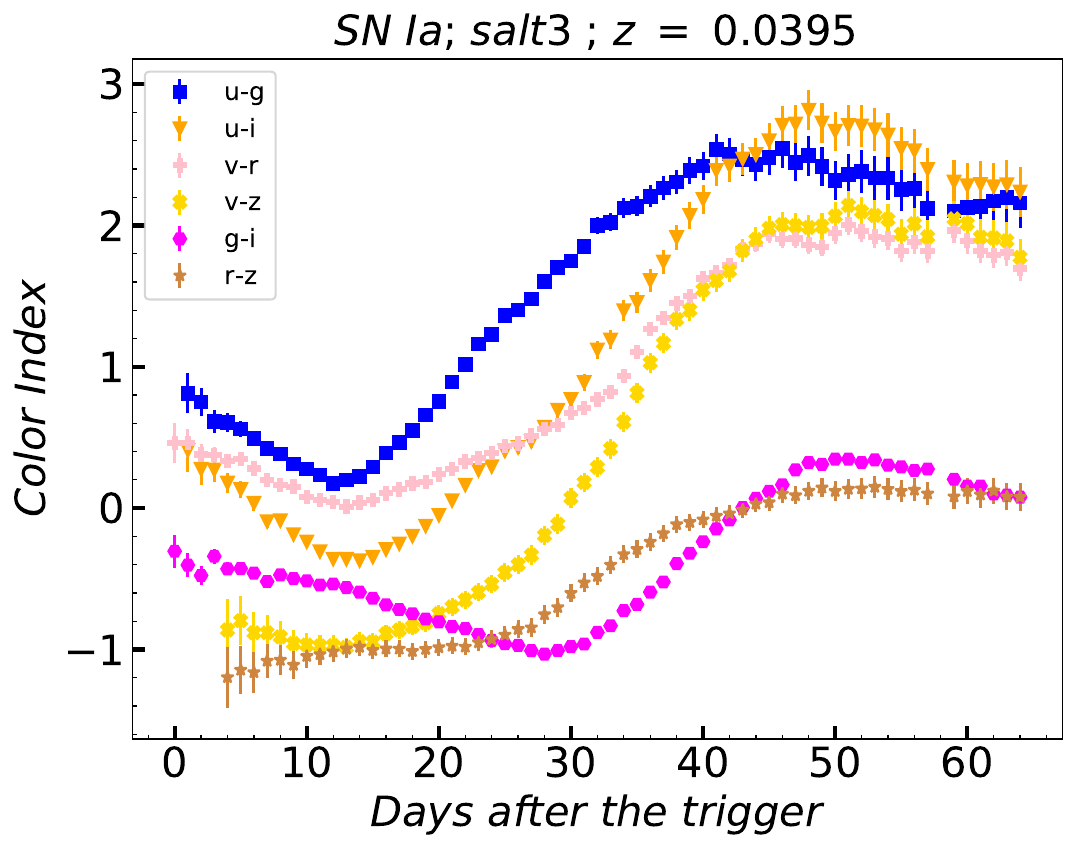}
  \end{minipage}
  \begin{minipage}[t]{0.435\textwidth}
  \centering
   \includegraphics[width=\textwidth]{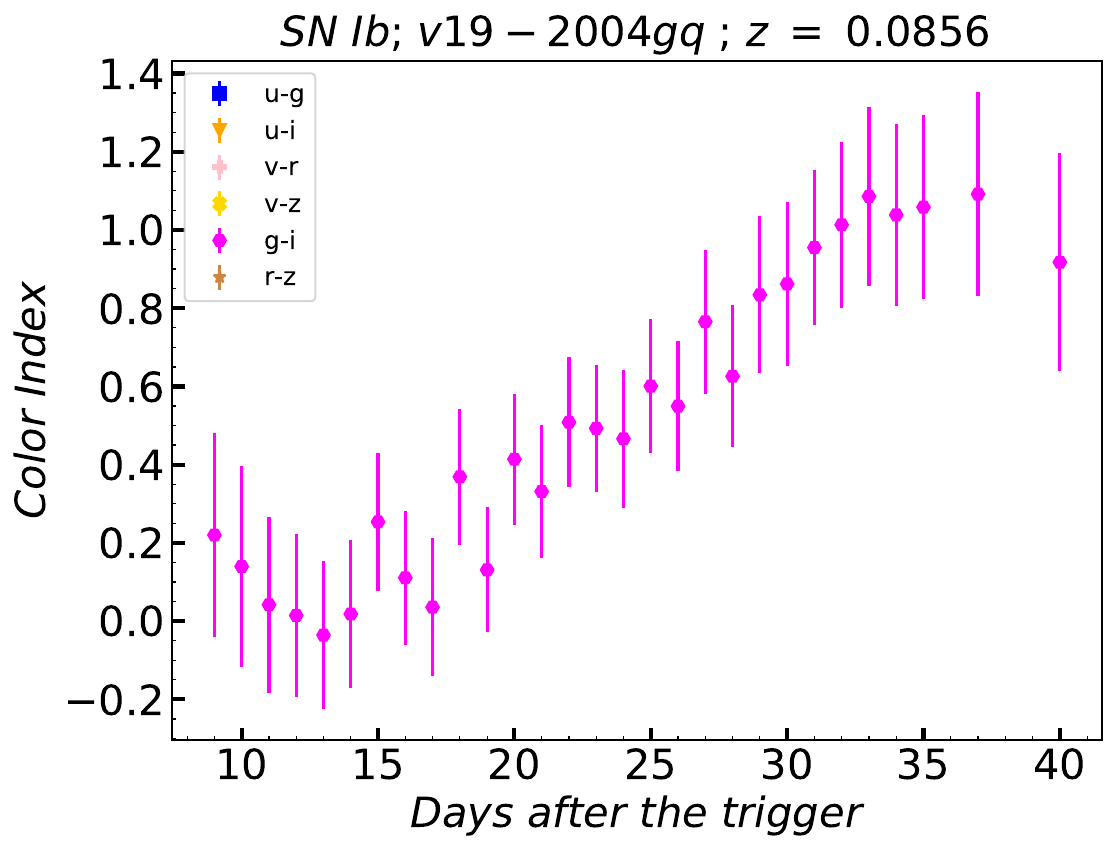}
  \end{minipage}
  \quad
  \begin{minipage}[t]{0.435\textwidth}
  \centering
   \includegraphics[width=\textwidth]{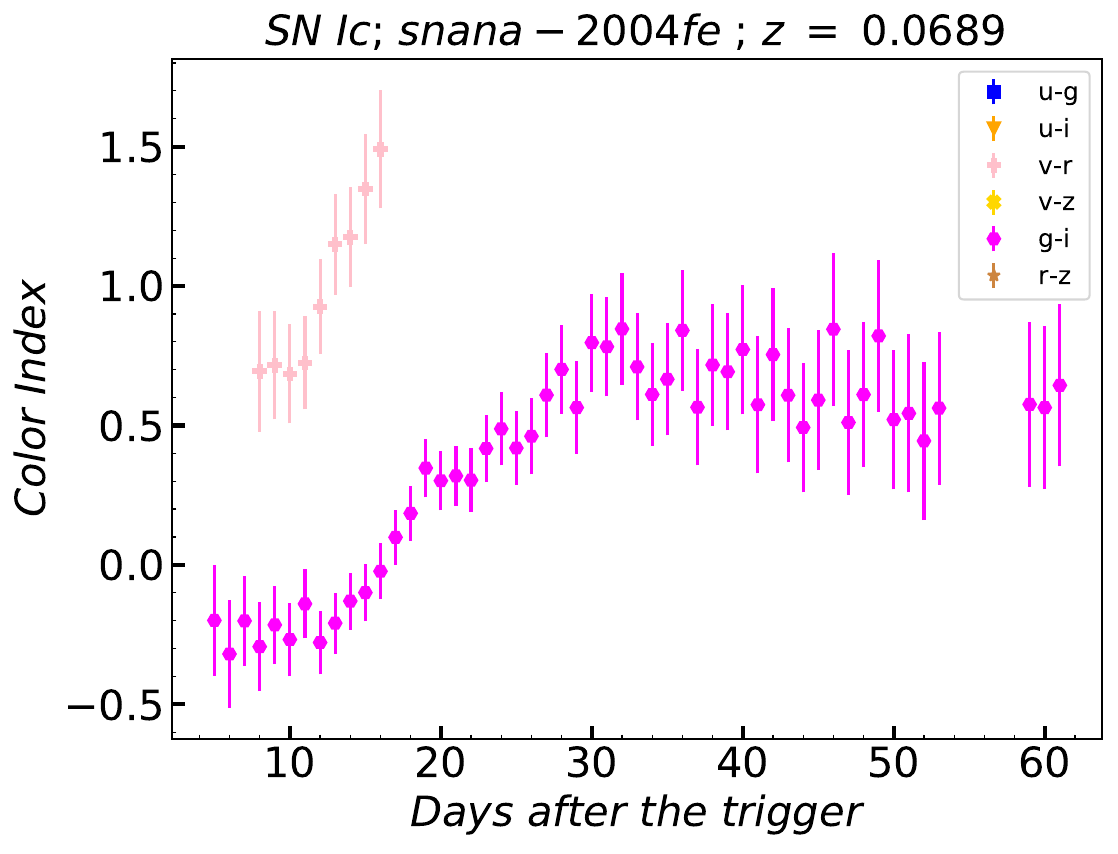}
  \end{minipage}
  \begin{minipage}[t]{0.435\textwidth}
  \centering
   \includegraphics[width=\textwidth]{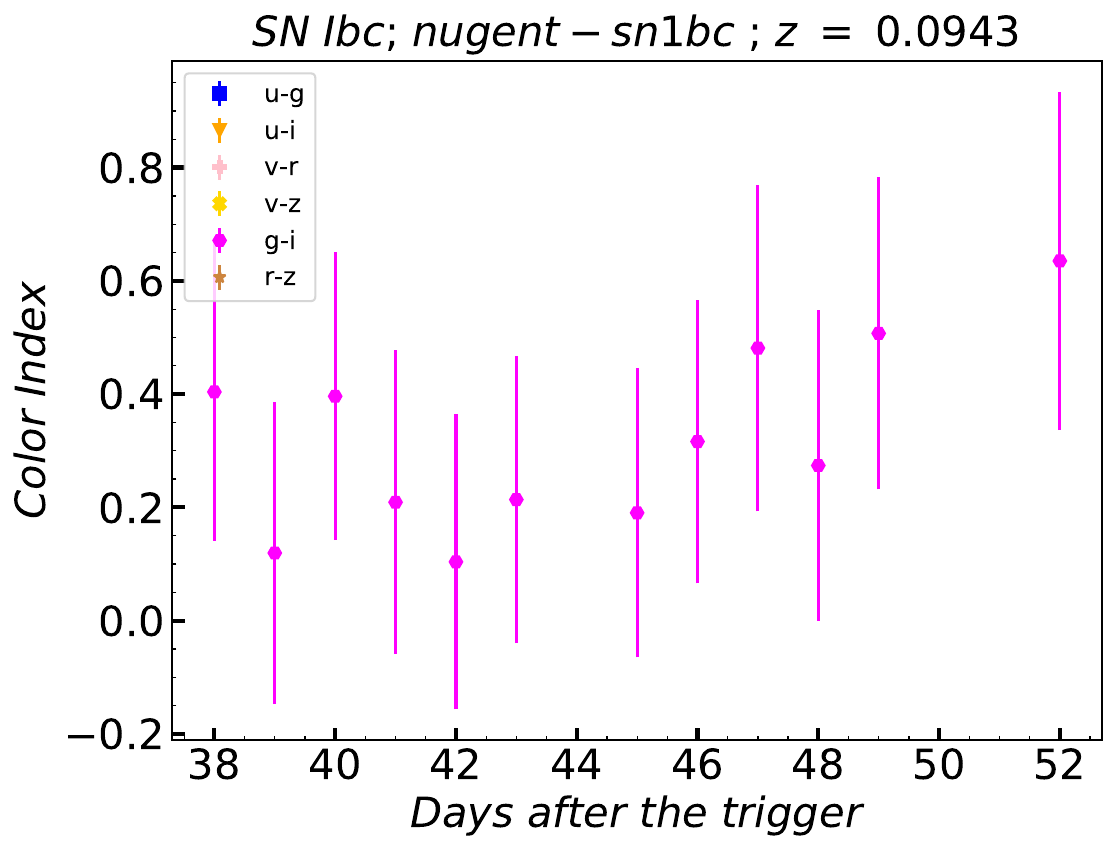}
  \end{minipage}%
  \quad
  \begin{minipage}[t]{0.435\textwidth}
  \centering
   \includegraphics[width=\textwidth]{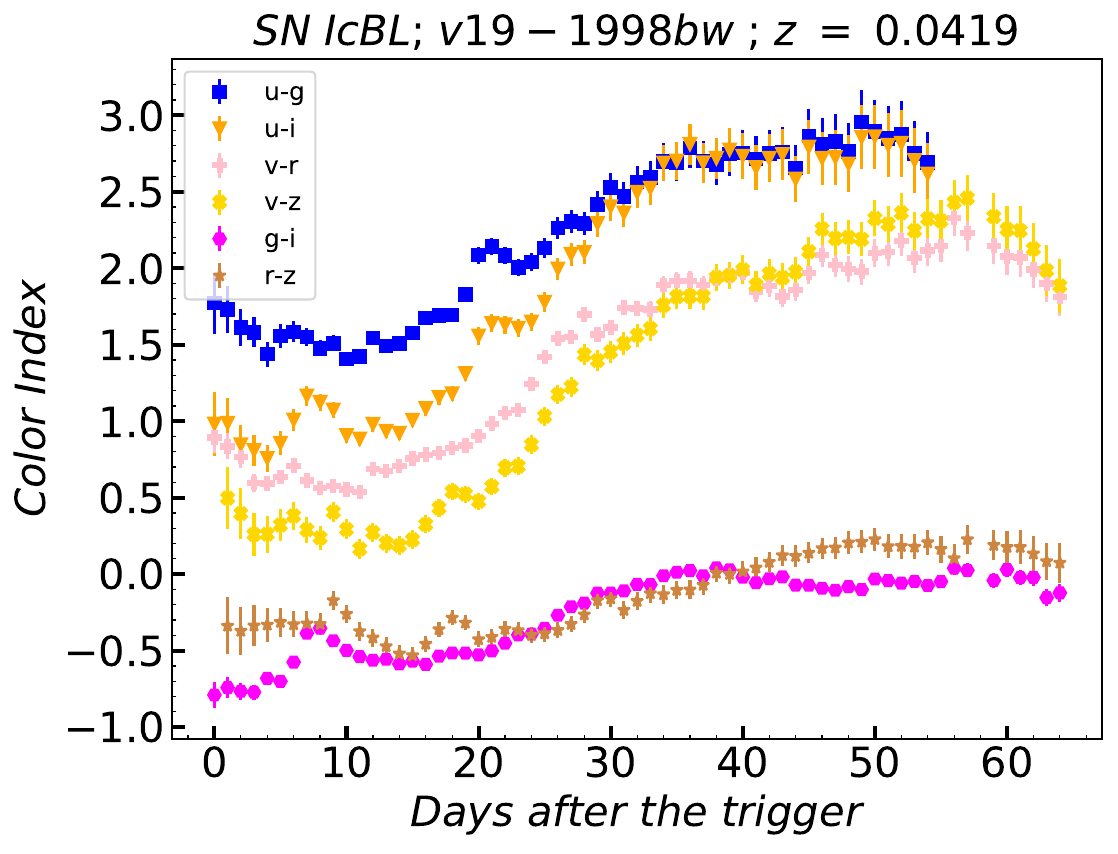}
  \end{minipage}
  \begin{minipage}[t]{0.435\textwidth}
  \centering
   \includegraphics[width=\textwidth]{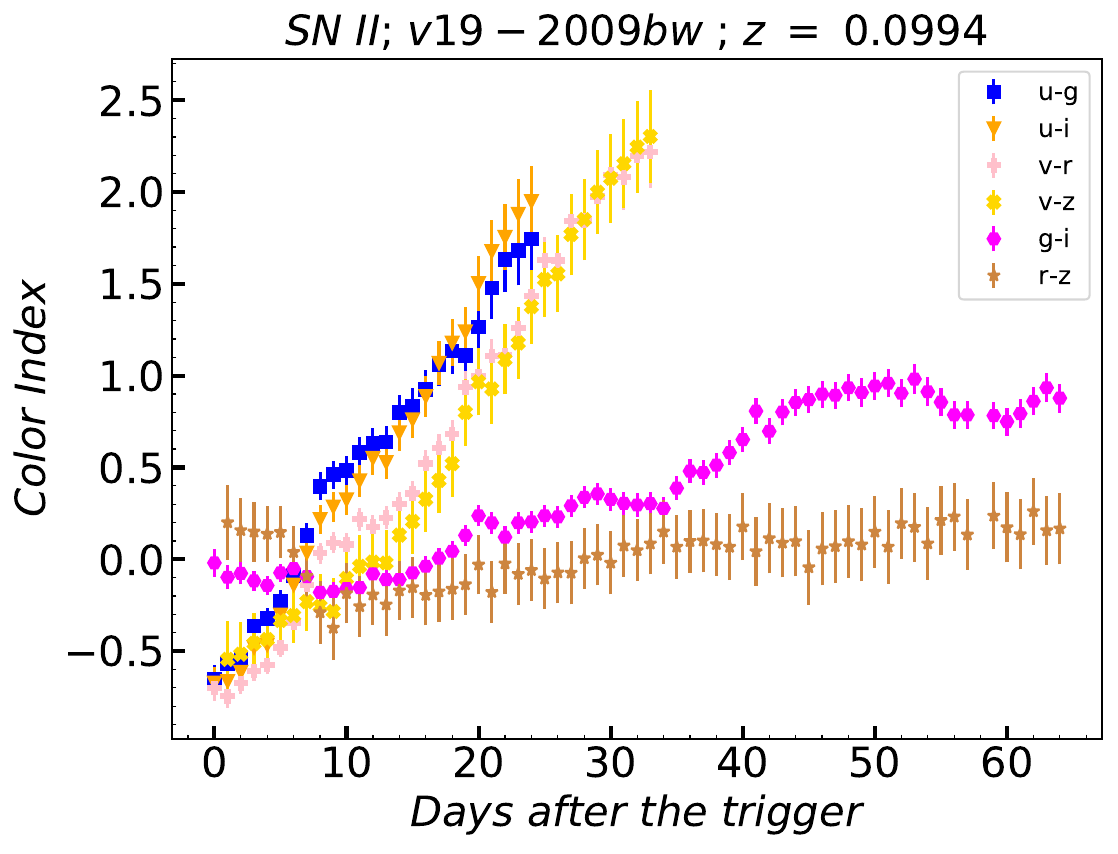}
  \end{minipage}
  \quad
  \begin{minipage}[t]{0.435\textwidth}
  \centering
   \includegraphics[width=\textwidth]{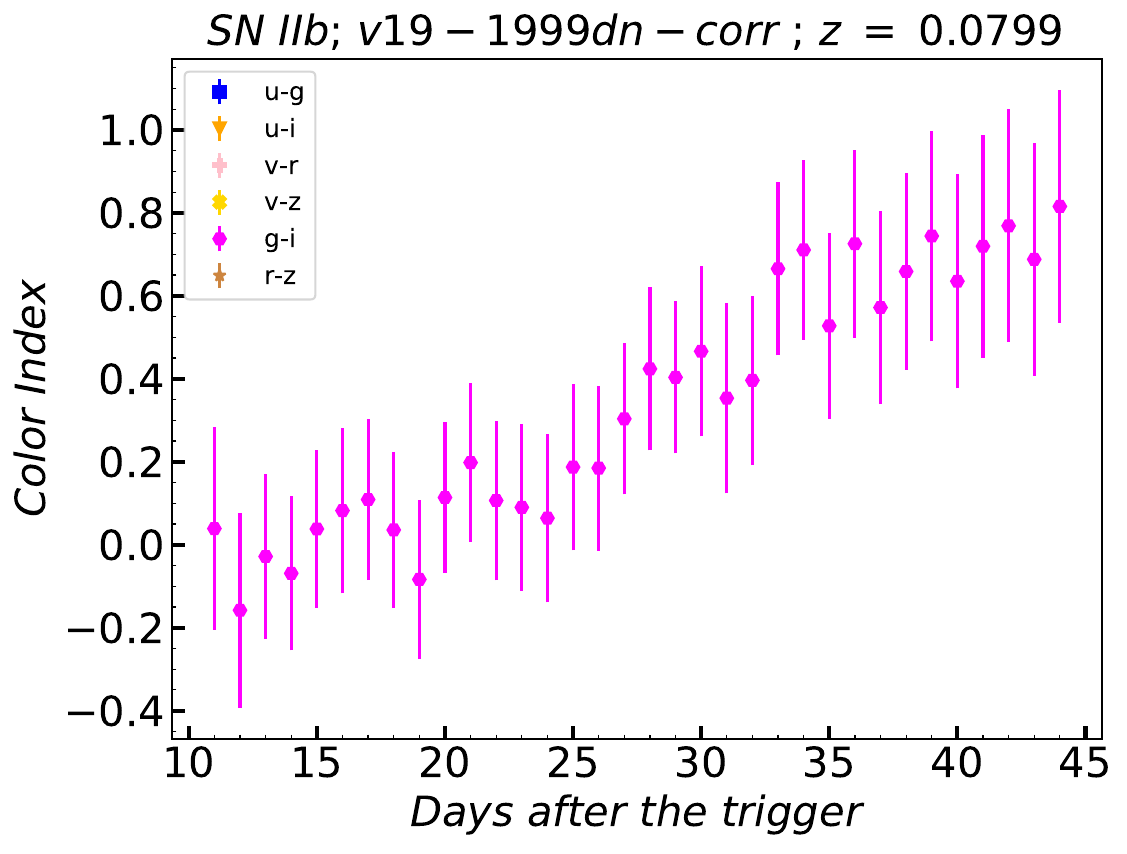}
  \end{minipage}
    \begin{minipage}[t]{0.435\textwidth}
  \centering
   \includegraphics[width=\textwidth]{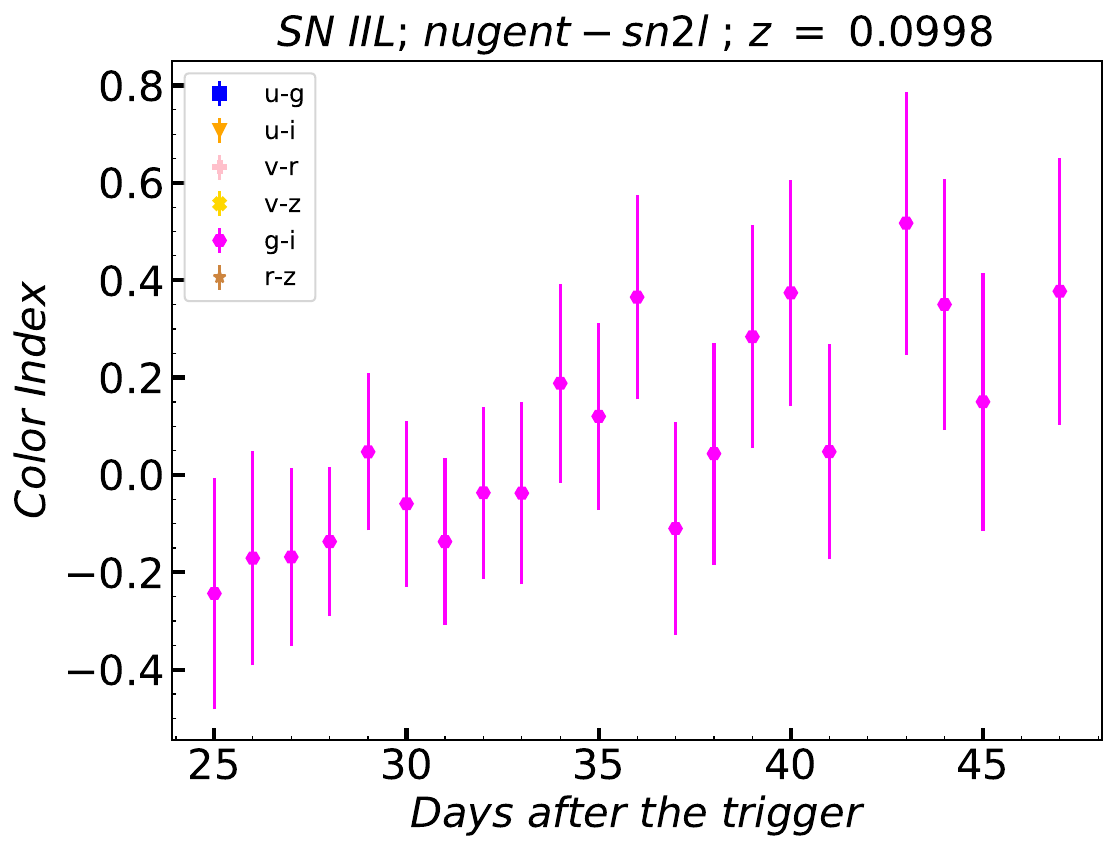}
  \end{minipage}
\caption{The evolution of color indices for different types of SNe are plotted with different symbols (see Section~\ref{sect:data})}
\label{fig:meph_color_lc}
\end{figure}

We concentrate on SNe with a redshift range of [$0$, $0.1$] owing to the limiting magnitude of the Mephisto. To generate the redshift values for the observed SNe, we utilize the built-in~\emph{zdist} function in SNCosmo.
The resulting redshift distribution is illustrated in the left panel of Fig.~\ref{fig:redshift}. We perform further selection for different train sub-samples. The right panel of Fig.~\ref{fig:redshift} shows that of the sample, which contains three days of observations before the peak of the light curve. The redshift distribution of other sub-samples does not show much difference. SNCosmo offers a rich variety of SN models, including $15$ sub-models specifically for SNe Ia, such as the widely used SALT2 and SALT3 models. In this simulation, we adopt the parameters of SALT2 as the fundamental parameter settings for all other models. Additionally, as our emphasis lies in the detection of early-time SNe Ia explosions, the necessity for extended light curve observations post-peak luminosity is obviated. In addition to SNe Ia, we also incorporate other types of SNe provided by SNCosmo, including SNe Ib, Ic, Ib/c, Ic-BL, IIP, IIL, IIb, and IIn.
By including these diverse types of SN models, the simulated light curves of SNe within the same class become more varied. The number of submodels corresponding to each type of SN is listed in the table~\ref{tab:para_samples} column \texttt{models}. As for SNe Ia, we adopt hsiao \citep{2007ApJ...663.1187H}, hsiao-subsampled \citep{2007ApJ...663.1187H}, salt2 \citep{2007A&A...466...11G,2008ApJ...674...51E,2010A&A...523A...7G,2014A&A...568A..22B,2021MNRAS.504.4111T}, salt2-extended \citep{2018PASP..130k4504P,2018ApJ...867...23H}, salt3 \citep{2021ApJ...923..265K}, salt2-extended-h17 \citep{2018ApJ...867...23H}, nugent-sn91bg \citep{2002PASP..114..803N}, mlcs2k2 \citep{2007ApJ...659..122J} models. After obtaining the simulated flux of the celestial object, we utilize equation~\ref{eq:flux2mag} for the conversion of flux to magnitude.
\begin{equation}
\bm{mag} = -2.5*\log(flux) + 25.0,
\label{eq:flux2mag}
\end{equation}
where it is assumed that the zero point is 25.0, the default value built-in SNCosmo. Based on the pilot observation and weather monitoring at Lijiang Observatory (GMG; ~\citealt{2020RAA....20..149X}), we simulate the measurement error in the light curve of SNe. The signal-to-noise ratio (SNR) is calculated by equation~\ref{eq:flux_err}.
\begin{equation}
\bm{SNR} = \frac{n_{target}}{\sqrt{n_{target}+n_{skybright}+n_{readout}+n_{dark}}},
\label{eq:flux_err}
\end{equation}
where $n_{target}$ represents target flux, $n_{skybright}$ is sky background noise flux, $n_{readout}$ represents readout noise flux and $n_{dark}$ is dark current noise flux. The $5\sigma$ limiting magnitude of $20$-seconds exposure in six filters of Mephisto are $20.07$, $20.23$, $21.09$, $21.22$, $20.91$, and $19.51$ respectively. To establish a comparatively realistic and comprehensive sample for training and testing purposes, we apply some selection criteria:
\begin{enumerate}
    \item Observations with low SNR ($<5$) or below the $5\sigma$ limiting magnitude are dropped. 
    \item At least 17 days of observations before the \emph{g}-band peak of the light curve.
    \item Samples with at least $20$ days of observations in total.
    \item Historical observational data statistics reveal that SNe Ia accounts for approximately $25$\% of the total SNe from OSC~\citep{2017ApJ...835...64G}. Thus, we construct our simulation sample with the same ratio, i.e., the number of SNe Ia to CCSNe is approximately $1$:$3$. 
\end{enumerate}
After applying the selection, a total of $8561$ samples of the simulated effective light curve are available, including $2141$ SNe Ia samples and $6420$ CCSNe samples (Table~\ref{tab:para_samples}). An instance of simulated light curves and color light curves can be seen in Fig.~\ref{fig:meph_lc} and Fig.~\ref{fig:meph_color_lc}\footnote{We select samples based on whether any one of \emph{u, g, i} bands meets the criteria, then padding in the missing data as inputs for the model. Fig.~\ref{fig:meph_color_lc} shows the color evolution obtained from the original photometric data, resulting in some missing data points (e.g., the second panel from the top in the right column in Fig.~\ref{fig:meph_color_lc}) in the color plot. Such a scenario does indeed occur.}, respectively. Here, we define~\emph{Days after the trigger} as the days with continuous following observations since the first detection of a Mephisto transient finder trigger. Dust in the Milky Way and host galaxy will affect the shape of an observed SN spectrum. It is important to take these effects into account in our model when fitting the model to observed data for more realistic. Consequently, we utilize the built-in SN models and dust parameters in SNCosmo for dust extinction and reddening, including dust extinction in the host galaxy and the Milky Way. We add both host galaxy dust and the Milky Way dust. Dust propagation effect reference~\citet{1999PASP..111...63F}, host effects and dust in the Milky Way referred from~\citet{2012ApJ...745...31B}. Another assumption that we must make clear is that host galaxy effects are not taken into account when identifying SNe Ia in this study.

\section{Method}
\label{sect:Methods}
\subsection{RNN, LSTM, GRU and BiLSTM}

\begin{figure}
\centering
\includegraphics[width=0.8\textwidth, angle=0]{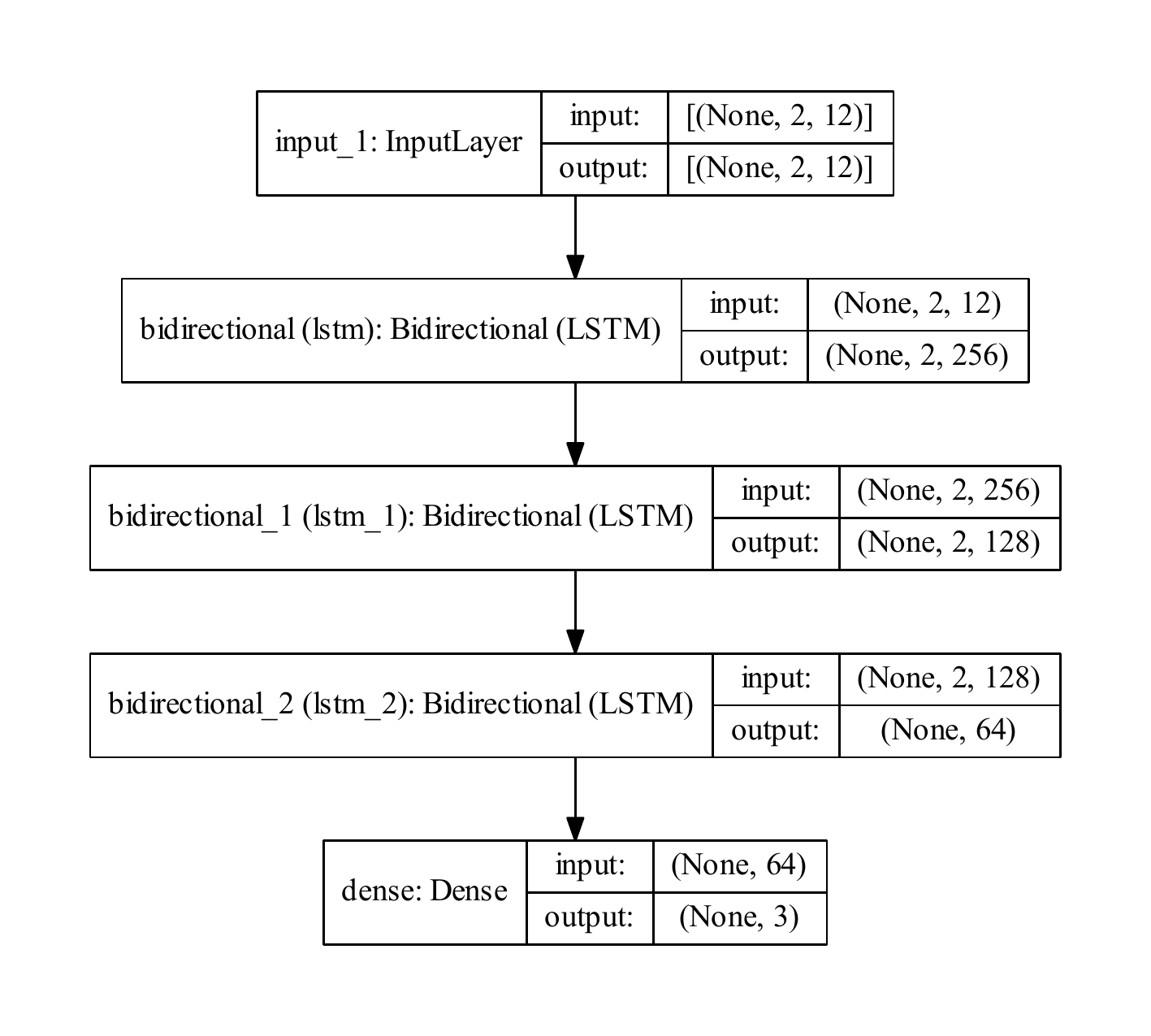}
\caption{{The flowchart of training procedures and parameters of} 
 \emph{\texttt{Mesiri}}.}
\label{fig:Mesiri}
\end{figure}

In time-series data classification, recurrent neural networks, such as LSTM, GRU, and BiLSTM, demonstrate superior performance and are widely applied. With the aim of selecting the optimal method for identifying early-time SNe Ia, in this work, we train several deep recurrent neural (DRNN) networks to classify the light curve of SNe Ia from CCSNe and name our approach~\emph{\texttt{Mesiri}}. The training procedures and parameters of ~\emph{\texttt{Mesiri}} are shown in Fig.~\ref{fig:Mesiri}.

\begin{figure}[h]
\begin{minipage}[t]{0.335\textwidth}  
  \centering
   \includegraphics[width=\textwidth]{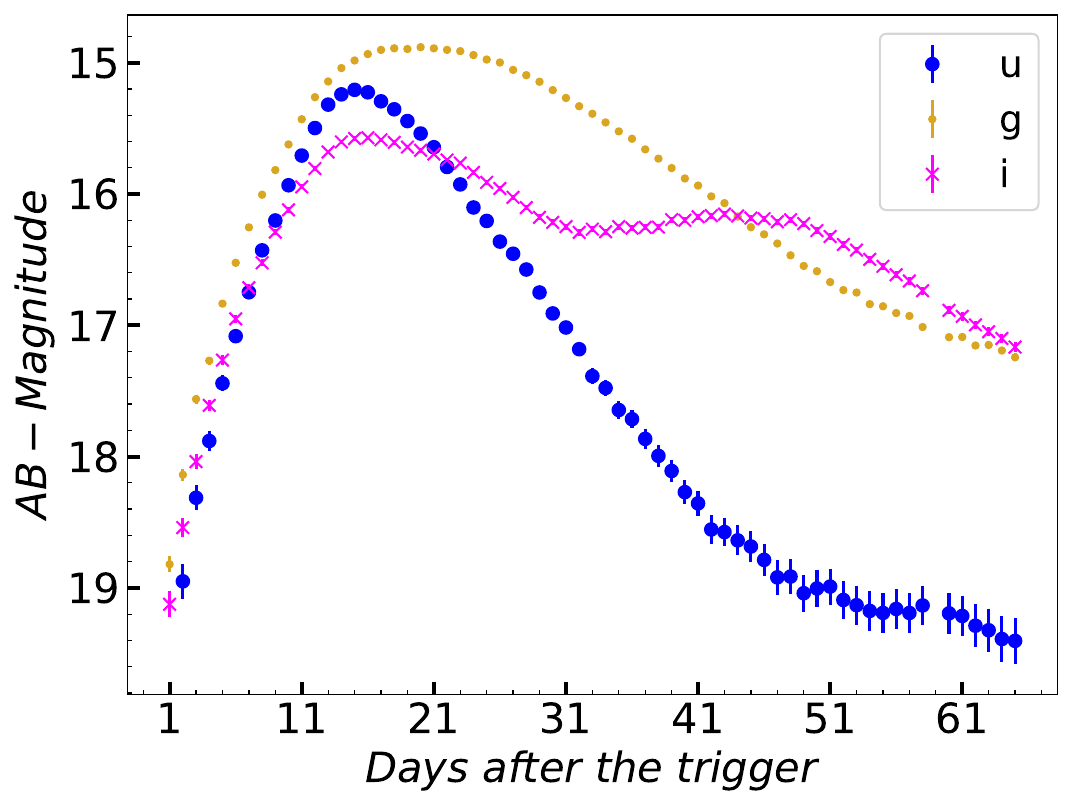}
  \end{minipage}
    \begin{minipage}[t]{0.335\textwidth}
  \centering
   \includegraphics[width=\textwidth]{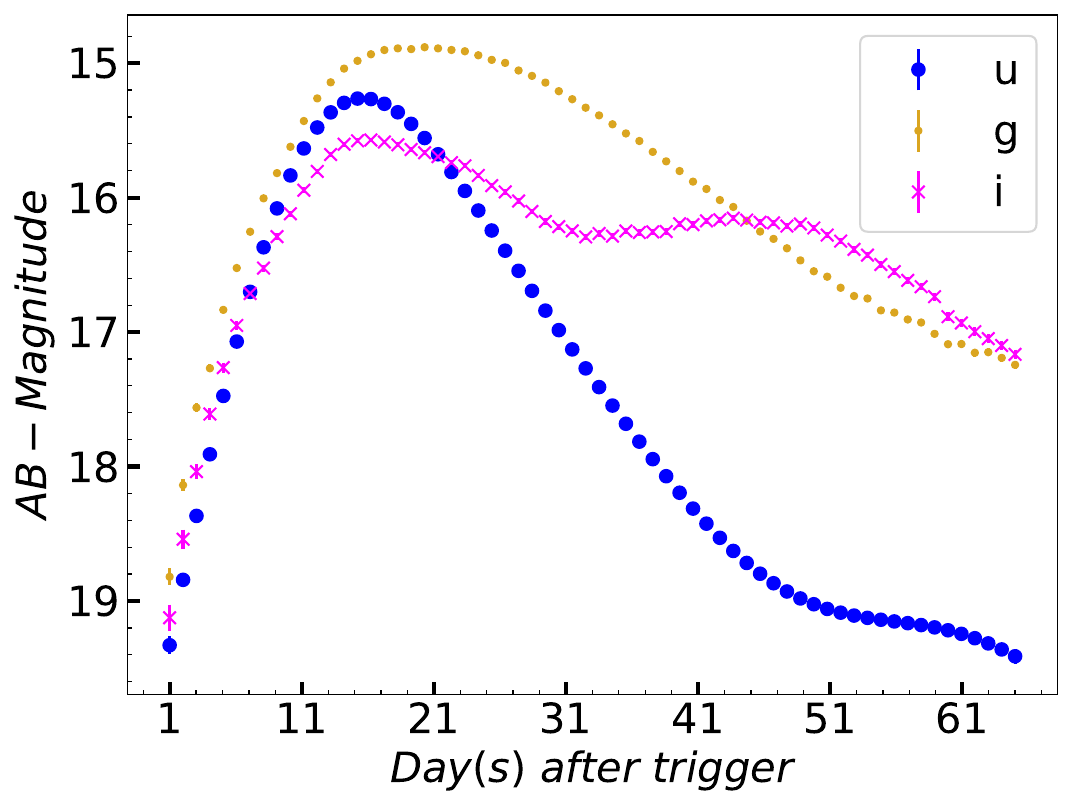}
  \end{minipage}
    \begin{minipage}[t]{0.345\textwidth}
      \centering
       \includegraphics[width=\textwidth]{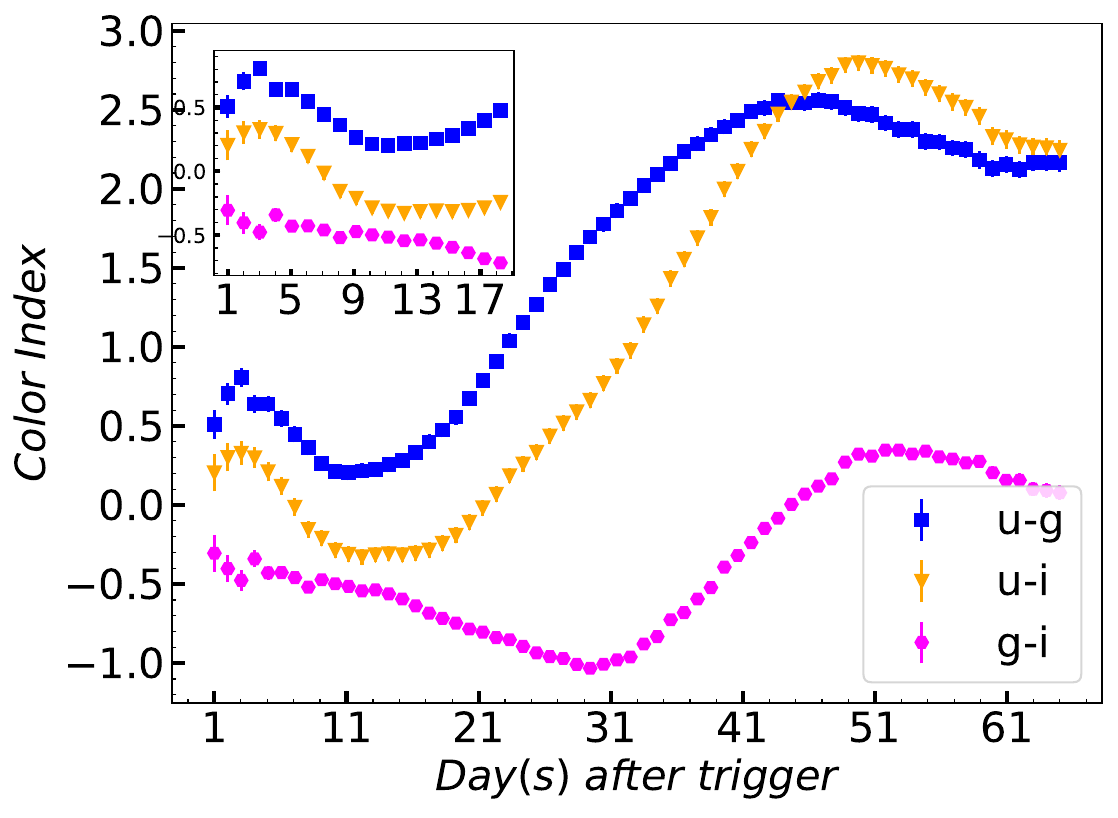}
    \end{minipage}
\caption{The left and middle panels represent the original observed light curves in ~\emph{u, g, i} filters and the GP interpolated light curves, respectively. The corresponding color curves are displayed in the right panel. The embedded subplot indicates the color evolution in the first 17 days after the trigger.}
\label{fig:gp_example}
\end{figure}

\begin{figure}
\centering
\includegraphics[width=\textwidth, angle=0]{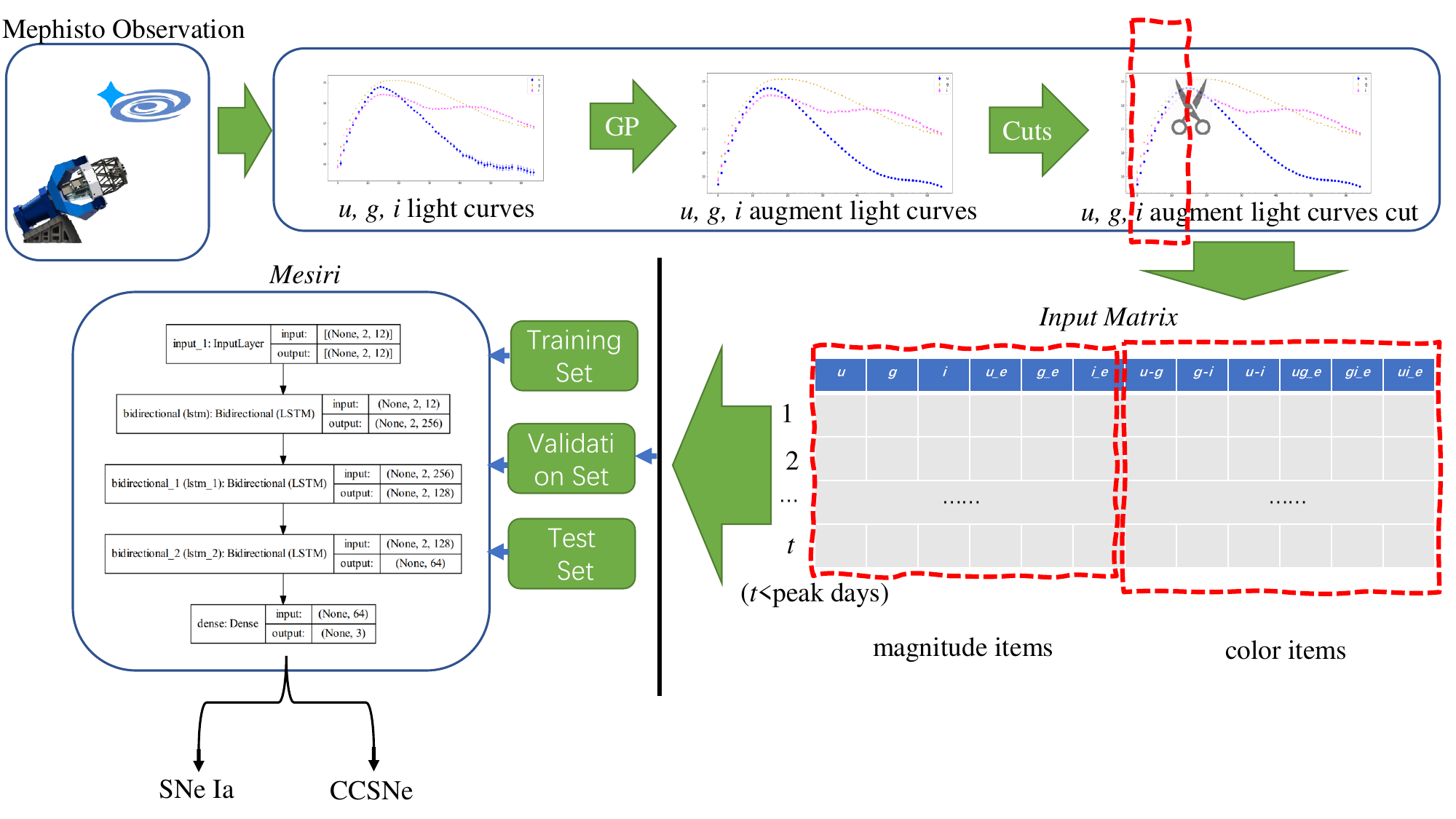}
\caption{The flowchart of \emph{\texttt{Mesiri}} pipeline. We show the BiLSTM-based cell unit as an example; it can be replaced by other cells in different tests.)}
\label{fig:Mesiri_pipeline} 
\end{figure}

The DRNN models a function that maps an input multi-band light curve matrix, $\bm{I}^{st}$, for transient $s$ up to a discrete-time $t$, onto an output probabilities over classes \{$c$=$0$,$1$,$2$\}\footnote{In the classification of recurrent neural networks, an additional category is typically added to the output classes, often referred to as the “unknown” or “other” category. The purpose of this additional category is to handle samples that the model cannot accurately classify into known categories, namely those that do not belong to any known category in the training set. This approach helps the model better deal with unknown samples, thereby improving the model’s generalization ability and the accuracy of classifying unknown samples.},
\begin{equation}
     \bm{y}^{st} = \bm{f}_t(\bm{I}^{st}; \bm{\theta}),
\end{equation}
where $\bm{\theta}$ stands for the parameters (e.g., weights and biases of the neurons) of our DRNN architecture. We define the input $\bm{I}^{st}$ as an $t \times 12$ (or $t \times 12$ +$1$, when redshift is considered) matrix representing the light curve up to a time-step $t$, where $12$ represents input features including ${u}^{st}$, ${g}^{st}$, ${i}^{st}$, ${u^{st}_{error}}$, ${g^{st}_{error}}$, ${i^{st}_{error}}$, ${(u-g)}^{st}$, ${(g-i)}^{st}$, ${(u-i)}^{st}$, ${(u-g)^{st}_{error}}$, ${(g-i)^{st}_{error}}$, ${(u-i)^{st}_{error}}$. Among them, the first six quantities represent the magnitude and magnitude error corresponding to \emph{u, g, i} bands, while the following six quantities represent the true color information and the associated color information errors. Here, the true color refers to color information obtained through simultaneous observations in multi-bands, as opposed to color information obtained through non-simultaneous multi-band observations. The output $\bm{y}^{st}$ is a probability vector with length $3$, where each element $y^{st}_c$ is the model's predicted probability of each class $c$, such that $y^{st}_c \ge 0$ and $\sum^{2}_{c=0} y^{st}_c = 1$. From Fig.~\ref{fig:meph_lc}, it is evident that in certain filters, we have limited data as a result of fluctuations in the throughput efficiency of different bands. To align the data points simultaneously observed with different filters, we utilize the Gaussian Process (GP) for interpolation~\citep{2023A&A...677A..16D}. We perform interpolation based on time points, i.e., starting from the earliest observed point and ending with the latest one. In most cases, the earliest data appears in the \emph{g} or \emph{r} bands since they have relatively high efficiency. A case in point is the observations in the ~\emph{u, g, i} filter if the~\emph{g}-band has $60$ observation points ($[0,59]$), while the observation time range for the~\emph{u}-band is [5, 35], we interpolate the values in the [0, 4] and [36, 59] intervals using the GP. The same interpolation technique is applied to the other bands. Moreover, there is a possibility that no observation data is available for a particular filter. In such cases, we adopt the commonly used padding technique in neural networks to fill in the values with a padding value of $-1$. The interpolated images before and after interpolation are shown in the left and middle panels of Fig.~\ref{fig:gp_example}, and the right panel shows the corresponding color light curves. The embedded subplot in the upper left corner of the right panel represents the color evolution in the first 17 days. Finally, the~\emph{\texttt{Mesiri}} pipeline is shown in Fig.~\ref{fig:Mesiri_pipeline}.

We define the global objective function as
\begin{equation}
\bm{Loss} = -{y} \log(p(y))+(1-y)\log{(1-p(y))},
\label{eq:objective}
\end{equation}
where we sum the weighted categorical cross-entropy overall $t$ time-steps in the training set, where,~$y$ belongs to \{0,1\} is the binary label,~$p(y)$ is the probability that the output belongs to $y$. We expect $p(y)$ to be as large as possible when $y$=1. Look at the ideal case: when $y$ is a positive case, $p(y)$=1, the loss is $0$; On the contrary, $p(y)$ tends to $0$, $\log p(y)$ tends to negative infinity, leading to a very large loss. We describe the architecture in detail, as follows:

\begin{description}
\item[\textbf{Input}:] The input is a $t \times 12$ matrix, or $t \times 12$ +$1$ when redshift is considered. However, as we are implementing a sequence classifier, we can consider the input at each time-step as being the vector of length $t \times 12$. Here input features including six magnitude items: ${u}^{st}$, ${g}^{st}$, ${i}^{st}$, ${u^{st}_{error}}$, ${g^{st}_{error}}$, ${i^{st}_{error}}$, and six color iterms: ${(u-g)}^{st}$, ${(g-i)}^{st}$, ${(u-i)}^{st}$, ${(u-g)^{st}_{error}}$, ${(g-i)^{st}_{error}}$, ${(u-i)^{st}_{error}}$, corresponding to the abbreviations in Fig.\ref{fig:Mesiri_pipeline} \emph{Input matrix} with magnitude items: ${u}$, ${g}$, ${i}$, ${u\_e}$, ${g\_e}$, ${i\_e}$, and six color iterms: ${ug}$, ${gi}$, ${ui}$, ${ug\_e}$, ${gi\_e}$, ${ui\_e}$.

\item[\textbf{First\ Layer}:] We apply all RNN/LSTM/GRU/BiLSTM unit cells in this study, as they offer significantly shorter training times without deterioration in classification performance. They can capture dependencies in time-varying data by controlling the information remembered at each step of the light curve. In the initial layer, each input sequence is encoded into a higher-dimensional representation one time-step at a time, utilizing $256$ units to generate an output vector of $t \times 256$ dimension.

\item[\textbf{Second\ Layer}:] The second RNN/LSTM/GRU/BiLSTM layer is conditioned on the input sequence. It takes the output of the first layer and generates an output sequence. We set up this layer with $128$ units, dropout and recurrent dropout layer in this layer to maintain the $t \times 128$ output shape.

\item[\textbf{Third\ Layer}:] The third RNN/LSTM/GRU/BiLSTM layer is conditioned on the second sequence. It takes the output of the second layer and generates an output sequence. Again, we establish this layer with $64$ units, dropout layer in this layer to maintain the $t \times 64$ output shape.

\item[\textbf{Dense\ Layer}:] In neural networks, a dense layer is the fundamental and simplest type of layer, often known as a fully connected layer. It establishes a connection between all $64$ neurons from the previous layer with $3$ neurons in the output layer by employing equation~\ref{eq:neuron}. In the problem of classification, the output vector includes all $2$ SNe categories.

\item[\textbf{Neurons}:]
The output of each neuron in a neural network layer can be formulated as the weighted sum of the connections to it from the preceding layer:
\begin{equation}
    \hat{y}_i = f\left(\sum\limits_{j=1}^{M} W_{ij} \, x_j + b_i \right),
    \label{eq:neuron}
\end{equation}
    
where $x_j$ are the different inputs to each neuron from the previous layer, $W_{ij}$ are the weights of the corresponding inputs, $b_i$ is a bias that is added to shift the threshold of where inputs become significant, $j$ is an integer running from $1$ to the number of connected neurons in the previous layer ($M$), and $i$ is an integer running from $1$ to the number of neurons in the next layer. For the dense layer, $\mathbf{x}$ is simply the $(1\times 64)$ matrix from the output of the second layer, $\mathbf{y}$ is made up of the $3$ output classes, $j$ runs from $1$ to $3$ and $i$ runs across the $64$ input neurons from the last layer. 

\item[\textbf{Dropout}:] We also implement dropout regularization to each layer of the neural network to reduce over-fitting during training. This is an important step that effectively ignores randomly selected neurons during training, such that their contribution to the network is temporarily removed. This process causes other neurons to more robustly handle the representation required to make predictions for the missing neurons, making the network less sensitive to the specific weights of any individual neuron. We set the dropout rate to $20$\% of the neurons present in the previous layer each time the dropout block appears in the DRNN in Fig.~\ref{fig:Mesiri}. 

\item[\textbf{Recurrent\ Dropout}:] Just as with regular dropout, recurrent dropout has a regularizing effect and can prevent over-fitting. We set the dropout recurrent rate to $20$\% of the neurons present in the previous layer each time the dropout block appears in the DRNN in Fig.~\ref{fig:Mesiri}.

\item[\textbf{Activation\ function}:]
Each neuron in a neural network applies an activation function to introduce non-linearity, which enables the network to handle a wide range of data. The most commonly used activation function for feed-forward networks is the hyperbolic tangent function, commonly abbreviated as tanh. 

\item[\textbf{Sigmoid\ regression}:] The sigmoid regression activation function is utilized in the final layer, specifically to handle binary classes. This function is applied to the dense layer output of each time step, normalizing the output vector between $0$ and $1$. Consequently, the sum of the values across all classes in each time step generates a total value of $1$. This process enables the output to be interpreted as a proportional likelihood that the input SN of each time step belongs to a certain class. The resulting vector of probabilities is generated as a result of this procedure.
\begin{equation}
\bm{y} = \mathrm{sigmoid}(\bm{\hat{y}}),
\end{equation}

is computed with a sigmoid activation function that is defined as
\begin{equation}
\mathrm{sigmoid}(\bm{x})_i = \frac{1}{1+e^{-x_j}}.
\end{equation}
We utilize the output sigmoid probabilities to prioritize the most fitting SN classes for each SN light curve at each time step.
\end{description}

\subsection{Evaluation}
\label{subsect:evaluation}

In evaluating performance, we utilized five commonly used metrics to evaluate our \emph{\texttt{Mesiri}}. The most straightforward metric is accuracy by employing equation \ref{eq:accuracy}, which refers to the proportion of correctly classified SNe in each class to the total number of SNe in each class. The precision, also known as purity, is the ratio of true positive predictions to the total number of positive predictions for each class defined by equation \ref{eq:precision}. The recall, also known as completeness, is similar to the true positive rate. It is a measure of the number of correct predictions in each class compared to the total number of that class in the testing set and is defined as equation \ref{eq:recall}. The F1 score is the harmonic mean of precision and recall, combining their trade-offs defined by equation \ref{eq:f1_score}. It ranges from $0$ to $1$, with a higher value indicating better model performance. The F1 score can be used as a comprehensive evaluation metric for the performance of classification models. The AUC represents the area under the Receiver Operating Characteristic (ROC) curve, which measures the accuracy of a binary classification model’s predictions. AUC represents the area under the ROC curve and measures the accuracy of the model’s classification at different thresholds. AUC ranges from $0$ to $1$, with a higher value indicating better model performance. By observing AUC, we can also understand the model’s performance at different thresholds, helping us choose the optimal classification threshold. These metrics can help evaluate the performance of classification models and measure their accuracy, coverage, and stability from different perspectives. Apart from the aforementioned evaluation metrics, the confusion matrix is a common approach to visualize the performance of a classification model, defined as in table~\ref{tbl:cm}.

\begin{equation}
    \mathrm{Accuracy = \frac{TP + TN}{TP + TN + FP + FN}},
    \label{eq:accuracy}
\end{equation}

\begin{table*}
    \centering
    \caption{Confusion Matrix}
    \label{tbl:cm}
    \begin{tabular}{l  c  c}
        \hline
        & Predicted Positive& Predicted Negative \\
        \hline 
        True Positive & TP & FN\\
        \hline
        True Negative & FP & TN\\
    \hline
    \end{tabular}
\end{table*}

\begin{equation}
    \mathrm{Precision/Purity = \frac{TP}{TP+FP}},
    \label{eq:precision}
\end{equation}

\begin{equation}
    \mathrm{Recall/Completeness = \frac{TP}{TP+FN}},
    \label{eq:recall}
\end{equation}

\begin{equation}
    \mathrm{F1\ score = 2 * \frac{Precision * Recall}{Precision + Recall}}.
    \label{eq:f1_score}
\end{equation}

Here TP, TN, FP, and FN represent the number of true positive samples, the number of true negative samples, the number of false positive samples, and the number of false negative samples, respectively.

\section{Result}
\label{sect:results}

The variation of light curves in the Mephisto \emph{u, v, g, r, i, z} bands is attributed to transmission efficiency. Therefore, we employ Gaussian Process (GP) and padding interpolation to interpolate the original light curve data, ensuring uniform light curve length across each band.

\begin{figure}[h]
  \begin{minipage}[t]{0.495\linewidth}
  \centering
   \includegraphics[width=\textwidth]{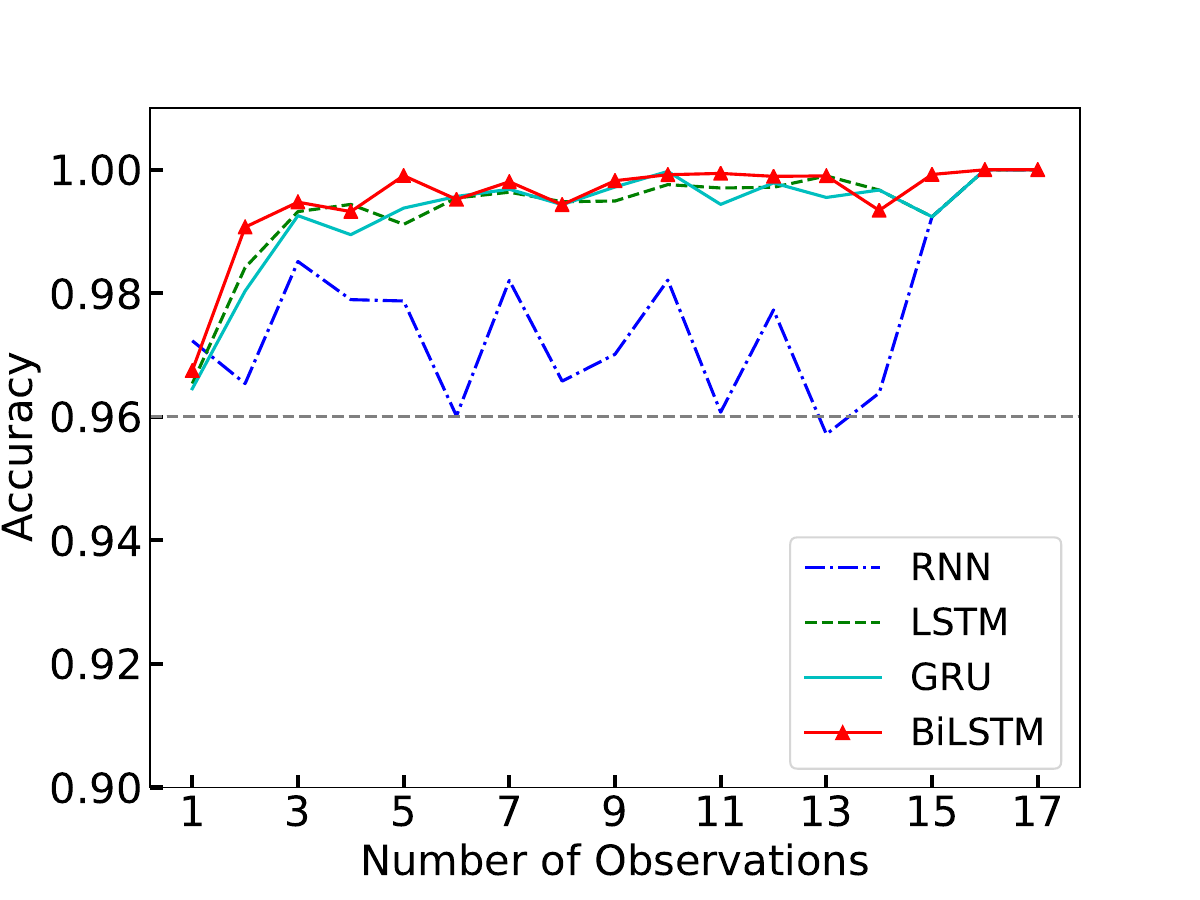}
  \end{minipage}
  \begin{minipage}[t]{0.495\textwidth}
  \centering
   \includegraphics[width=\textwidth]{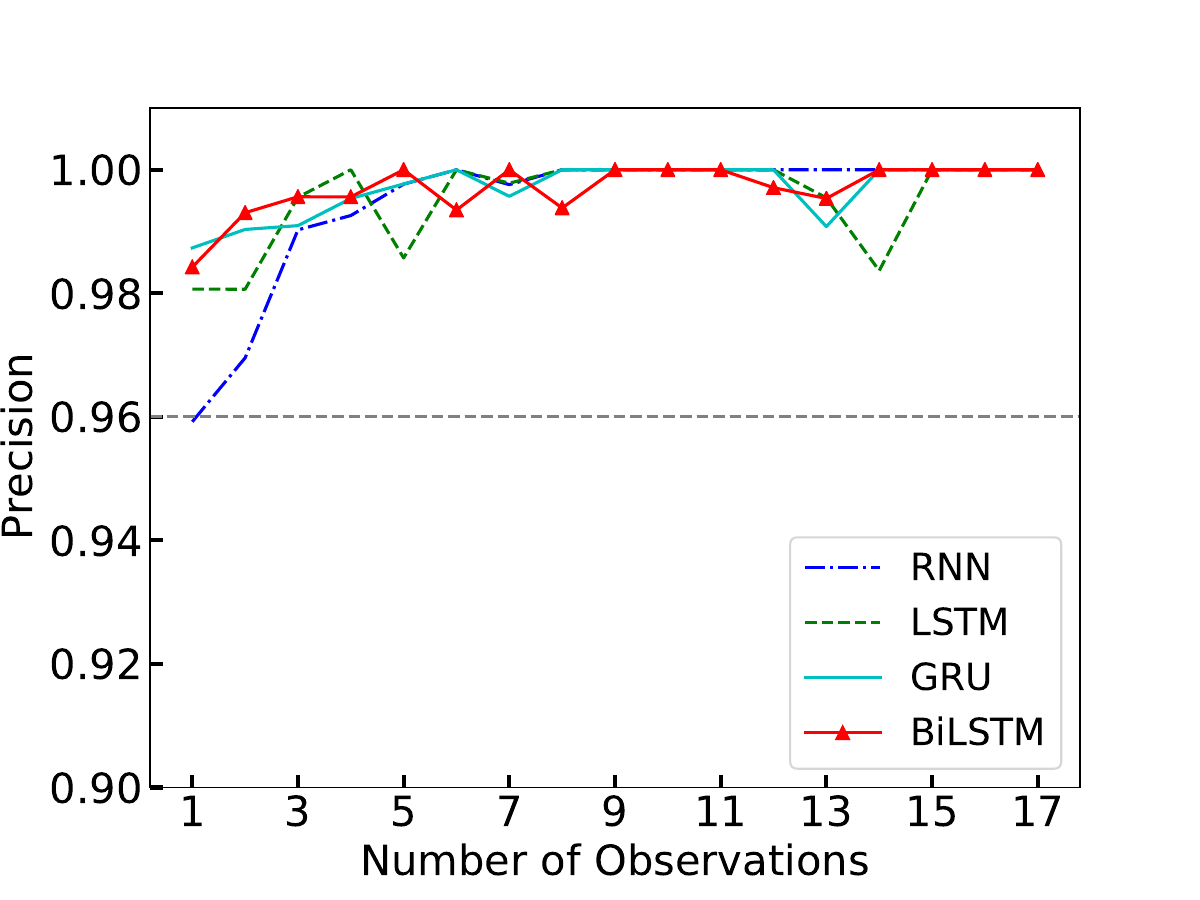}
  \end{minipage}
  \quad
  \begin{minipage}[t]{0.495\linewidth}
  \centering
   \includegraphics[width=\textwidth]{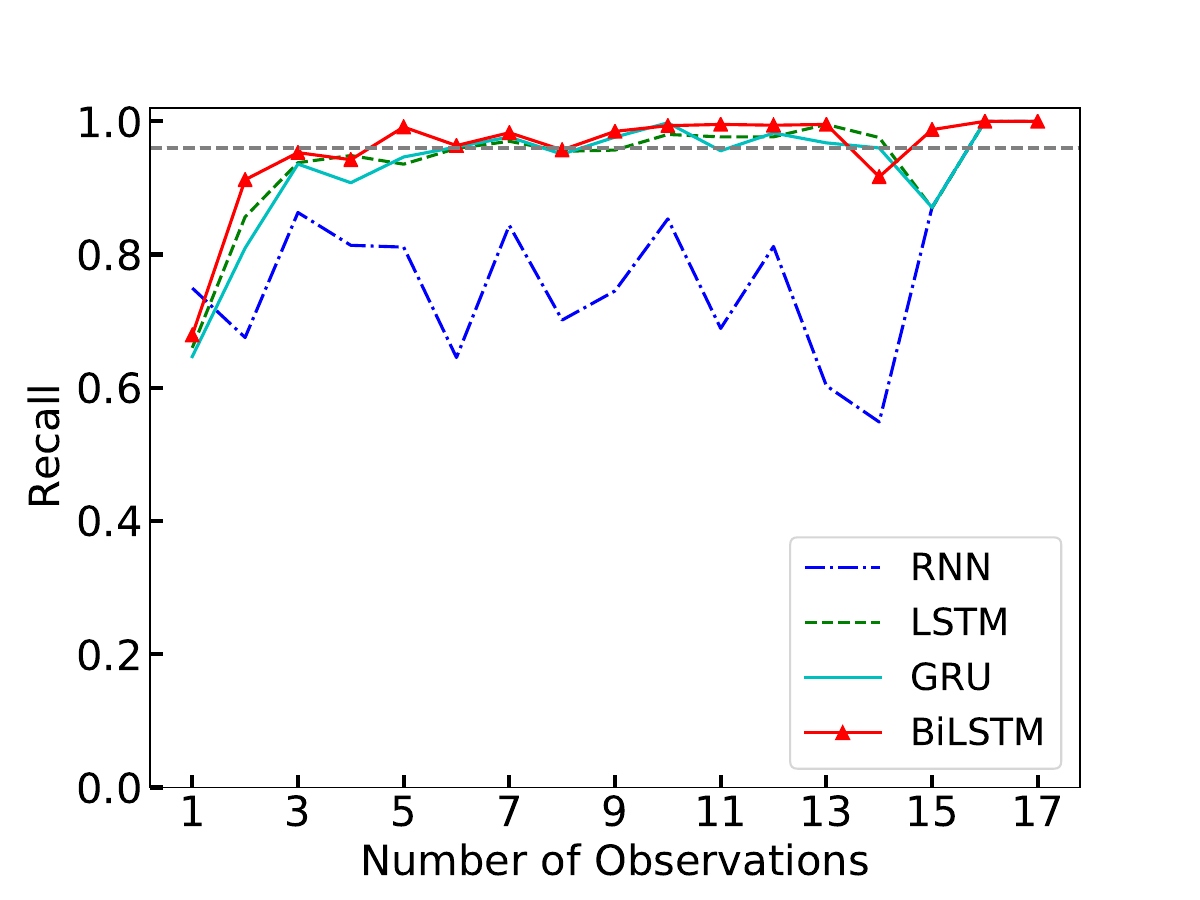}
  \end{minipage}
  \begin{minipage}[t]{0.495\textwidth}
  \centering
   \includegraphics[width=\textwidth]{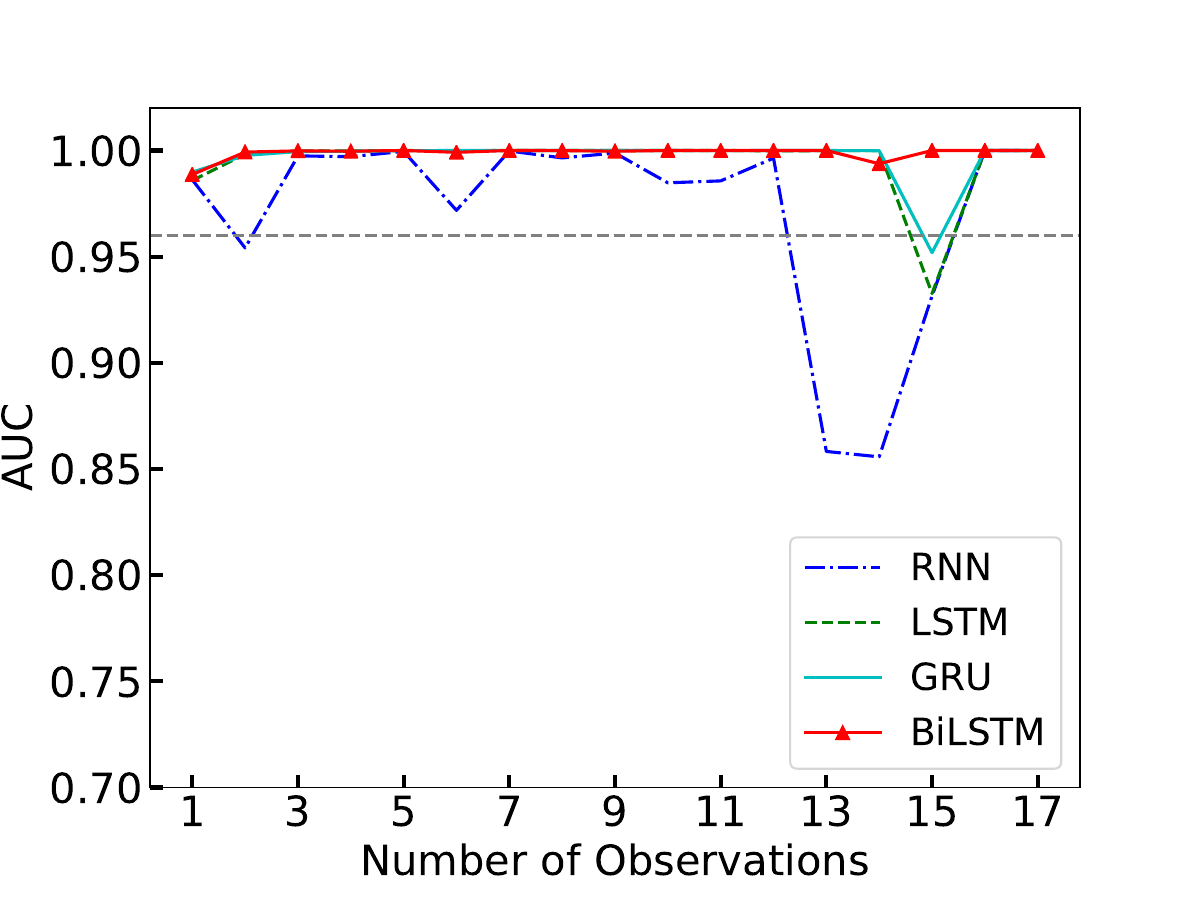}
  \end{minipage}
  \quad
  \begin{minipage}[t]{0.495\linewidth}
  \centering
   \includegraphics[width=\textwidth]{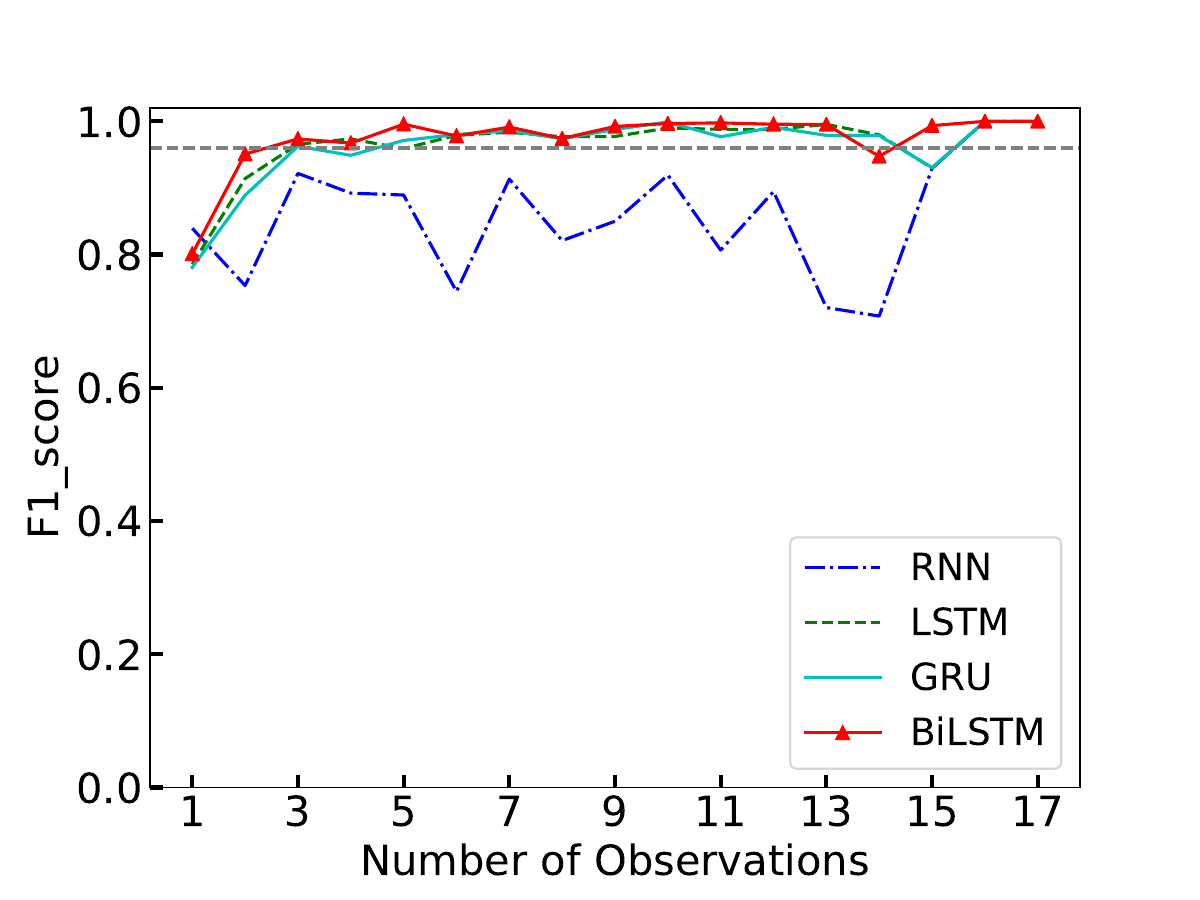}
  \end{minipage}
\caption{Comparison of models for five evaluation metrics in each cell-based model. The cell name is labeled in each panel. The gray horizontal dashed line corresponds to a value of $96$\%. It can be observed that all evaluation metrics have achieved good results with the inclusion of real-time color. The network architecture based on BiLSTM units shows the best performance.}
\label{fig:all_evaluations}
\end{figure}

During the model training process, we divide the data into training, validation, and testing sets. For the purpose of enhancing the objectivity of performance evaluation and mitigating the risk of over-fitting, we employ a five-fold cross-validation approach during model training, i.e., the training is repeated five times, and each time, the testing set is differently selected from the simulation sample. 
Multiple models are trained using distinct sets of simulated data points at the early stages of the SN explosion, e.g., different numbers of data points before the peak. This allows us to achieve a diverse set of models tailored to specific data scenarios for improved classification matching. The experimental results, as shown in Fig.~\ref{fig:all_evaluations} and Table~\ref{tbl:no-z-acc}, present the accuracy of identifying early-time SNe Ia under different scenarios without redshift information. Here, we emphasize that since the uncertainties associated with each evaluation are derived from the standard deviation computed five-fold cross-validation, consequently, some of these values (c.f. Table~\ref{tbl:no-z-acc} and Table~\ref{app:evaluations}) may exceed 100\% after incorporating the standard deviation. The similar scenario is also depicted in Section~\ref{sect:discussion} from Fig.~\ref{fig:vs_morearly} to Fig.~\ref{fig:vs_shallow}. 
The first column in Table~\ref{tbl:no-z-acc} represents the neural network base cell units utilized in \emph{\texttt{Mesiri}}, while the subsequent column of time denotes the observational epochs of pre-maximum luminosity. Most of the results show acceptable performance, e.g., accuracy greater than 96\%. And our models based on BiLSTM are not only lightweight but also highly accurate. It is worth noting that the identification accuracy of SNe Ia from a light curve, when first discovered/triggered without redshift, is $97.15\pm0.63$\%.  Further details can be found in Section \ref{subsec:veryearly}. 
The detailed identification results for each scenario can be found in Appendix \ref{app:evaluations} and Table \ref{tab:evaluations}. Furthermore, Appendix \ref{app:cm} and Fig.~\ref{app:cm} present all confusion matrices. We also adopt a normalized version of the confusion matrix for easier interpretation.

\begin{table*}
    \centering
    \caption{Accuracies for light curves with $1, 3, 6, 9, 17$ epochs (day(s)) of pre-maximun without redshift information. These averages and standard deviations were computed from 5 independent runs. Note that each evaluation, when combined with uncertainties represented by standard deviations, may exceed 100\%.}
    \label{tbl:no-z-acc}
    \setlength{\tabcolsep}{0.6mm}
    \begin{tabular}{l c c c c c}
        \hline
        Base cell& 1 day & 3 days & 6 days & 9 days & 17 days \\
        \hline RNN&$97.23\pm0.69$\%&$97.89\pm0.82$\%&$98.21\pm0.62$\%&$98.21\pm0.87$\%&$100.0\pm0.0$\%\\        LSTM&$96.54\pm0.94$\%&$99.44\pm0.17$\%&$99.64\pm0.10$\%&$99.76\pm0.25$\%&$100.0\pm0.0$\%\\        GRU&$96.46\pm0.41$\%&$98.95\pm0.47$\%&$99.69\pm0.14$\%&$99.97\pm0.05$\%&$100.0\pm0.0$\%\\        BiLSTM&$96.75\pm0.79$\%&$99.32\pm0.61$\%&$99.81\pm0.12$\%&$99.92\pm0.06$\%&$100.0\pm0.0$\%\\
    \hline
    \end{tabular}
\end{table*}

The increasing accumulation of observational data demonstrates the classification accuracy of the simultaneous observation of three bands with colored features, which is comparable to that of non-simultaneous observation of three bands with colored features. This observation underscores the positive impact of increased observational data on classification accuracy. Through tests conducted on a Windows 11 operating system with an Intel® Core™ i7-10700 CPU @ 2.90GHz processor running a classifier in Jupyter Notebook, testing was performed on 31 epochs of 5 observed target sources. The average time to classify each instance was approximately within 2 seconds. It is relevant to mention that our network architecture is implemented using \texttt{TensorFlow}~\citep{tensorflow2015-whitepaper} and~\texttt{Keras}~\citep{chollet2015_keras}. The entire pipeline is implemented using the~\texttt{Python} programming language and~\texttt{scikit-learn}~\citep{scikit-learn}.

\section{Further comparison}
\label{sect:discussion}

Given that the findings indicate the superior performance of the network architecture utilizing BiLSTM, all subsequent analyses will be conducted with a focus on the BiLSTM model.

\subsection{Early observation and training sample}
\label{subsec:veryearly}

In the following, the `very early' and `early' observations in the context of the training sample are described. The `very early' observation scenario is considered as the observation at a very early time (just after the explosion/discovery) and with a good cadence. For instance, if a source has observational data from Mephisto for 17 days before peak brightness across all three bands, then `very early time' is taken as [$1$, $2$, $3$, $4$, $5$, \dots,$17$] sequentially. The second scenario (i.e., `early') means our observations are comparatively not so early, and the data acquisition is random before the maximum luminosity (e.g., 3, 7, 10, 15,...., days after the explosion/discovery).

\begin{figure}
\centering
\includegraphics[width=0.8\textwidth, angle=0]{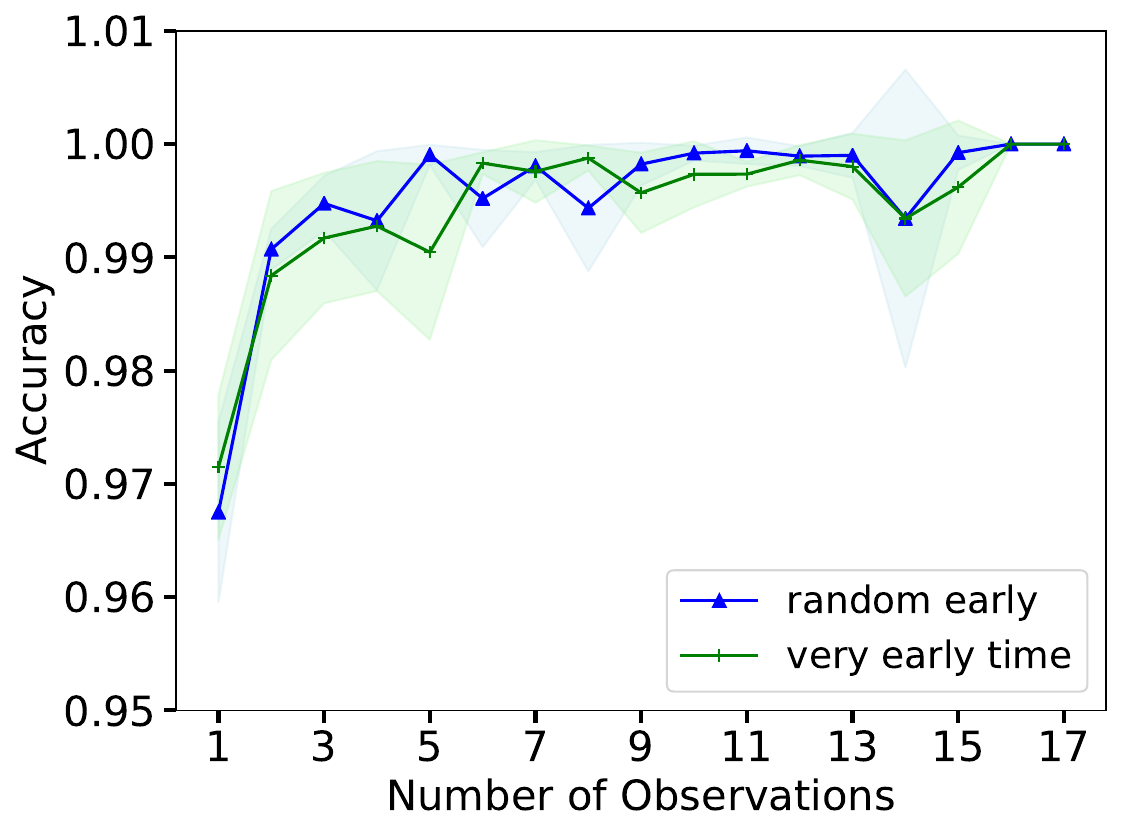}
\caption{The accuracy of `very early' and `early' time classification models is compared. 
The blue triangle and shaded region represent the results with random epochs of observation before the maximum light (peak) and the standard deviation of model errors after five-fold cross-validation, respectively. The green symbols exhibit the results of the early time explosion using sequential epochs of observation.}
\label{fig:vs_morearly} 
\end{figure}

In Fig.~\ref{fig:vs_morearly}, the accuracy of early and very early time classification models is compared. The blue triangle line represents the results of an early SN explosion with random epochs of observation, while the light blue shaded region represents the standard deviation of model errors after five-fold cross-validation. On the other hand, the green cross line and green shaded region represent the results of early explosion using sequential epochs of observation. Upon examining the results, it is evident that the two trained models produce nearly similar outcomes. Here, the fact to be noted that the identification accuracy, precision, and AUC of `very early' SNe Ia, when the first discovery/trigger without redshift, are $97.15\pm0.63$\%, $92.86\pm4.10$\%, $98.5\pm0.75$\%, respectively.

\subsection{With redshift versus without redshift}

\begin{figure}
\centering
\includegraphics[width=0.8\textwidth, angle=0]{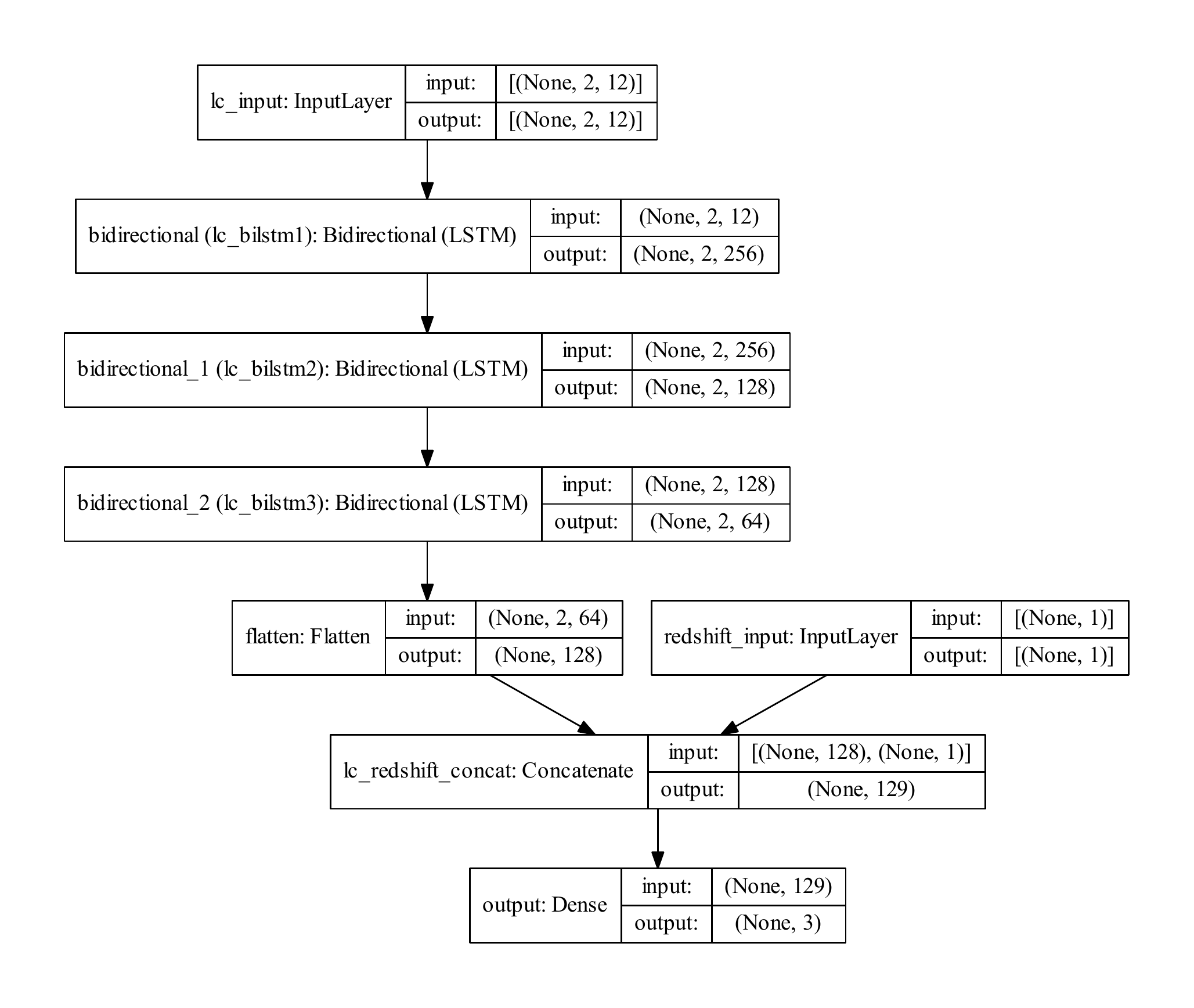}
\caption{\emph{\texttt{Mesiri}} with input features that incorporate redshift information between the third and fourth layers.}
\label{fig:Mesiri_z} 
\end{figure}

\begin{figure}
\centering
\includegraphics[width=0.8\textwidth, angle=0]{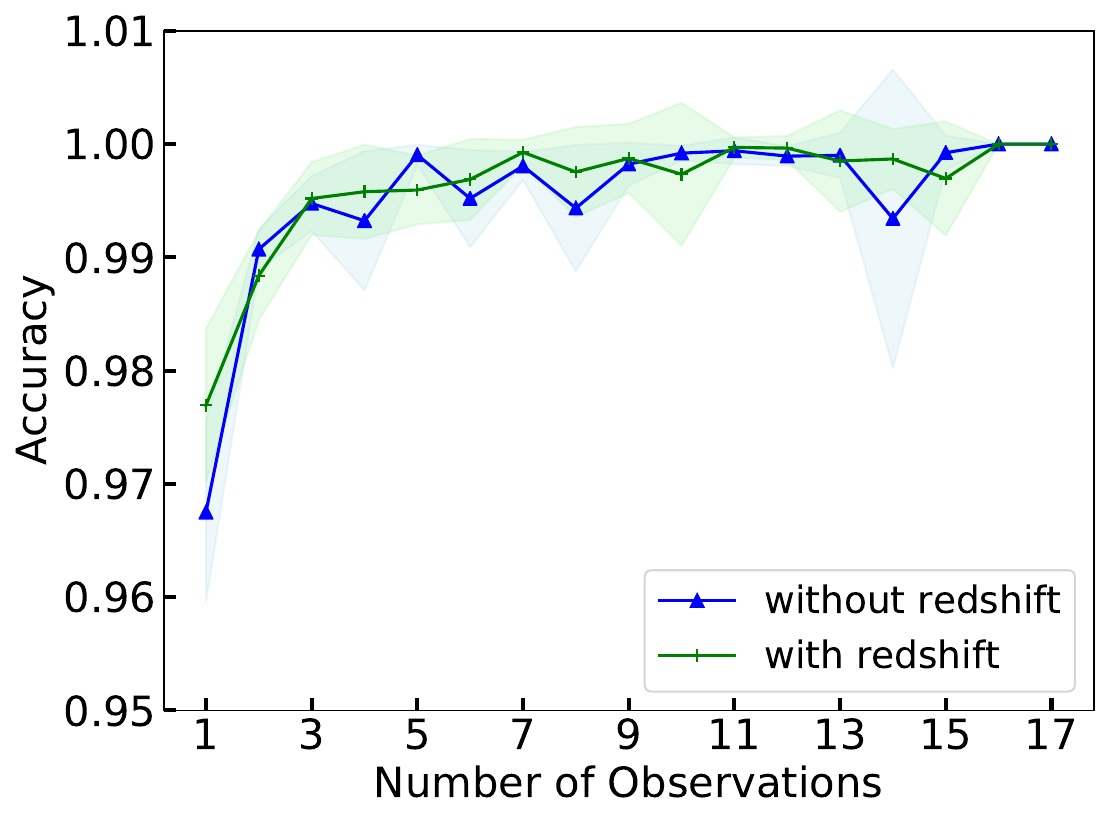}
\caption{The accuracy of early classification models with and without redshift is compared. The blue triangle line represents the results of an early SN explosion with random epochs of observation, while the light blue shaded region represents the standard deviation of model errors after five-fold cross-validation. The results of early explosion with redshift are represented by the green cross line and green region.}
\label{fig:vs_withredshift} 
\end{figure}

\cite{2016ApJS..225...31L} and \cite{2024AAS...24326129D} observe that redshift is not notably influential in the classification of SNe at low redshifts, whereas \cite{2022AJ....163...57Q} demonstrates that the inclusion of redshift information enhances performance at all epochs. The spectroscopic redshift is time-consuming and difficult to obtain at the early stage. In reality, using deep learning for photometric redshift is an active area of research \citep[e.g.][]{2021FrASS...8...70B,2022RAA....22k5017Z}. 
Therefore, we implement a comparative analysis to assess the impact of redshift. The backbone architecture is shown in Fig.~\ref{fig:Mesiri_z}, which is similar to \emph{\texttt{Mesiri}}, with the only difference being the addition of redshift information between the third and fourth layers.
Fig.~\ref{fig:vs_withredshift} demonstrates that our constructed classification model remains insensitive to the inclusion of redshift information during the initial phase of the SN explosion. With the redshift information, the variation of the estimate, i.e., the shaded region, shrinks slightly.

\subsection{Simultaneously versus not simultaneously}

To differentiate Mephisto's performance from other telescopes under the condition of non-simultaneous observation data, we process the simulated data to obtain approximately non-simultaneous observation data. We then implement a comparative analysis using the BiLSTM. It is to emphasize that this data is still based on Mephisto's observation scheme. Specifically, the data were processed as follows: first, the early-time data of the \emph{u, g, i} bands were randomly selected for multiple observations. For example, if we want to study the classification with three observations, we randomly select three observations before the peak luminosity in the \emph{u}-band, and the same for the~\emph{g} and \emph{i} bands. The selected \emph{u, g, i} data would be aligned. It is important to note that this alignment does not refer to aligning the data in the time domain. Subsequently, the color information and corresponding color error information will be calculated. Nevertheless, here the color information is not the true color obtained through simultaneous observations. Finally, the selected data would be inserted into the model for training.

\begin{figure}[h]
  \begin{minipage}[t]{0.495\linewidth}
  \centering
   \includegraphics[width=\textwidth]{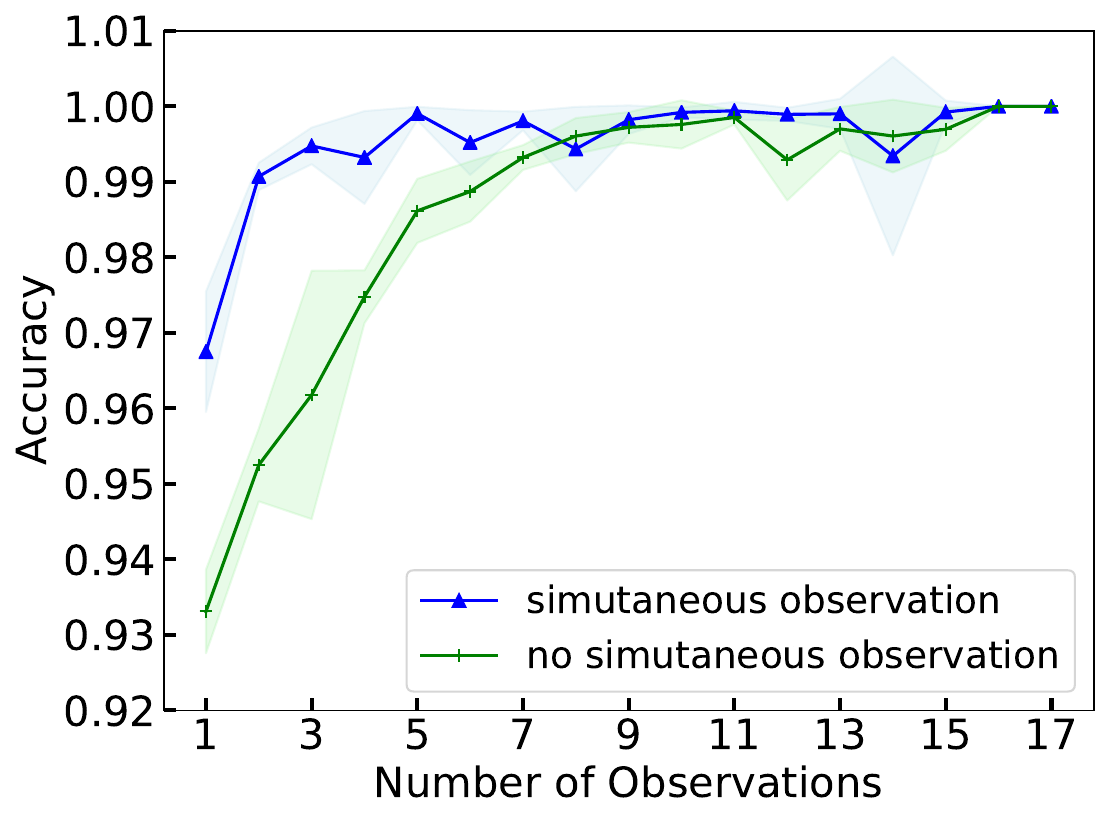}
  \end{minipage}
  \begin{minipage}[t]{0.495\textwidth}
  \centering
   \includegraphics[width=\textwidth]{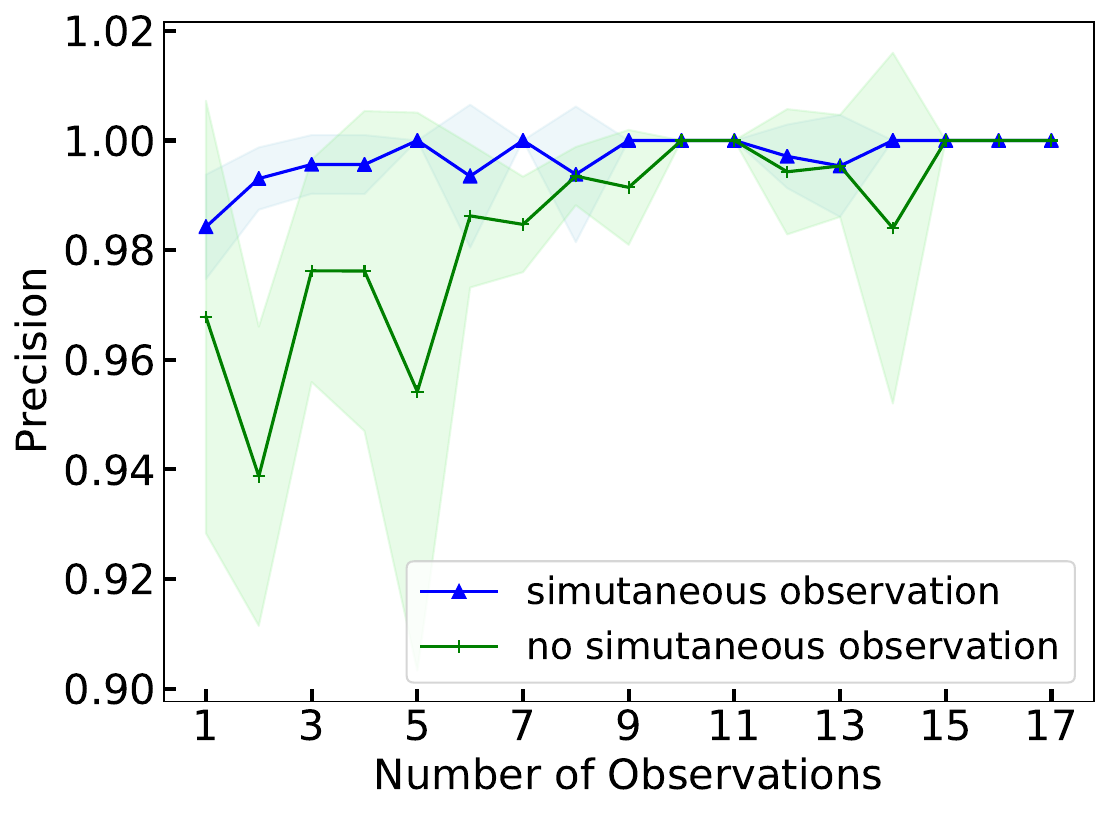}
  \end{minipage}
  \quad
  \begin{minipage}[t]{0.495\linewidth}
  \centering
   \includegraphics[width=\textwidth]{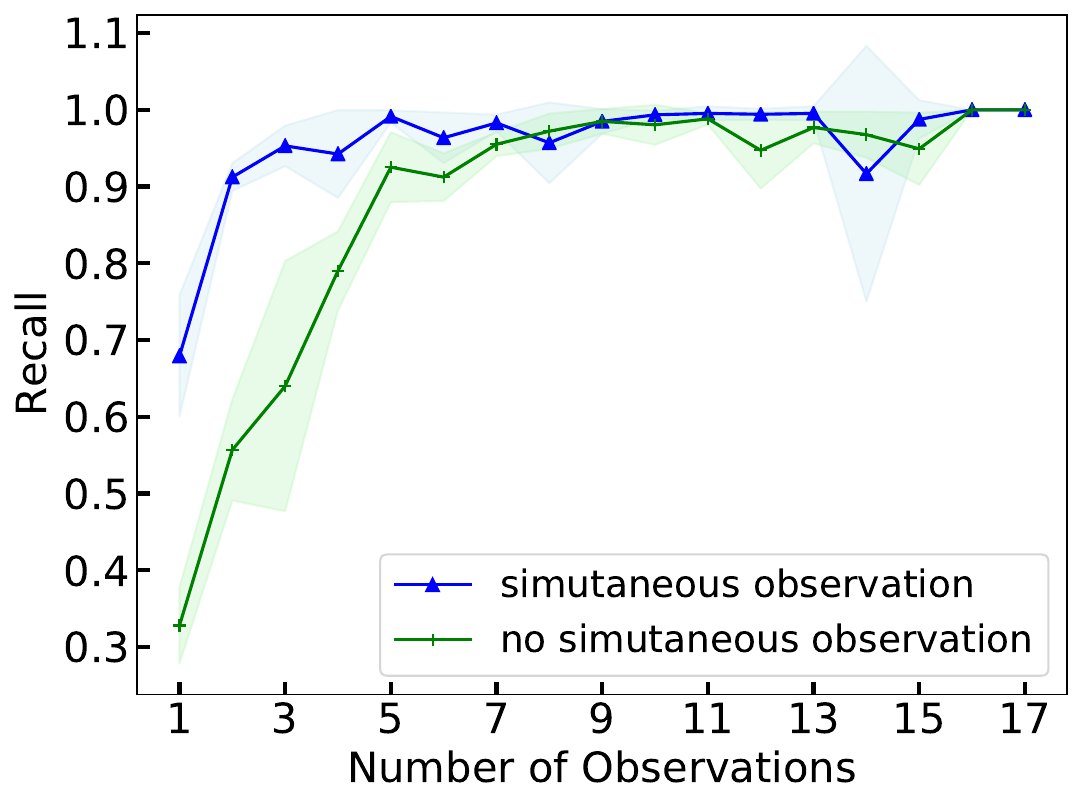}
  \end{minipage}
  \begin{minipage}[t]{0.495\textwidth}
  \centering
   \includegraphics[width=\textwidth]{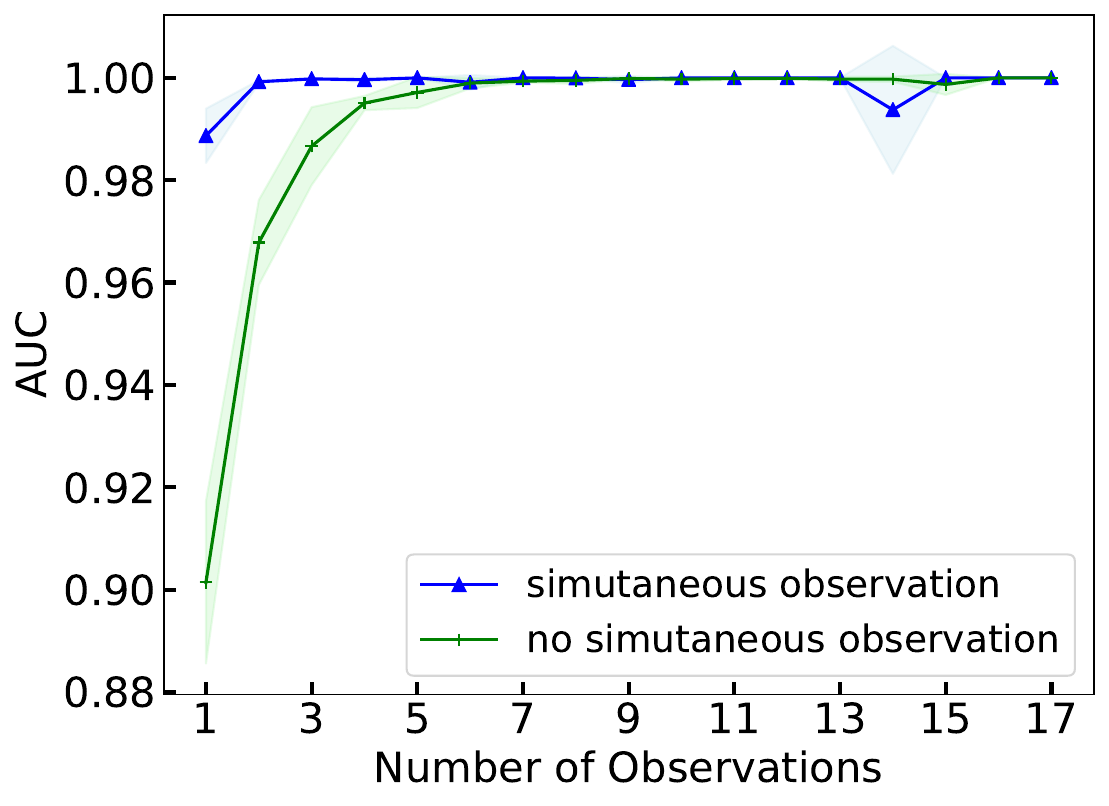}
  \end{minipage}
  \quad
  \begin{minipage}[t]{0.495\linewidth}
  \centering
   \includegraphics[width=\textwidth]{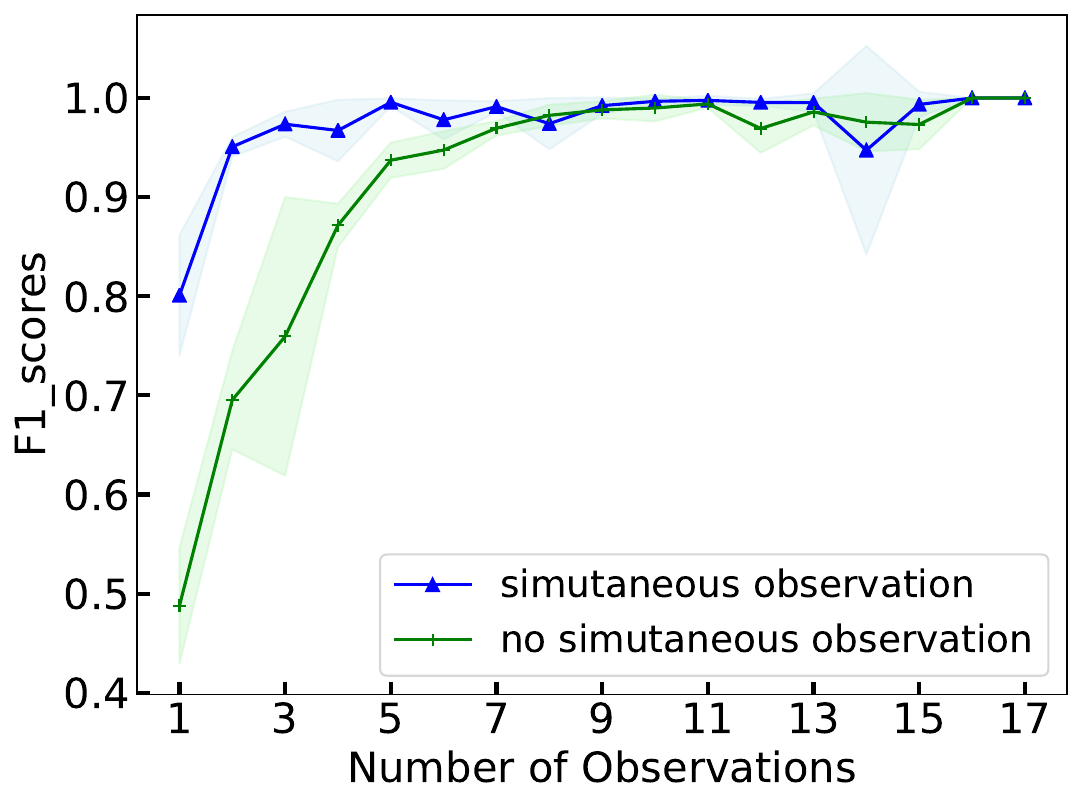}
  \end{minipage}
\caption{Comparison between simultaneous/non-simultaneous observation for five evaluation metrics. The blue triangle solid line with the corresponding light blue area represents the results obtained from simultaneous observations, and the green cross line represents those obtained from non-simultaneous observations.}
\label{fig:vs_nomephisto}
\end{figure}

In Fig.~\ref{fig:vs_nomephisto}, one can see that our network demonstrates good performance in the classification of SNe, e.g. the AUC reaches nearly $100\%$ after three days of observation. Almost in all cases, the simultaneous observation outperforms that of no simultaneous one, especially when only a few days of observations. 
With the accumulated sufficient data, \emph{\texttt{Mesiri}} achieves similar high performance in the early time identification of SNe Ia in both simultaneous and non-simultaneous observation scenarios.

\subsection{With color versus without color}

\begin{figure}
\centering
\includegraphics[width=0.8\textwidth, angle=0]{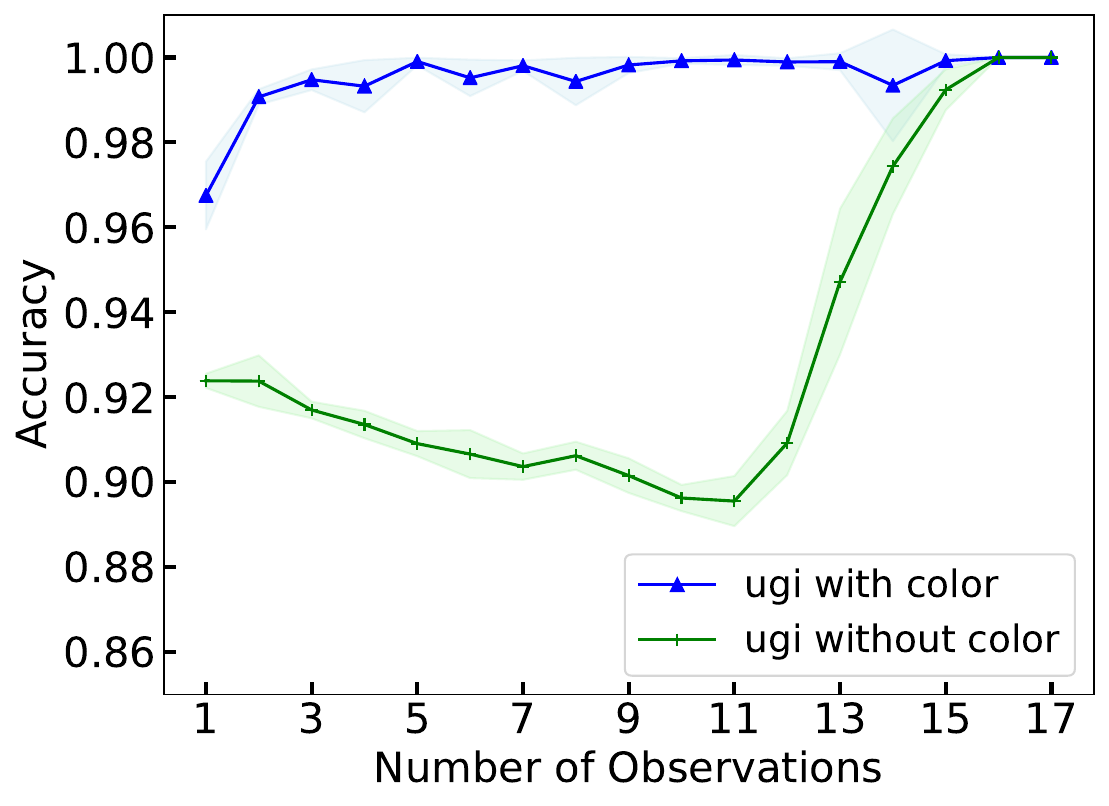}
\caption{The accuracy of classification models, with or without color, is compared in this plot. The results of an early SN explosion with random epochs of observation are represented by the blue triangle line, while the light blue region represents the standard deviation of model errors after five-fold cross-validation. On the other hand, the results of early explosion without color information are represented by the green cross line and green region.}
\label{fig:vs_withcolor} 
\end{figure}

We emphasize that each additional computation incurs a time cost, even in the present era with remarkable computing capabilities. To assess the model’s capacity to deduce color information, we undertook a test to determine if explicitly incorporating color information into the data improved the model's performance.

The experimental results from Fig.~\ref{fig:vs_withcolor} demonstrate that the accuracy of observed data explicitly containing color information exceeds $96$\%. Starting from the accumulation of second observation data, the identification accuracy of SNe Ia surpasses $98$\%. This is depicted by the solid lines in the light blue triangle region of Fig.~\ref{fig:vs_withcolor}. Conversely, observed data without explicit color information exhibits lower accuracy, below $94$\% for the majority of early-time SN eruptions. This is shown by the dashed lines in the light green region of Fig.~\ref{fig:vs_withcolor}. Consequently, it can be inferred that the accuracy of observed data explicitly containing color information significantly surpasses the accuracy achieved when color information is not explicitly displayed. It is worth noting that even without explicitly displayed color information, the combined observations of the~\emph{u, g, i} bands using our BiLSTM-based model architecture result in an identification accuracy of SNe Ia is still above $89.55\pm0.59$\%. This accuracy surpasses the precision attained in existing research on early-time identification of SNe Ia.

\subsection{\emph{u, g, i} versus~\emph{v, r, z}}

\begin{figure}
\centering
\includegraphics[width=0.8\textwidth, angle=0]{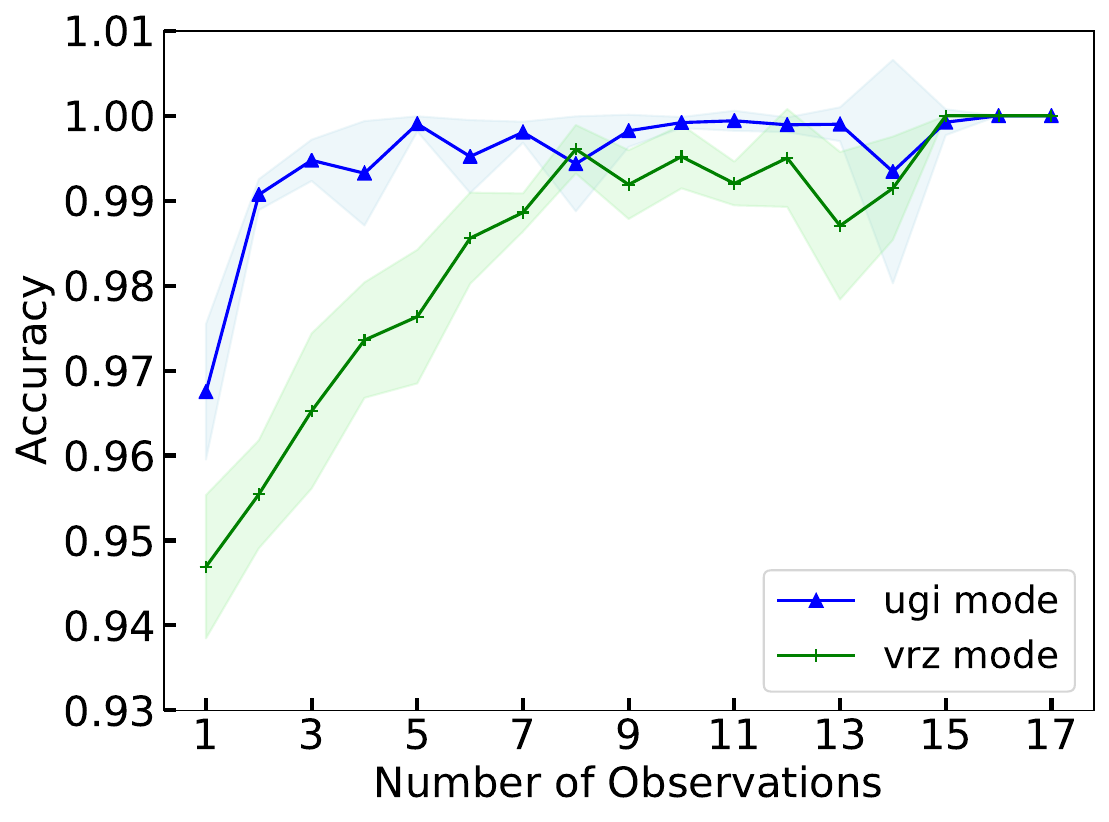}
\caption{The accuracy of early classification models is compared with respect to~\emph{v, r, z}. The blue triangle line represents the results of early SN explosion based on epochs of observation~\emph{u, g, i}, while the light blue region indicates the standard deviation of model errors after performing five-fold cross-validation. Conversely, the results of early explosion using~\emph{v, r, z} mode epochs of observation are represented by the green cross line and green region.}
\label{fig:vs_vrz} 
\end{figure}

On account of the scheduled survey mode of Mephisto, which is divided into filter combinations of \emph{u, g, i} and \emph{v, r, z}, this section examines the efficacy of identifying SNe Ia in the early stages of their explosion under the \emph{v, r, z} filter combination mode. We selected the Bidirectional Long Short-Term Memory (BiLSTM) method, which had the best comprehensive evaluation in the above methods, as the final basic unit method to compare the performance of SNe Ia in the \emph{u, g, i} and \emph{v, r, z} bands. The comparative evaluation criteria employ the same metrics as \emph{{u, g, i}}, specifically encompassing accuracy, precision, recall, AUC, and F1 score. Since these five metrics yield comparable findings upon comparison, only the accuracy comparison figure~\ref{fig:vs_vrz} is shown. The classification accuracy, shown in Fig.~\ref{fig:vs_vrz}, indicates that the~\emph{u, g, i} filters yield slightly better accuracy than the~\emph{v, r, z} filters when the observed data spans one or two days. In this regard, we compared the early-time spectra data of SNe Ia and CCSNe by selecting and combining them with the Mephisto total efficiency curve. We found that the variation of emission and absorption lines in the \emph{u, g, i} bands of SNe Ia explosions are much larger than those of CCSNe during the early time explosion. Furthermore, the variation of emission and absorption lines in the \emph{u, g, i} bands during the early time of SNe Ia explosions are stronger than those in the \emph{v, r, z} bands. In spite of that, both the~\emph{u, g, i} and~\emph{v, r, z}-based methods achieve an accuracy of over $94$\% in the early time identification of SNe Ia.

\subsection{Shallow learning}
\begin{figure}[h]
  \begin{minipage}[t]{0.495\linewidth}
  \centering
   \includegraphics[width=80mm]{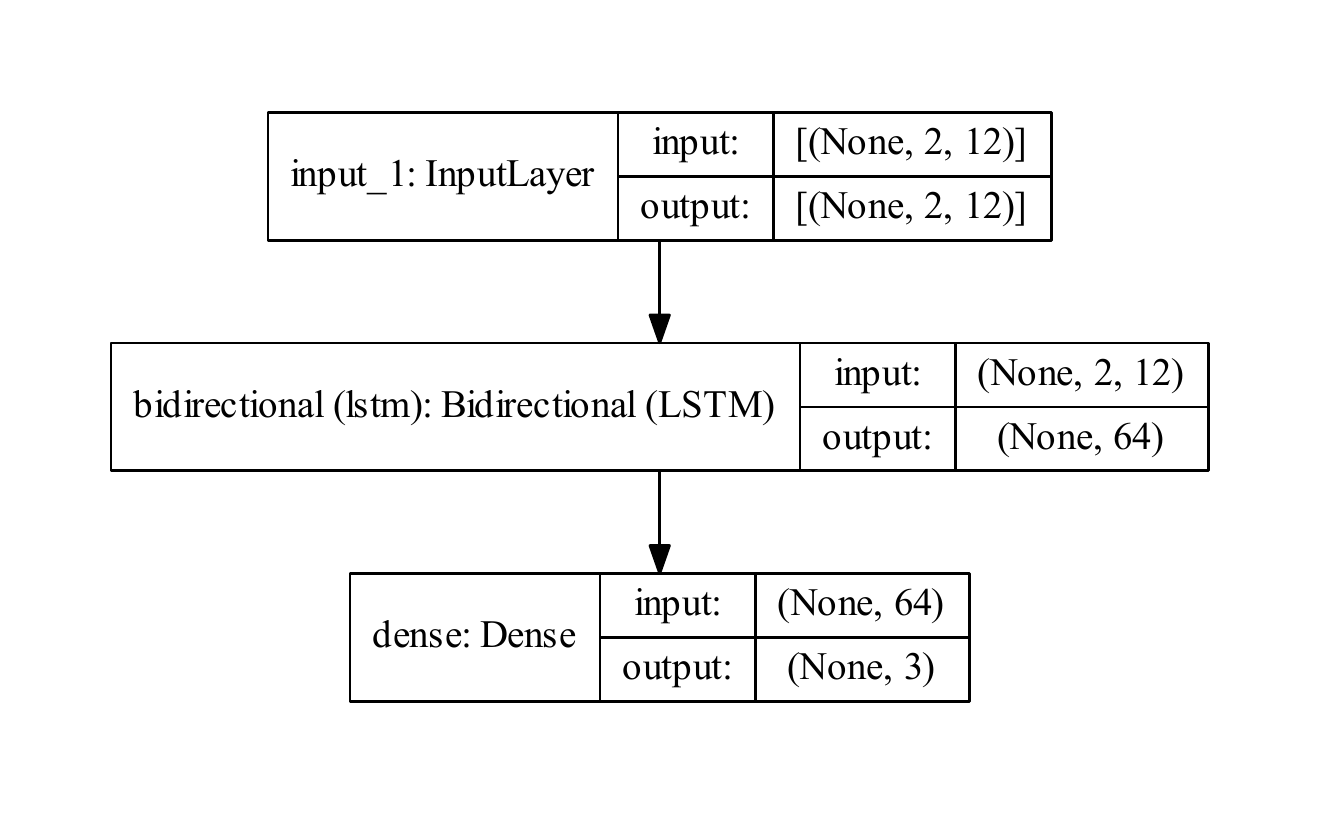}
  \end{minipage}
  \begin{minipage}[t]{0.495\textwidth}
  \centering
   \includegraphics[width=\textwidth]{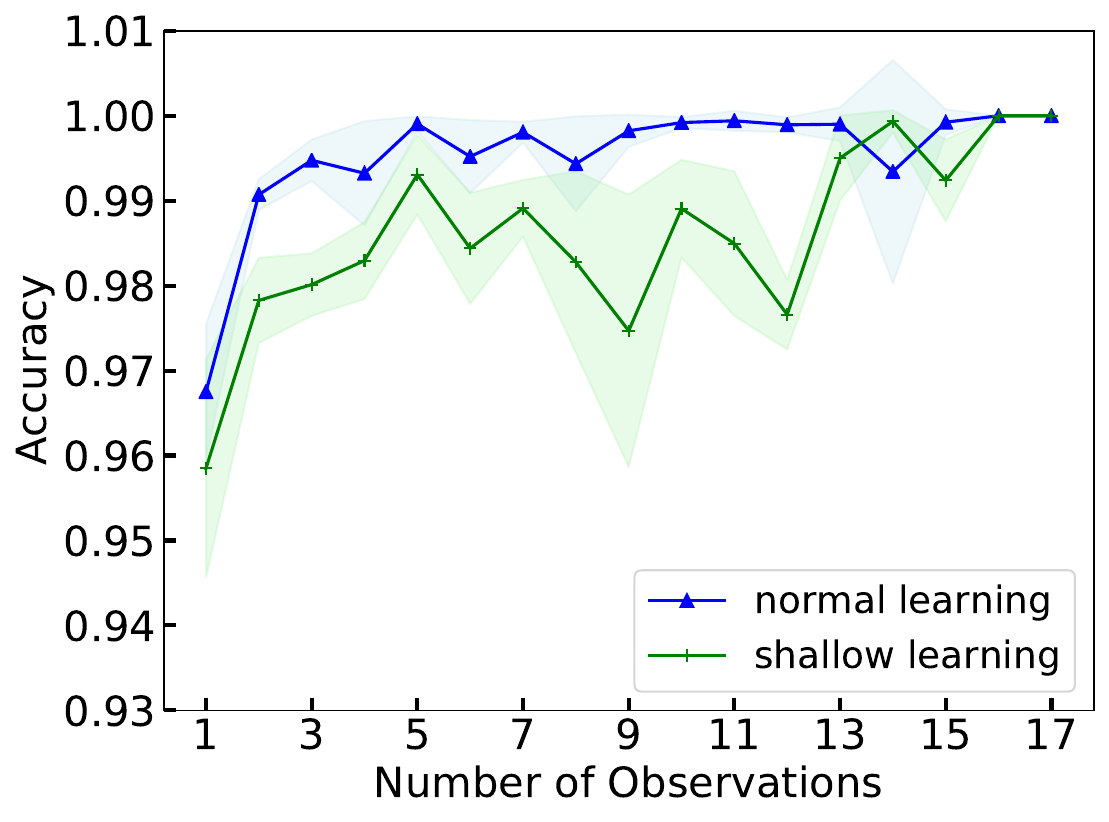}
  \end{minipage}
\caption{Shallow learning architecture, which is composed of a single BiLSTM layer of 64 units, is shown in the left panel. A comparison of the accuracy between normal learning architecture (blue triangle line and light blue region) and shallower learning (green cross line and light green region) is displayed in the right panel. Following five-fold cross-validation, the standard deviation of model errors is represented by the light shadow region.}
\label{fig:vs_shallow}
\end{figure}

\citet{2023ApJ...954....6G} achieves comparable or superior results to the leading classification algorithms with a simpler network architecture with the photometric redshift, extinction,
and host galaxy photometry (shallow learning) and achieves an overall accuracy of $82\pm2$\% and an accuracy of $87\pm5$\% at both early (within $3$ days of an event’s discovery) and
late phases (within $30$ days of discovery). We also construct a shallow learning architecture with only light curve with true color information, by way of contrast, shown in the left panel in Fig.~\ref{fig:vs_shallow}. It consists of a single BiLSTM layer of 64 units. From the right panel in Fig.~\ref{fig:vs_shallow}, which is a comparison of the accuracy between \emph{\texttt{Mesiri}} and shallower learning. Upon scrutiny, it becomes evident that, despite its reduced computational time, the shallow learning model does not demonstrate the same level of effectiveness as the \emph{\texttt{Mesiri}}. 

\section{Summary}
\label{sect:summary}

The Multi-channel Photometric Survey Telescope (Mephisto) has established itself as a powerful facility for time-domain astronomy owing to its capacity to capture the real-time color of various celestial objects, including transients. The real-time color of the SNe explosion can provide vital information about the progenitor stars and the explosion mechanism, especially during the early phase. The real-time color with better accuracy can be used for the early classification of transients such as various types of SNe or TDEs, etc as well. We take advantage of simultaneous three-band photometry by the Mephisto facility and study the classification between type Ia SNe and core-collapse SNe. We developed an identifier, \emph{\texttt{Mesiri}} based on the recurrent neural network. The training samples are simulated according to the observing features of Mephisto (c.f. simultaneous multi-band observations), the weather conditions of its location, and 1-day cadence observational mode data. We focus on both real-time and non-real-time observational data with random cadence. The identifier \emph{\texttt{Mesiri}} can efficiently identify early observed SNe Ia with accuracy, precision, and an AUC above $96\%$. And the accuracy, precision, and AUC reached 96.75\%, 98.42\%, and 98.87\%, respectively, when real-time data was considered (i.e. once single epoch observational data we have). Specifically, the identification accuracy, precision, and AUC of SNe Ia when the first discovery/trigger without redshift are $97.15\pm0.63$\%, $92.86\pm4.10$\%, $98.5\pm0.75$\%. Our result has a better significance than the previous studies of a similar kind and emphasizes the crucial information by the true color. 
The key points of our study are summarized below:

\begin{itemize}
    \item The classification utilizing true color demonstrates superior performance compared to that employing pseudo-color, particularly during the initial phases characterized by limited observations.
    
    \item In this study, several neural networks are compared, including RNN, LSTM, GRU, and BiLSTM. Our results show that BiLSTM performs best in the early classification of SNe Ia explosions with true color information. Therefore, BiLSTM has been employed in successive evaluations (see Section~\ref{sect:discussion}).

    \item \emph{\texttt{Mesiri}} remains insensitive to the inclusion of redshift for the low redshift SNe.

    \item Our calculation indicates that although the identification accuracy of SNe Ia is higher (96\%) when the explicit color information (i.e., $(u-g)$, ${g-i)}$, ${(u-i)}$, ${(u-g)_{error}}$, ${(g-i)}_{error}$, ${(u-i)}_{error}$) is utilized. However, if we use only three bands and their corresponding errors (i.e., $u$, $g$, $i$, ${{u}_{error}}$, ${{g}_{error}}$, ${{i}_{error}}$), the identification accuracy is good enough ($\sim$90\%) but consuming less computation time. 

    \item The~\emph{u, g, i} filters yield slightly better accuracy than the~\emph{v, r, z} filters when the observed data spans one or two days. Nevertheless, both the~\emph{u, g, i} and~\emph{v, r, z}-based achieve an accuracy better than $94$\% in the early time identification of SNe Ia.
\end{itemize}

Additionally, one of the major contamination is the extinction and reddening from the Milky Way and the host galaxy. We perform extra comparisons with test samples including extinction \citep{2024AJ....167...31O}. 
By applying the Mesiri classifier to light curves with extinctions, we observed that the identification accuracy decreases to $\sim$88\% in average. Such effect will be incorporated in our future study.

Moreover, it is noteworthy that the training sample employed in our study was not subjected to data balancing, resulting in an imbalanced sample ratio between SNe Ia and CCSNe during model training. Despite this, our model can achieve unprecedented performance with true color information. However, in our study, we acknowledge the following limitations. Firstly, we carry out our research using simulated data, which cannot reflect the real observational noise or other systematics.
A larger training sample with real observations, especially data from the target telescope, i.e., Mephisto, is required. Secondly, the study of SNe Ia has also revealed their inherent diversity, which was not taken into account in our simulated data. The classification of sub-types of CCSNe as well as other transients, such as Tidal Disruption Events and Kilonovae, will be included in our classification scheme using true color information in future studies. It is worth noting that Mephisto is presently in the commissioning phase, and it's already providing good scientific data \citep[e.g.][]{2024Natur.625..253C,2024ApJ...971L...2C,2024ApJ...969..126Y}. 

\begin{acknowledgements}
Mephisto is developed at and operated by the South-Western Institute for Astronomy Research of Yunnan University (SWIFAR-YNU), funded by the ``Yunnan University Development Plan for World-Class University'' and ``Yunnan University Development Plan for World-Class Astronomy Discipline''. The authors acknowledge support from the ``Science \& Technology Champion Project'' (202005AB160002) and from two ``Team Projects'' -- the ``Innovation Team'' (202105AE160021) and the ``Top Team'' (202305AT350002), all funded by the ``Yunnan Revitalization Talent Support Program''. This research was also supported by Yunnan Fundamental Research Projects (grant NO.202301AU070006). We would like to extend our sincere appreciation to the referee for his/her insightful comments and constructive suggestions, which have significantly enhanced the quality of the manuscript.
\end{acknowledgements}

\appendix                  
\section{\emph{\texttt{Mesiri}} Early Classification Evaluations}
\label{app:evaluations}
 
In section~\ref{sect:results}, to concisely present the experimental findings, we have selected evaluation results for five scenarios of the \emph{\texttt{Mesiri}} as listed in Table~\ref{tbl:no-z-acc}. Here, we enumerate all scenario evaluation results obtained from our experiments. Table~\ref{tab:evaluations} presents an elaborate account of diverse evaluation metrics following a five-fold cross-validation on SNe Ia using units, taking into account the quantity of observed data from the commencement of the SN explosion.

\begin{table*}
\centering
\caption[]{Utilizing \emph{\texttt{Mesiri}} for the early-time identification of Type Ia SNe across all scenarios in the initial stages of SN explosions. Note that each evaluation, when combined with uncertainties represented by standard deviations, may exceed 100\%.}
\label{tab:evaluations}
\setlength{\tabcolsep}{3.3pt} 
\begin{tabular}{p{1.3cm}p{2.2cm}p{2.2cm}p{2.2cm}p{2.2cm}p{2.2cm}}
    \hline
    ObsDay(s)  &  Accuracy $ $ & Precision & Recall & F1\_score  & AUC \\
  \hline
1&$96.75\pm0.79$\%&$98.42\pm0.95$\%&$67.9\pm7.93$\%&$80.08\pm6.07$\%&$98.87\pm0.53$\%\\
2&$99.07\pm0.18$\%&$99.3\pm0.56$\%&$91.23\pm1.81$\%&$95.08\pm0.99$\%&$99.92\pm0.02$\%\\
3&$99.48\pm0.24$\%&$99.56\pm0.53$\%&$95.3\pm2.64$\%&$97.36\pm1.26$\%&$99.98\pm0.00$\%\\
4&$99.32\pm0.61$\%&$99.56\pm0.53$\%&$94.24\pm5.72$\%&$96.73\pm3.09$\%&$99.96\pm0.04$\%\\
5&$99.9\pm0.08$\%&$100.0\pm0.0$\%&$99.14\pm0.80$\%&$99.57\pm0.40$\%&$100.0\pm0.00$\%\\
6&$99.52\pm0.42$\%&$99.35\pm1.30$\%&$96.36\pm3.29$\%&$97.8\pm1.96$\%&$99.91\pm0.14$\%\\
7&$99.81\pm0.12$\%&$100.0\pm0.0$\%&$98.29\pm1.08$\%&$99.14\pm0.55$\%&$100.0\pm0.00$\%\\
8&$99.42\pm0.55$\%&$99.38\pm1.23$\%&$95.7\pm5.26$\%&$97.41\pm2.59$\%&$99.99\pm0.01$\%\\
9&$99.82\pm0.18$\%&$100.0\pm0.0$\%&$98.49\pm1.60$\%&$99.22\pm0.82$\%&$99.96\pm0.07$\%\\
10&$99.92\pm0.06$\%&$100.0\pm0.0$\%&$99.35\pm0.53$\%&$99.67\pm0.26$\%&$100.0\pm0.00$\%\\
11&$99.94\pm0.11$\%&$100.0\pm0.0$\%&$99.53\pm0.93$\%&$99.76\pm0.47$\%&$100.0\pm0.0$\%\\
12&$99.89\pm0.08$\%&$99.71\pm0.57$\%&$99.42\pm0.71$\%&$99.56\pm0.35$\%&$100.0\pm0.0$\%\\
13&$99.9\pm0.19$\%&$99.53\pm0.93$\%&$99.53\pm0.93$\%&$99.53\pm0.93$\%&$100.0\pm0.00$\%\\
14&$99.33\pm1.31$\%&$100.0\pm0.0$\%&$91.67\pm16.6$\%&$94.74\pm10.5$\%&$99.38\pm1.25$\%\\
15&$99.92\pm0.15$\%&$100.0\pm0.0$\%&$98.75\pm2.5$\%&$99.35\pm1.29$\%&$100.0\pm0.0$\%\\
16&$100.0\pm0.0$\%&$100.0\pm0.0$\%&$100.0\pm0.0$\%&$100.0\pm0.0$\%&$100.0\pm0.0$\%\\
17&$100.0\pm0.0$\%&$100.0\pm0.0$\%&$100.0\pm0.0$\%&$100.0\pm0.0$\%&$100.0\pm0.0$\%\\
\hline
\end{tabular}
\end{table*}

\section{Sample Distribution}
\label{app:distprepeak}
Given the assumption in our sample selection that the rise time to peak luminosity of the SN explosion is 17 days, practical observational constraints often prevent the availability of complete observational data. To simulate realistic scenarios, each scenario assumes a different duration of observational data. Consequently, we employed a scenario-specific model training approach to better illustrate the sample distribution for the pre-maximum epochs in each scenario during the model training process. Fig.~\ref{fig:distprepeak}  and Fig.~\ref{fig:distprepeak2} depict the sample distribution for the pre-maximum epochs in each scenario (the pre-maximum epochs are provided in the title description of each subplot).

\begin{figure}[h]
  \begin{minipage}[t]{0.495\linewidth}
  \centering
   \includegraphics[width=\textwidth]{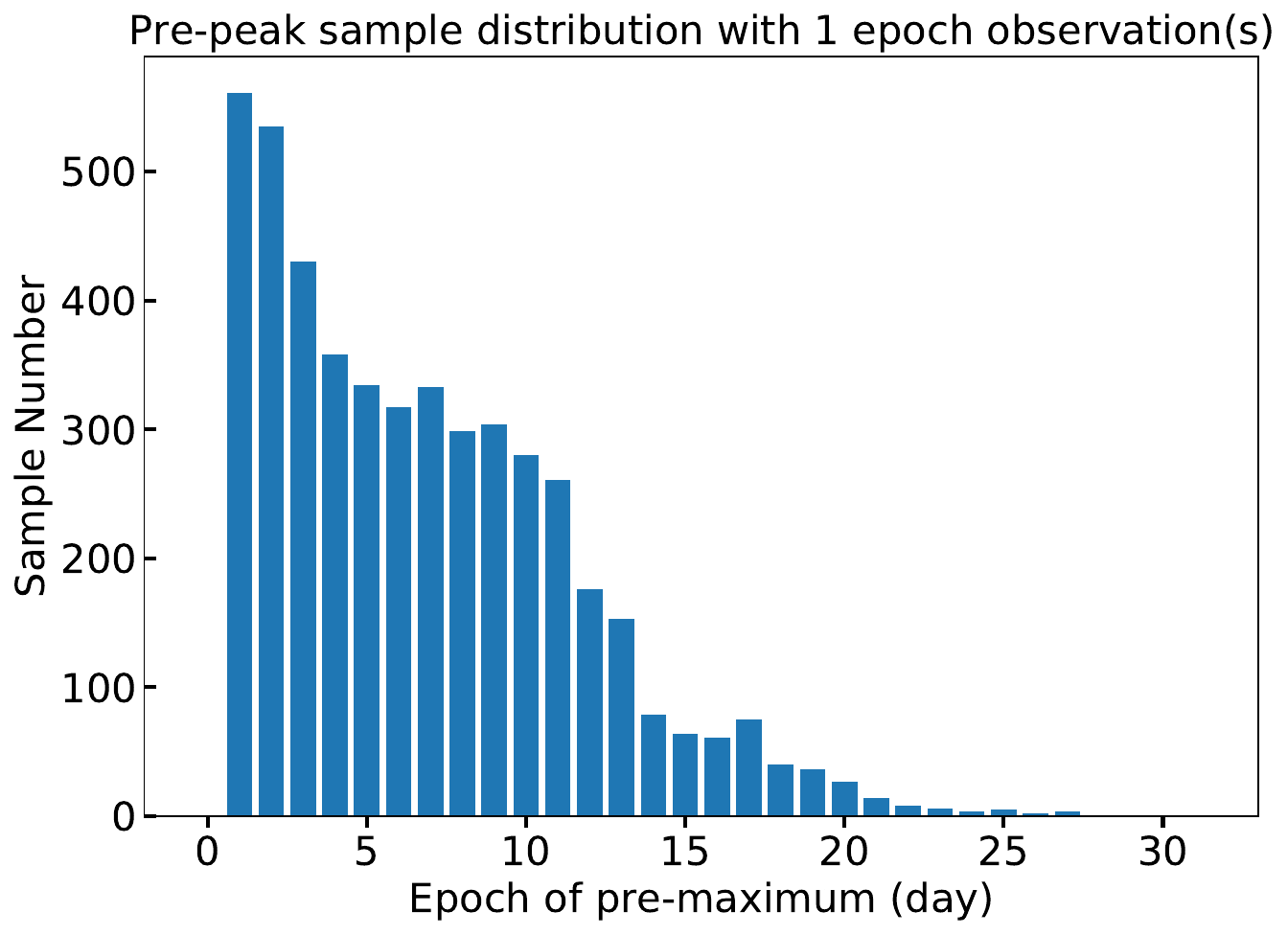}
  \end{minipage}
  \begin{minipage}[t]{0.495\textwidth}
  \centering
   \includegraphics[width=\textwidth]{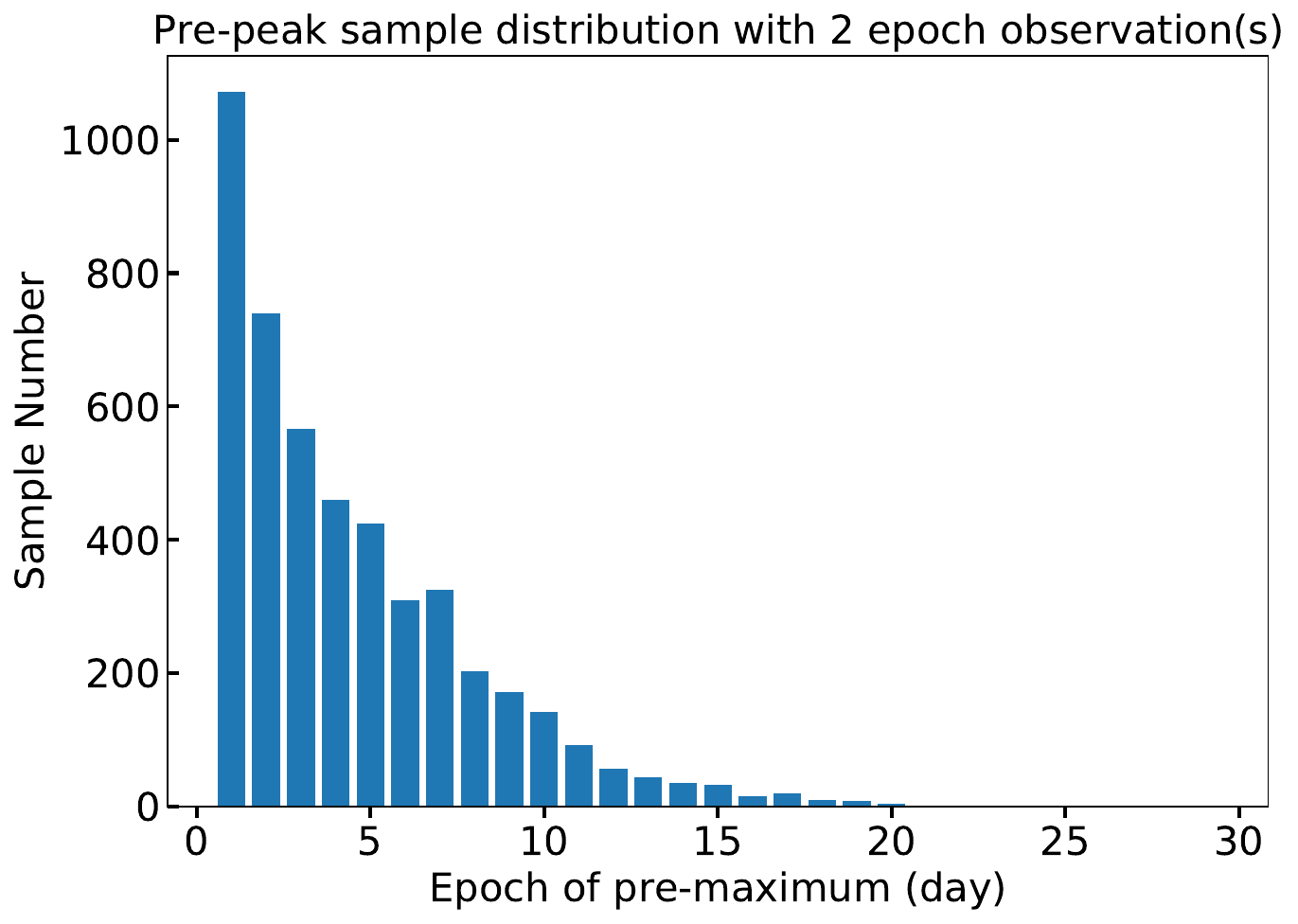}
  \end{minipage}
  \quad
  \begin{minipage}[t]{0.495\textwidth}
  \centering
   \includegraphics[width=\textwidth]{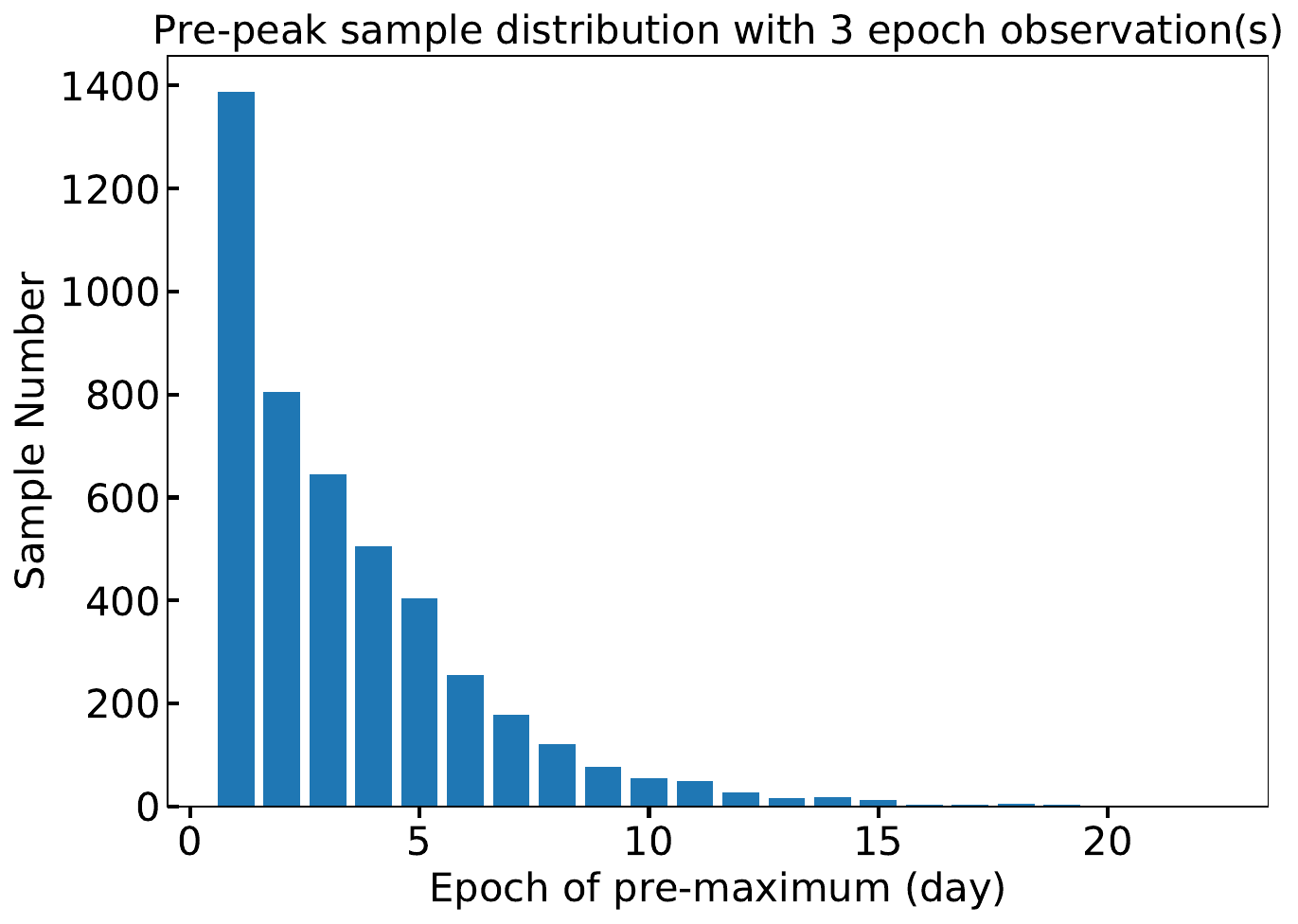}
  \end{minipage}
  \begin{minipage}[t]{0.495\textwidth}
  \centering
   \includegraphics[width=\textwidth]{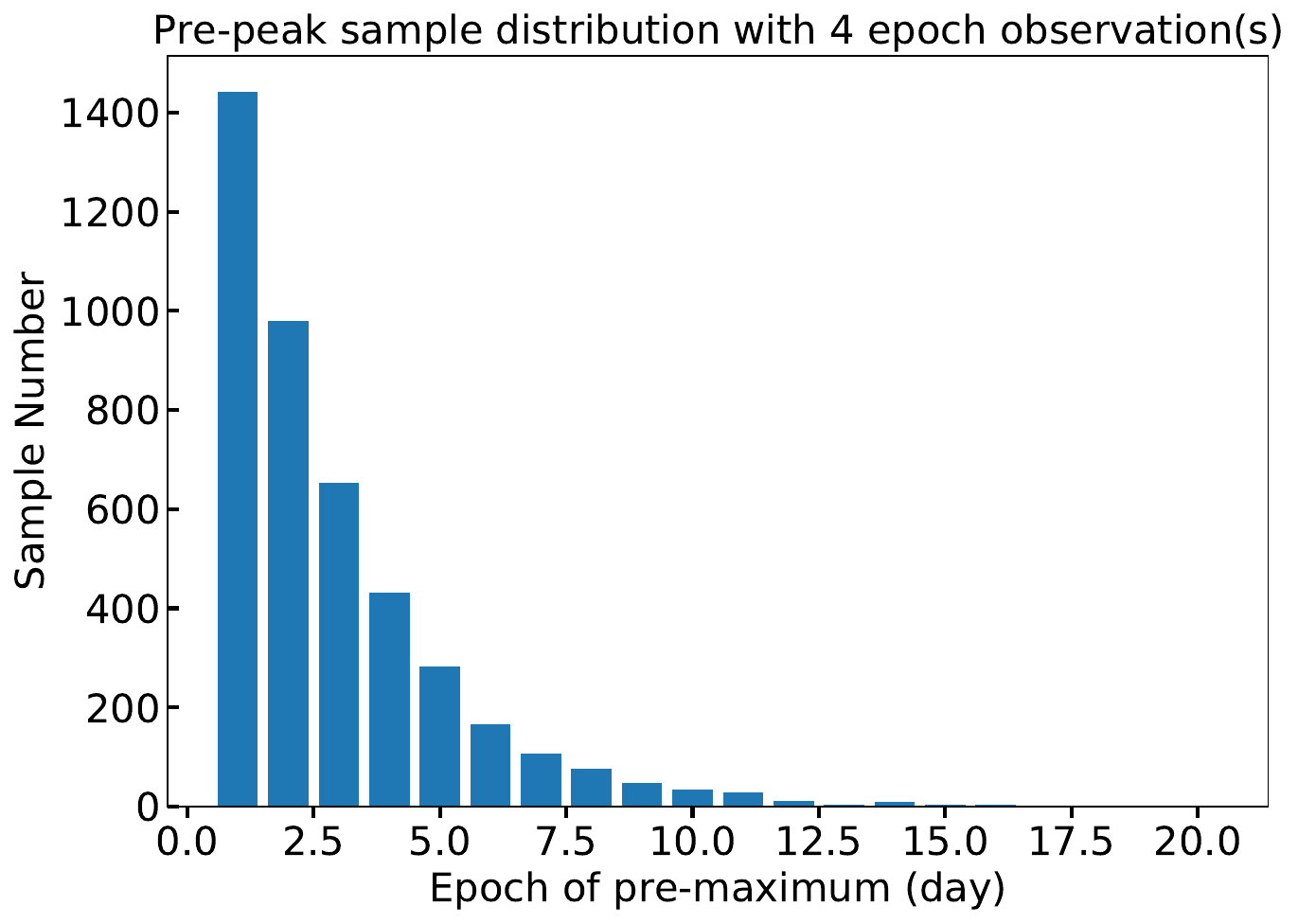}
  \end{minipage}
  \quad
  \begin{minipage}[t]{0.495\textwidth}
  \centering
   \includegraphics[width=\textwidth]{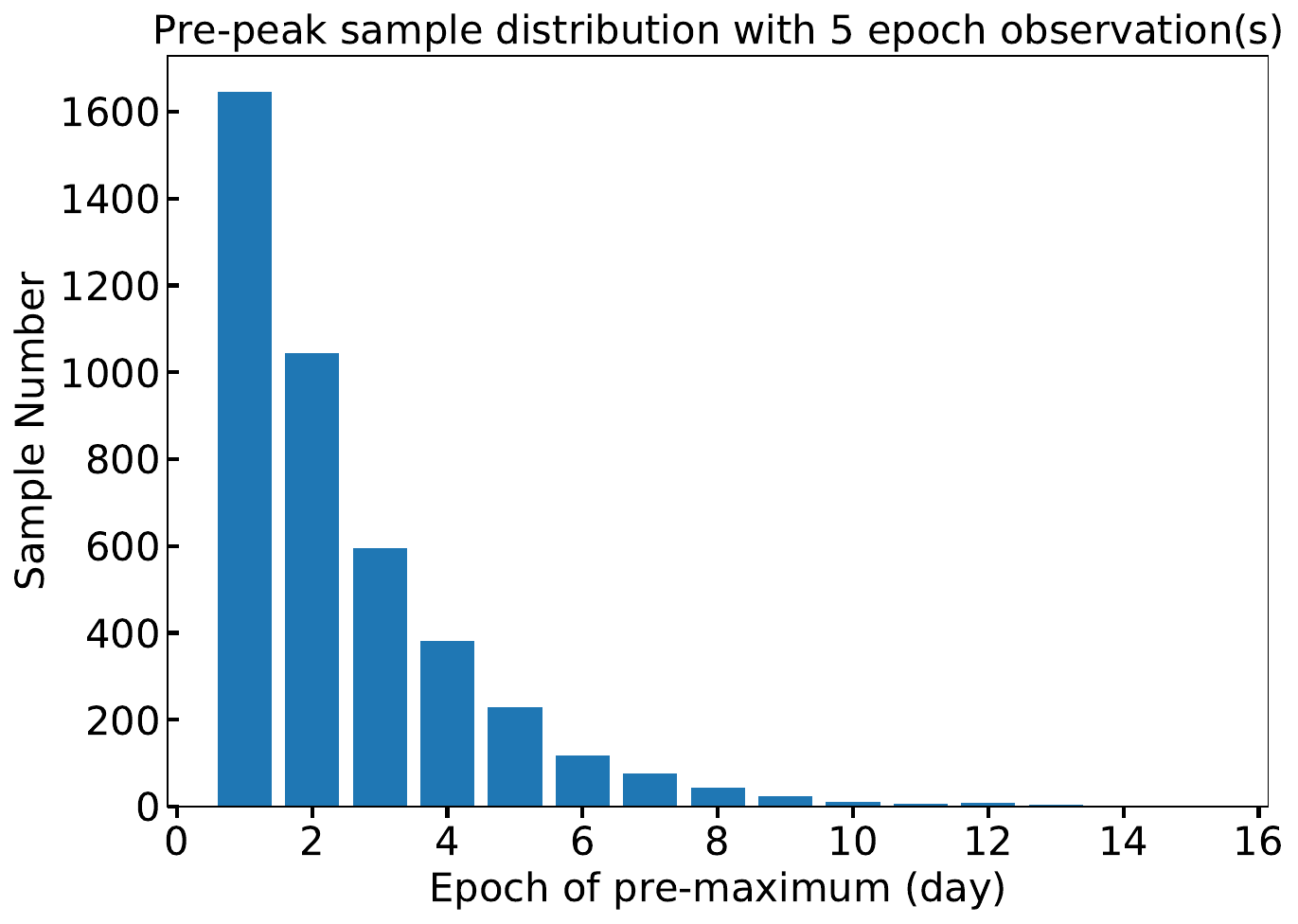}
  \end{minipage}
  \begin{minipage}[t]{0.495\textwidth}
  \centering
   \includegraphics[width=\textwidth]{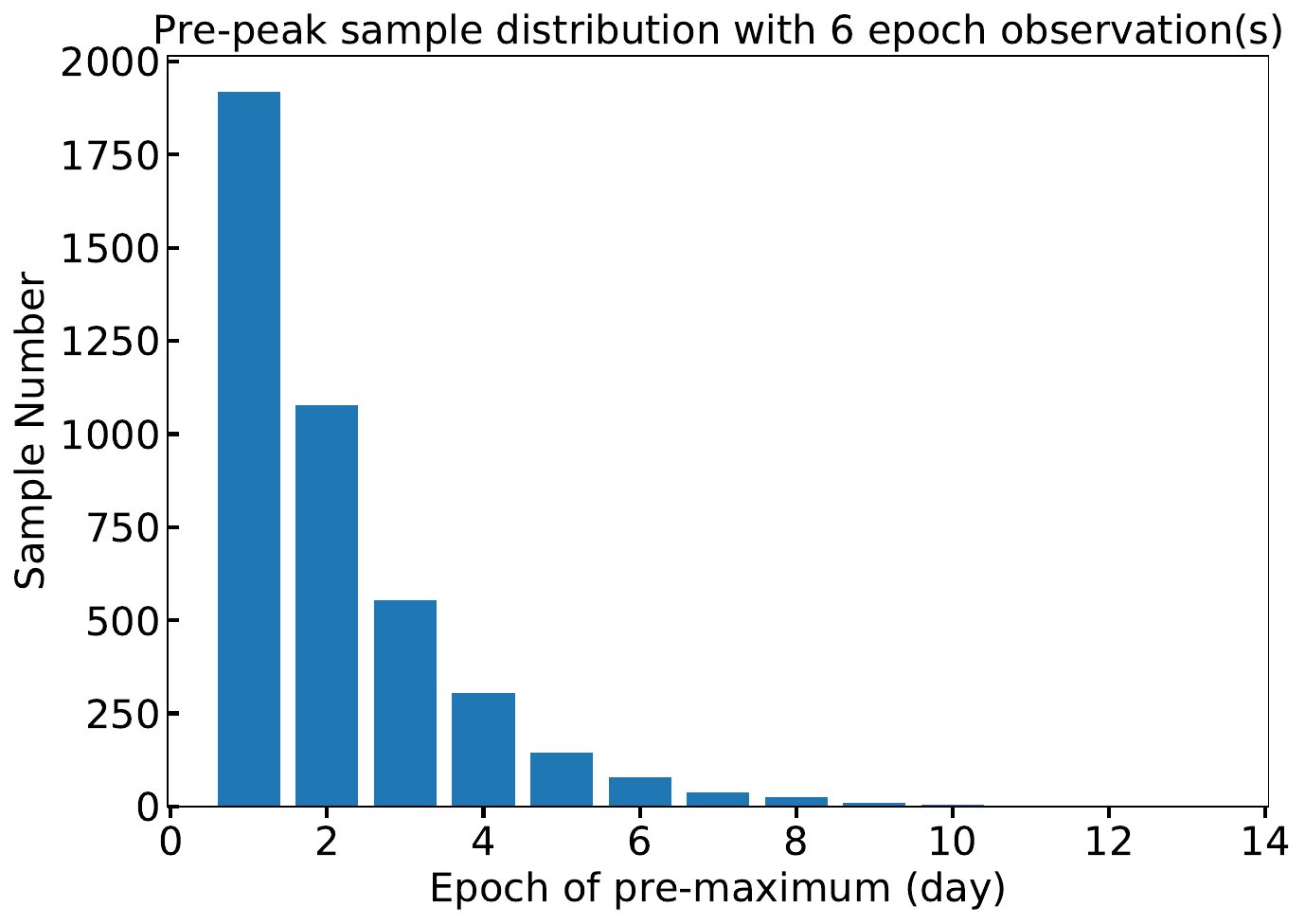}
  \end{minipage}
  \quad
  \begin{minipage}[t]{0.495\textwidth}
  \centering
   \includegraphics[width=\textwidth]{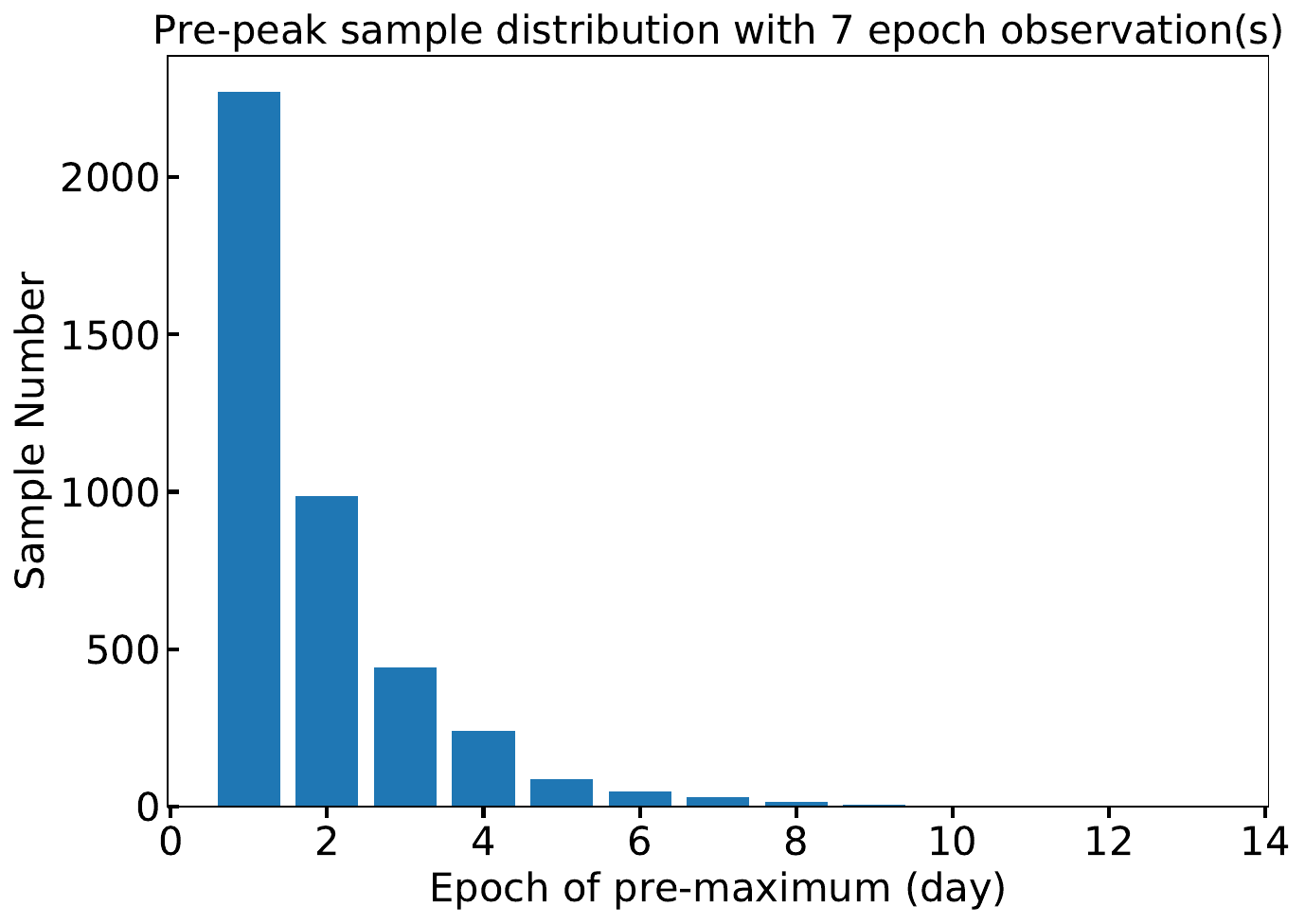}
  \end{minipage}
    \begin{minipage}[t]{0.495\textwidth}
  \centering
   \includegraphics[width=\textwidth]{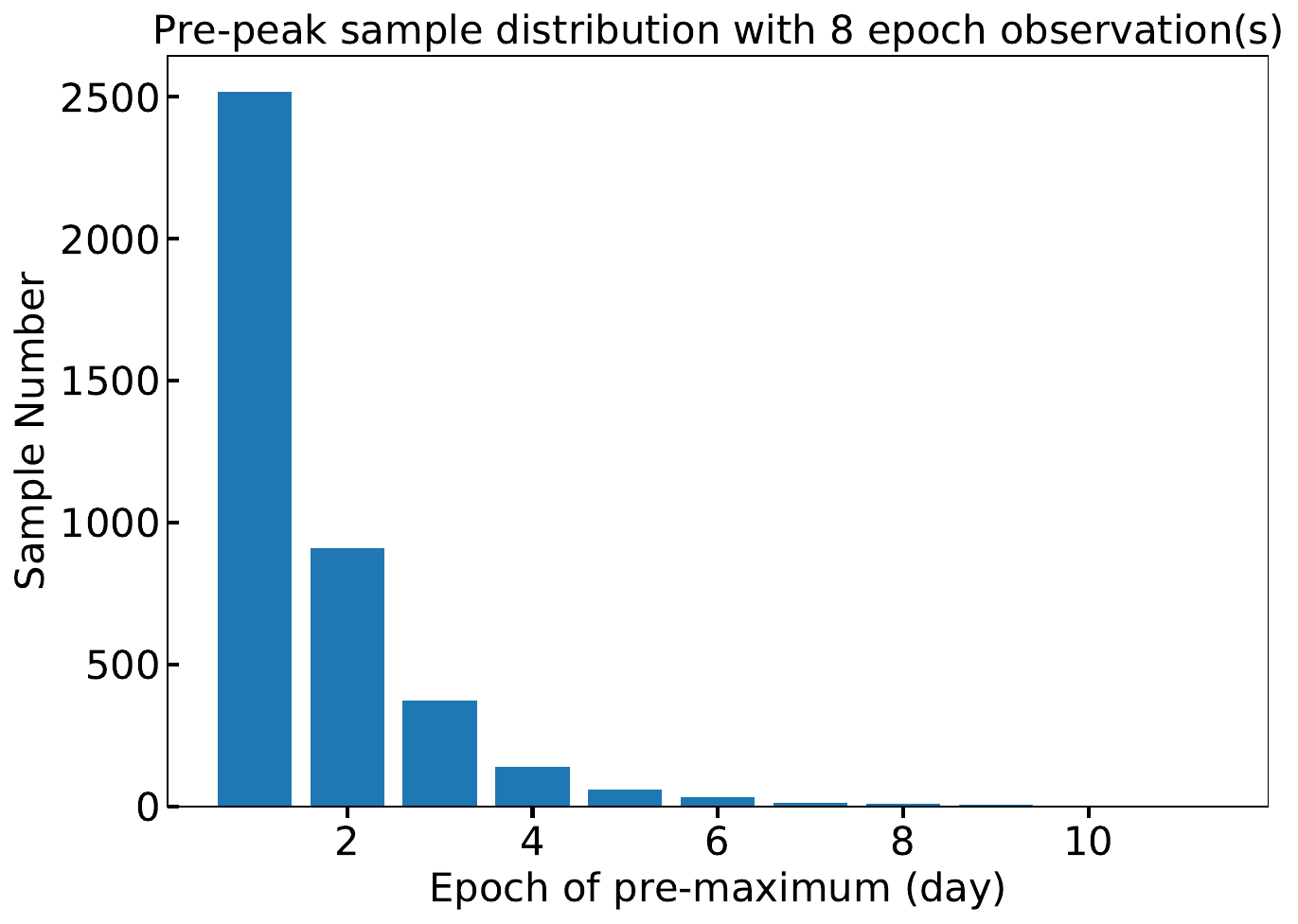}
  \end{minipage}%
\caption{The sample distribution for the pre-maximum epochs in each scenario (the pre-maximum epochs are provided in the title description of each subplot).}
\label{fig:distprepeak}
\end{figure}

\begin{figure}[h]
  \begin{minipage}[t]{0.495\linewidth}
  \centering
   \includegraphics[width=\textwidth]{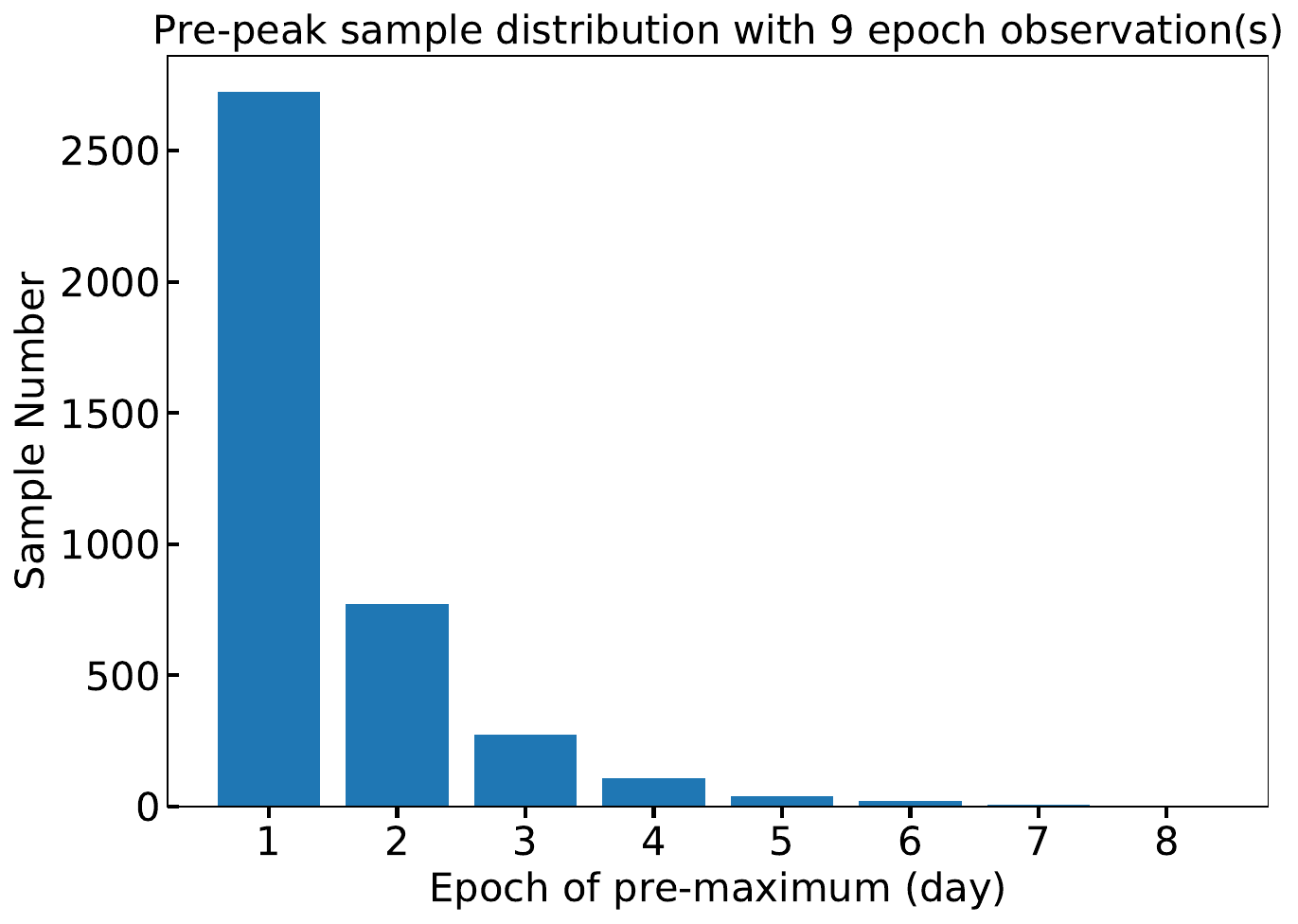}
  \end{minipage}
  \begin{minipage}[t]{0.495\textwidth}
  \centering
   \includegraphics[width=\textwidth]{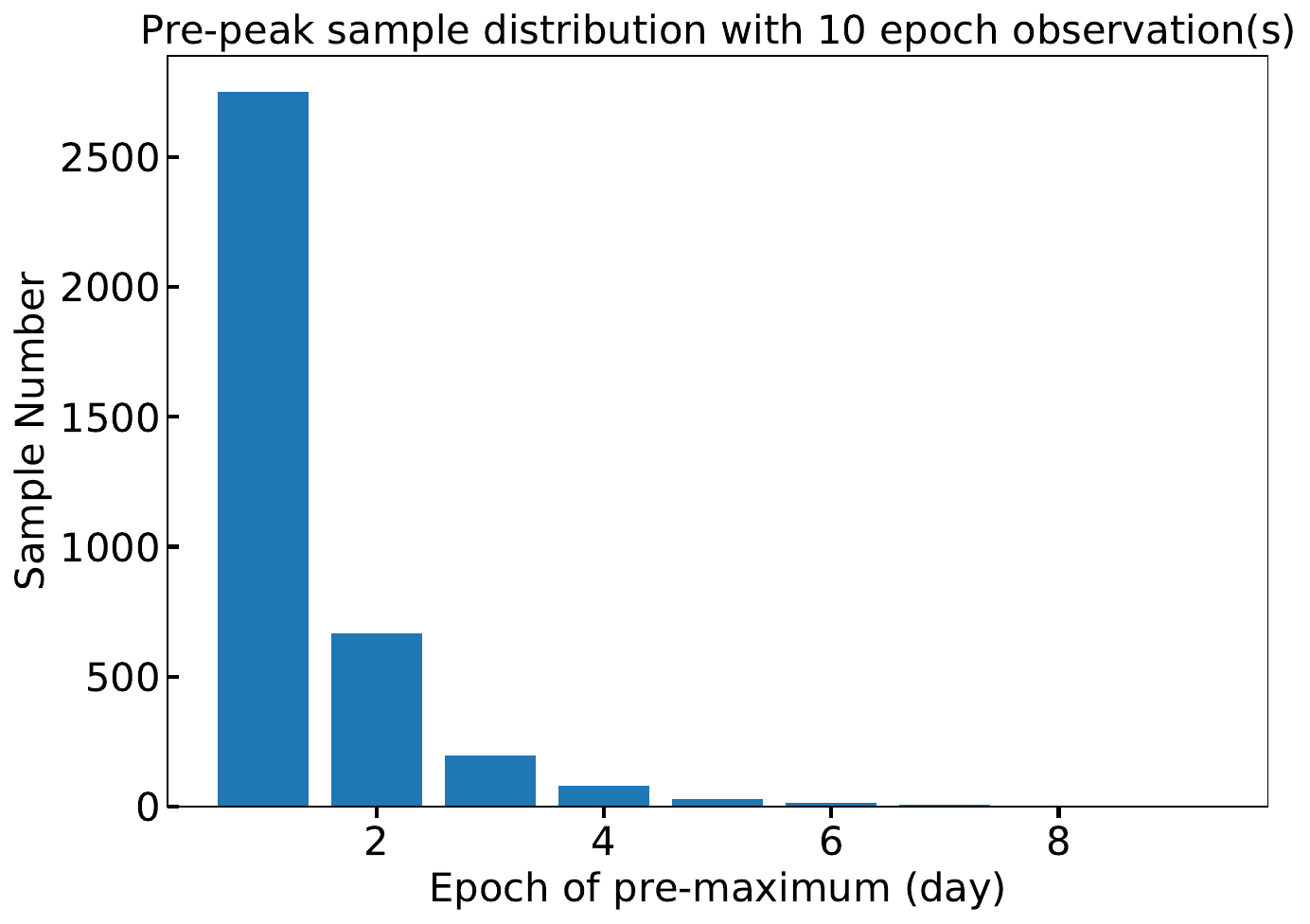}
  \end{minipage}
  \quad
  \begin{minipage}[t]{0.495\textwidth}
  \centering
   \includegraphics[width=\textwidth]{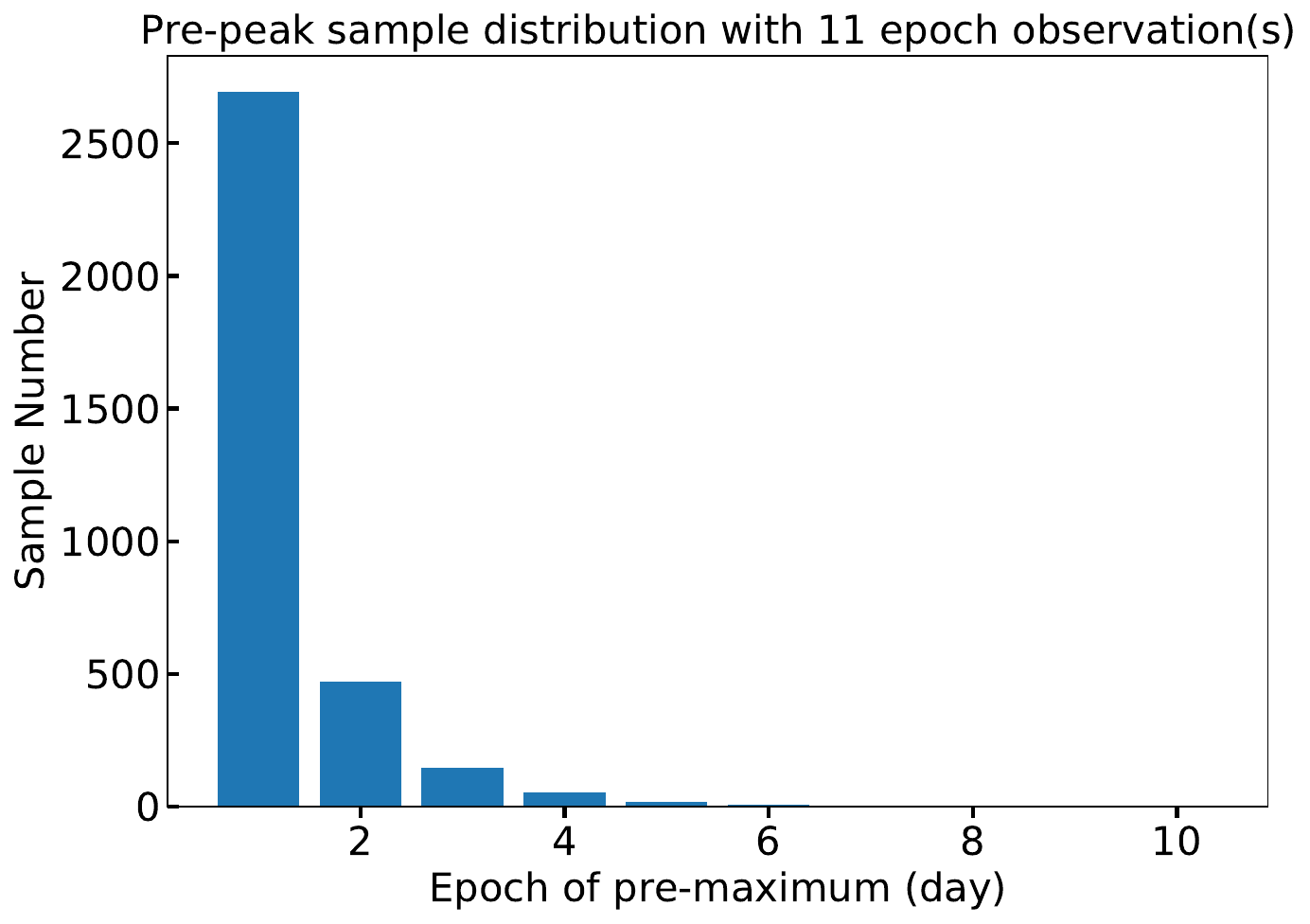}
  \end{minipage}
  \begin{minipage}[t]{0.495\textwidth}
  \centering
   \includegraphics[width=\textwidth]{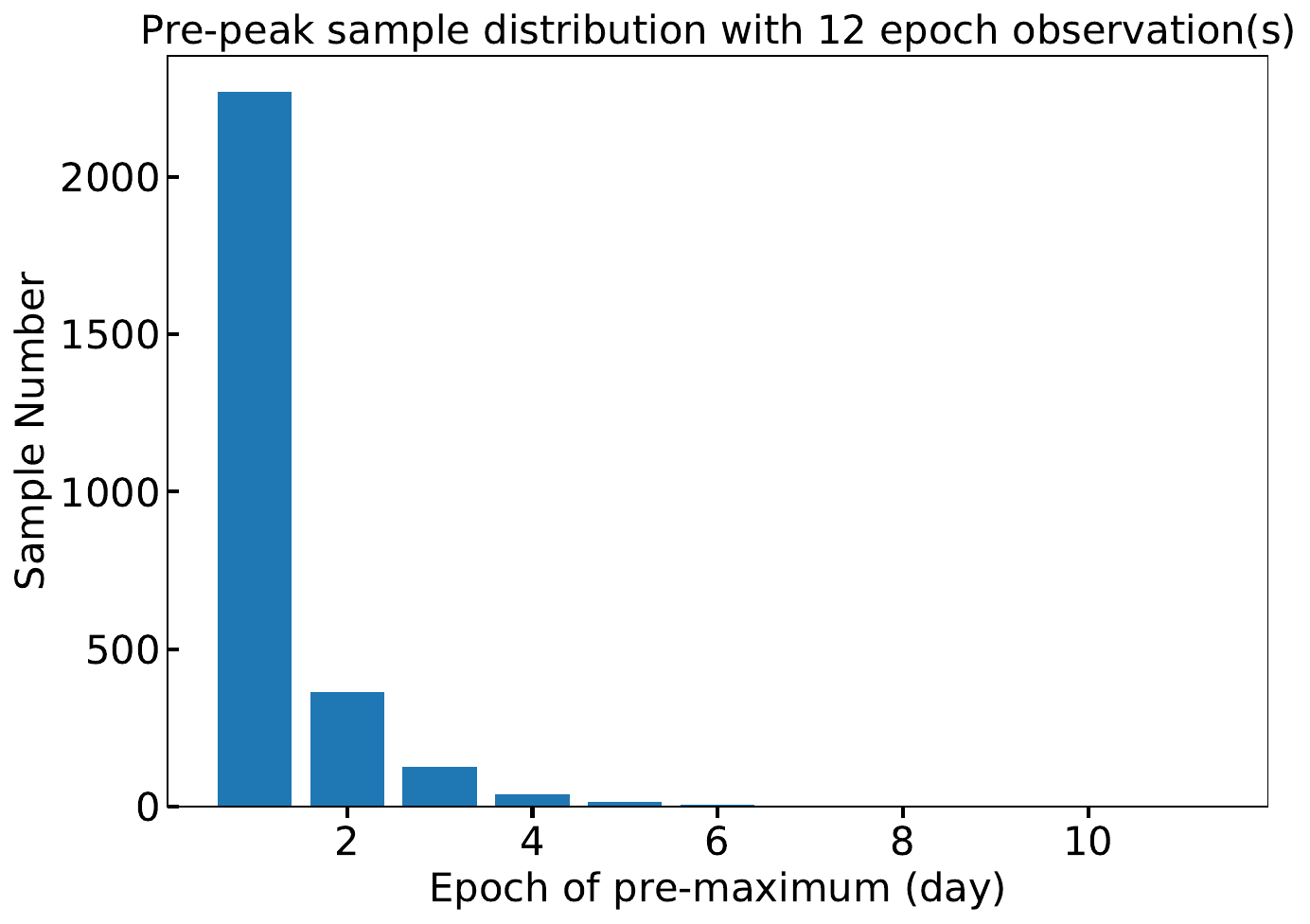}
  \end{minipage}
  \quad
  \begin{minipage}[t]{0.495\textwidth}
  \centering
   \includegraphics[width=\textwidth]{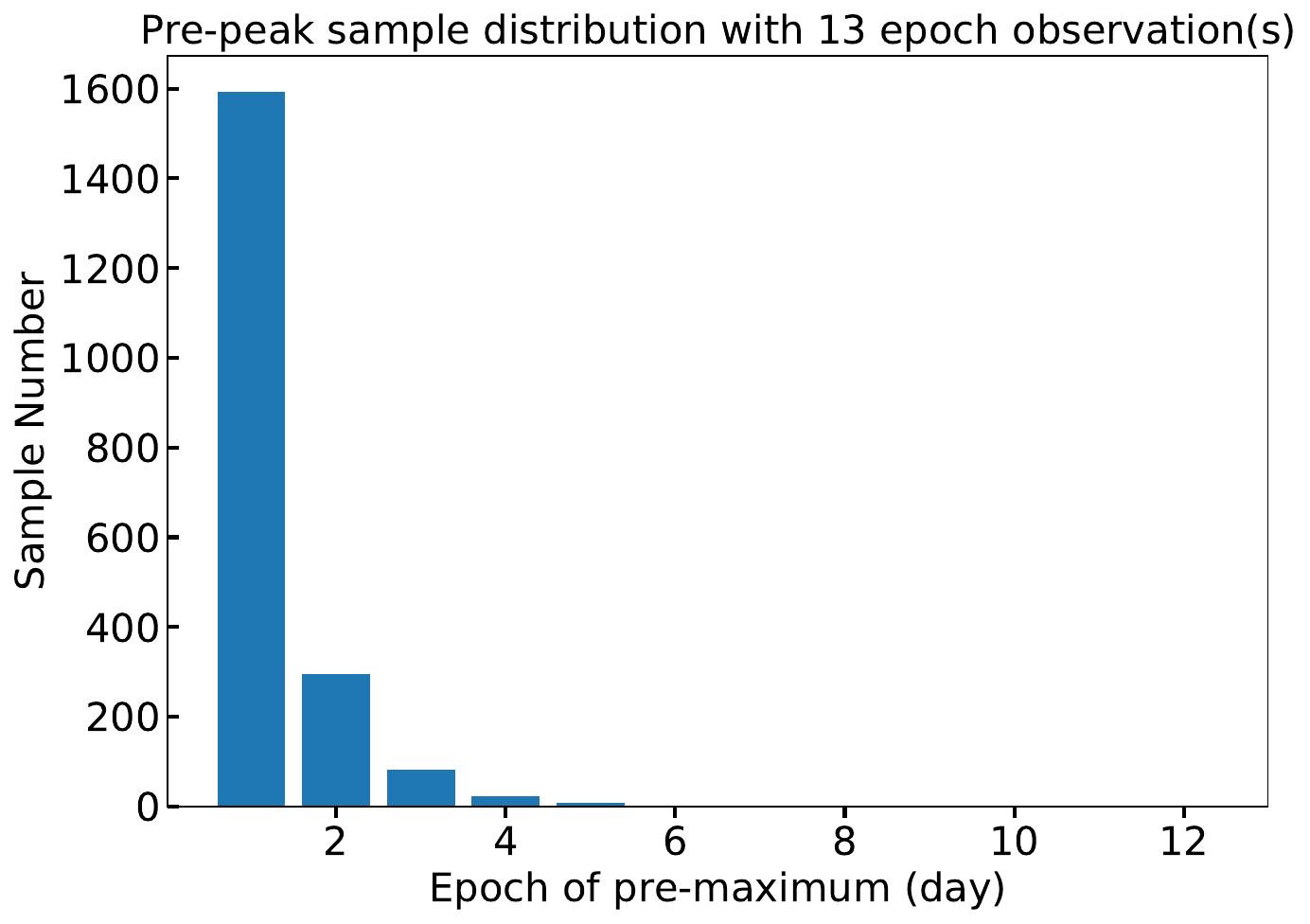}
  \end{minipage}
  \begin{minipage}[t]{0.495\textwidth}
  \centering
   \includegraphics[width=\textwidth]{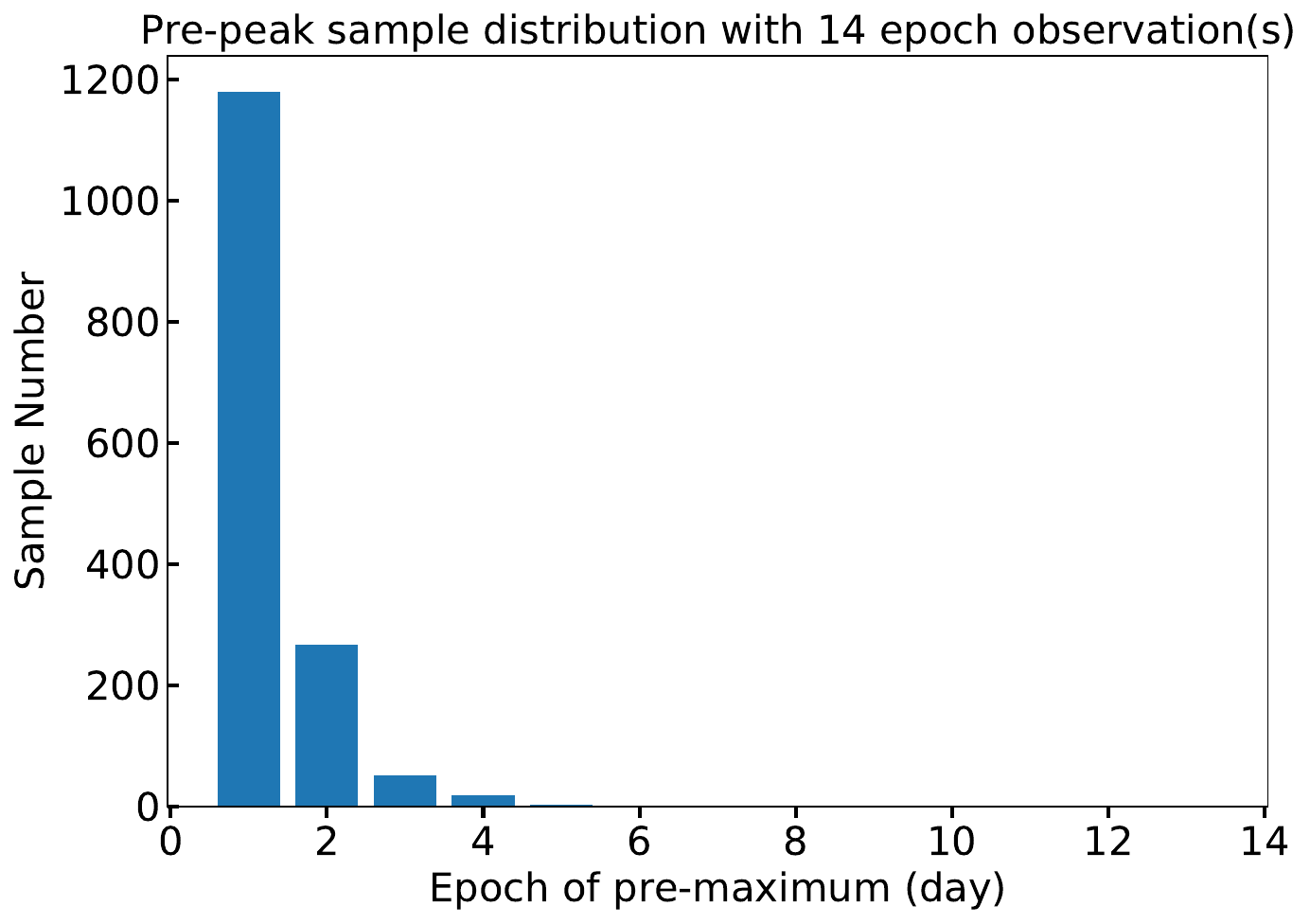}
  \end{minipage}
  \quad
  \begin{minipage}[t]{0.495\textwidth}
  \centering
   \includegraphics[width=\textwidth]{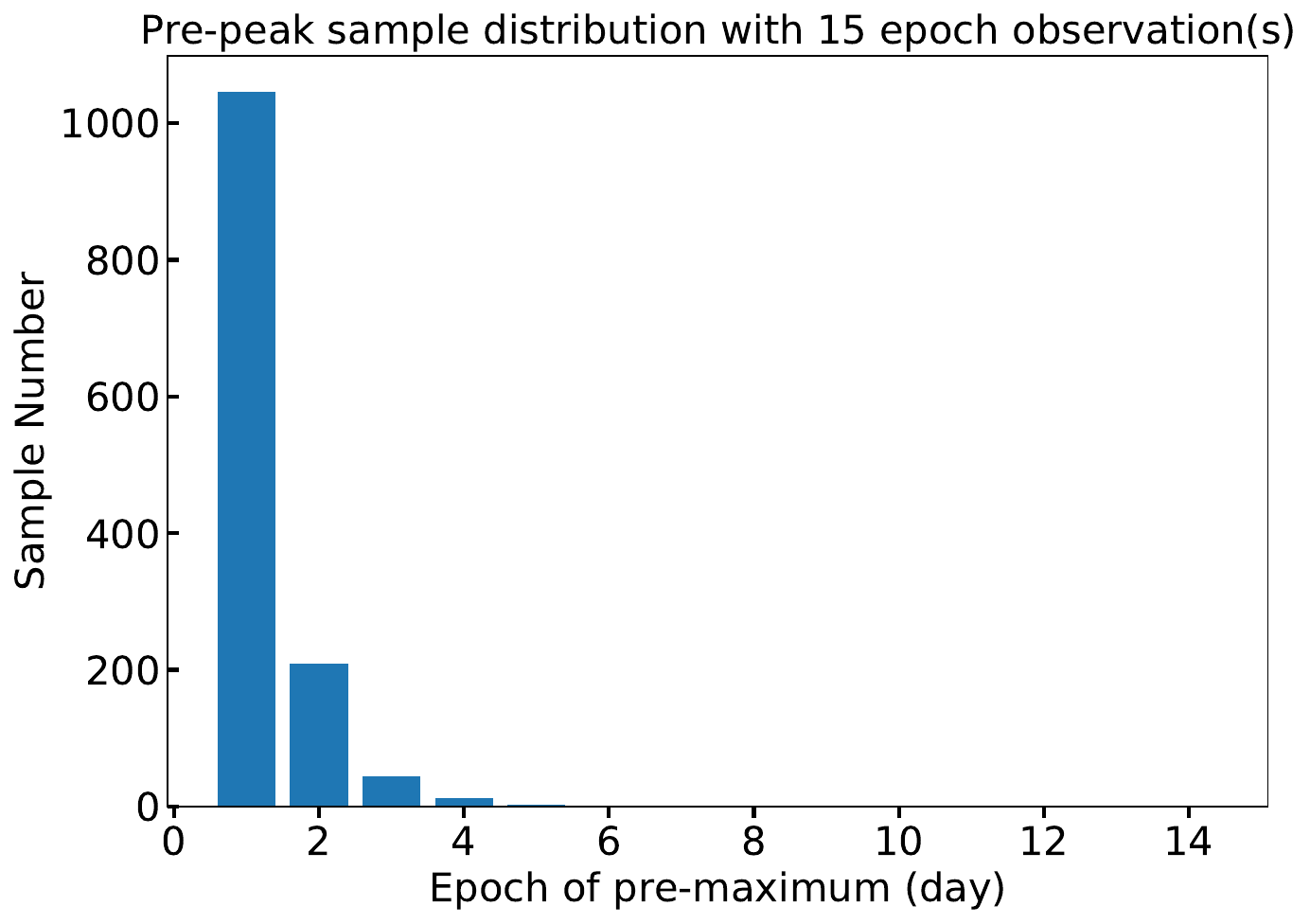}
  \end{minipage}
    \begin{minipage}[t]{0.495\textwidth}
  \centering
   \includegraphics[width=\textwidth]{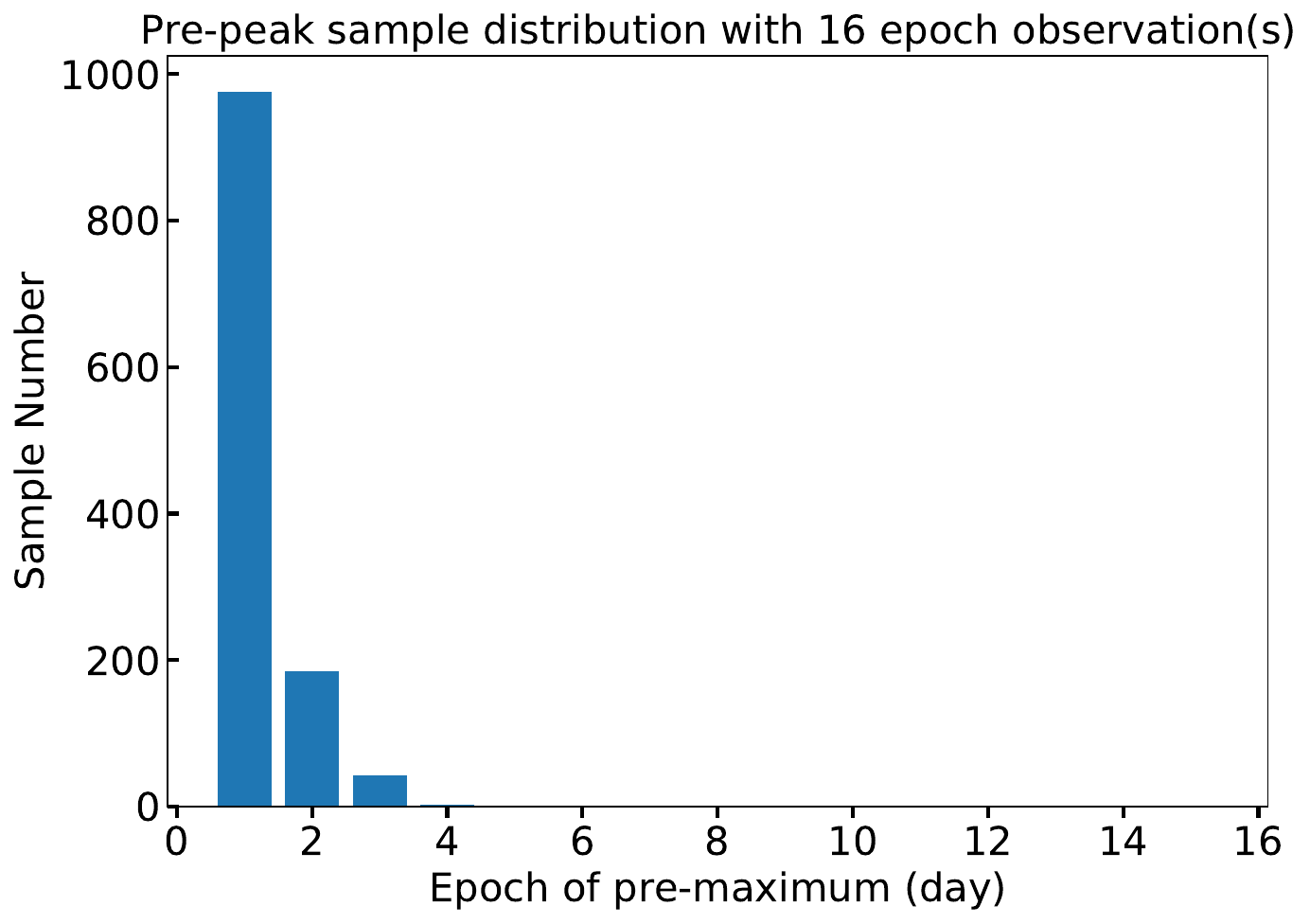}
  \end{minipage}
\caption{Continued~\ref{fig:distprepeak}, the sample distribution for the pre-maximum epochs in each scenario (the pre-maximum epochs are provided in the title description of each subplot).}
\label{fig:distprepeak2}
\end{figure}

\begin{figure}[h]
  \begin{minipage}[t]{0.495\textwidth}
  \centering
   \includegraphics[width=\textwidth]{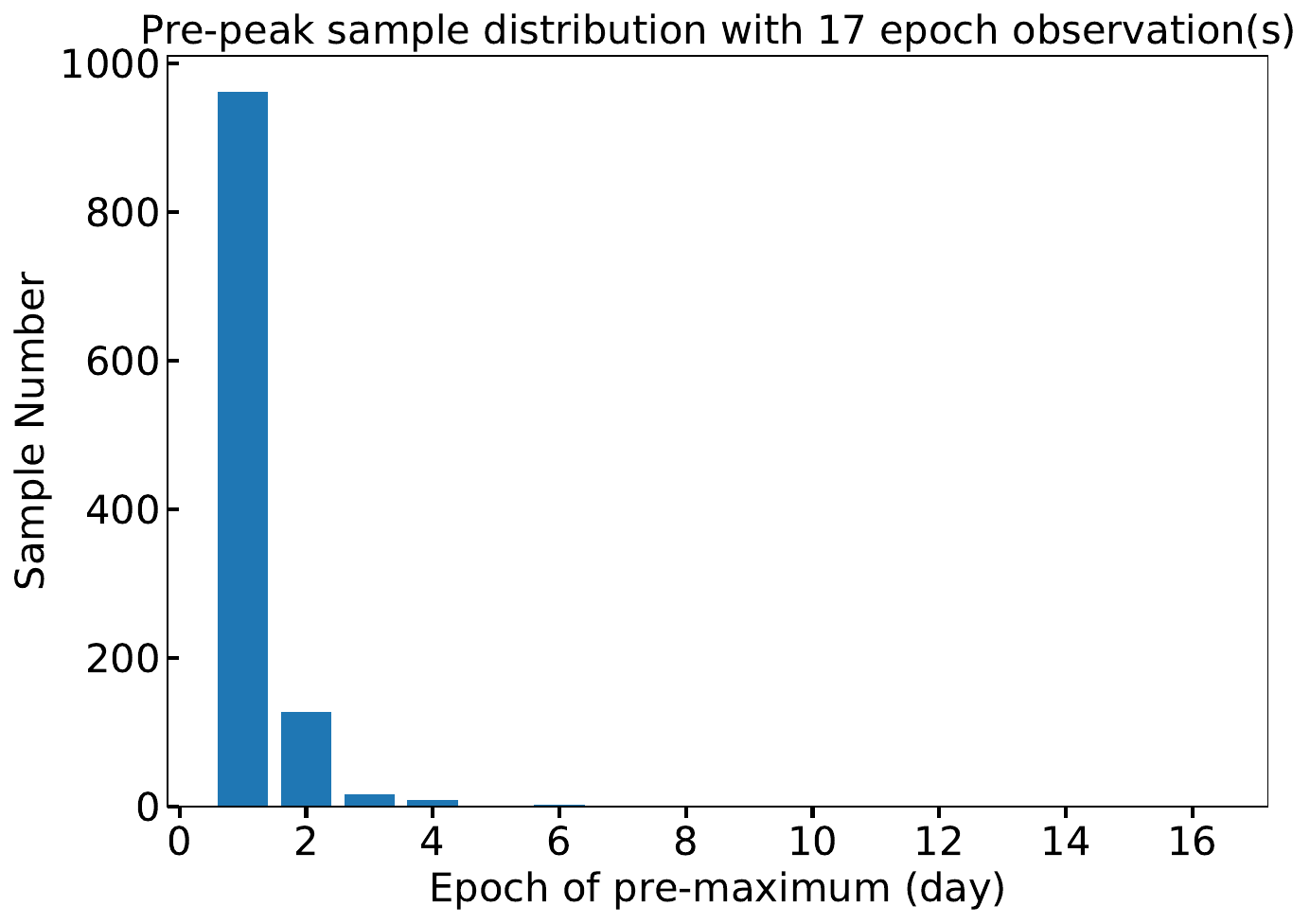}
  \end{minipage}
\caption{Continued~\ref{fig:distprepeak2}, the sample distribution for the pre-maximum epochs in each scenario (the pre-maximum epochs are provided in the title description of each subplot).}
\label{fig:distprepeak3}
\end{figure}

\section{confusion matrix}
\label{app:cm}

As highlighted in \ref{subsect:evaluation}, in addition to the five metrics for model evaluation provided after model training in the main text, there is also the commonly used evaluation metric of the confusion matrix. We employed \emph{\texttt{Mesiri}} for each scenario-specific training sample, as illustrated by the distribution of sample pre-maximum luminosity in Appendix~\ref{app:distprepeak}, and the corresponding confusion matrices for each scenario are provided. Fig.~\ref{fig:cm_distprepeak} and Fig.~\ref{fig:cm_distprepeak2} represent confusion matrices. From left to right and top to bottom, each subplot represents the confusion matrix of the corresponding samples (left panel) and is normalized in percentages (right panel) for the pre-maximum epoch of observations, respectively. Beneath each set of confusion matrices, descriptions are provided for the sampled scenarios that align with the model training.

\begin{figure}[h]
  \begin{minipage}[t]{0.495\linewidth}
  \centering
   \includegraphics[width=\textwidth]{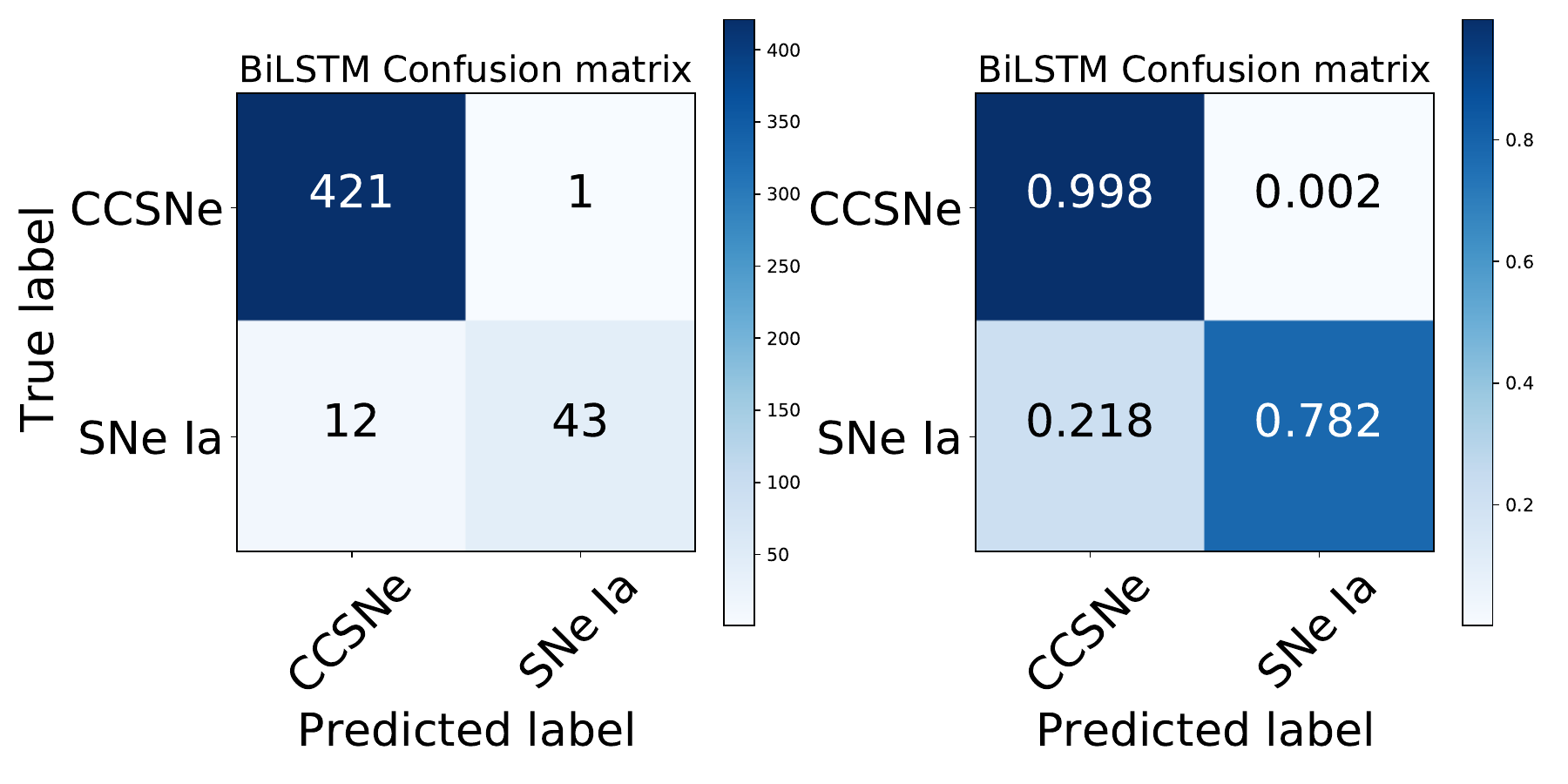}
   {\caption*{(1) Confusion matrix with 1 epoch early-time
   observation}}
  \end{minipage}
  \begin{minipage}[t]{0.495\textwidth}
  \centering
   \includegraphics[width=\textwidth]{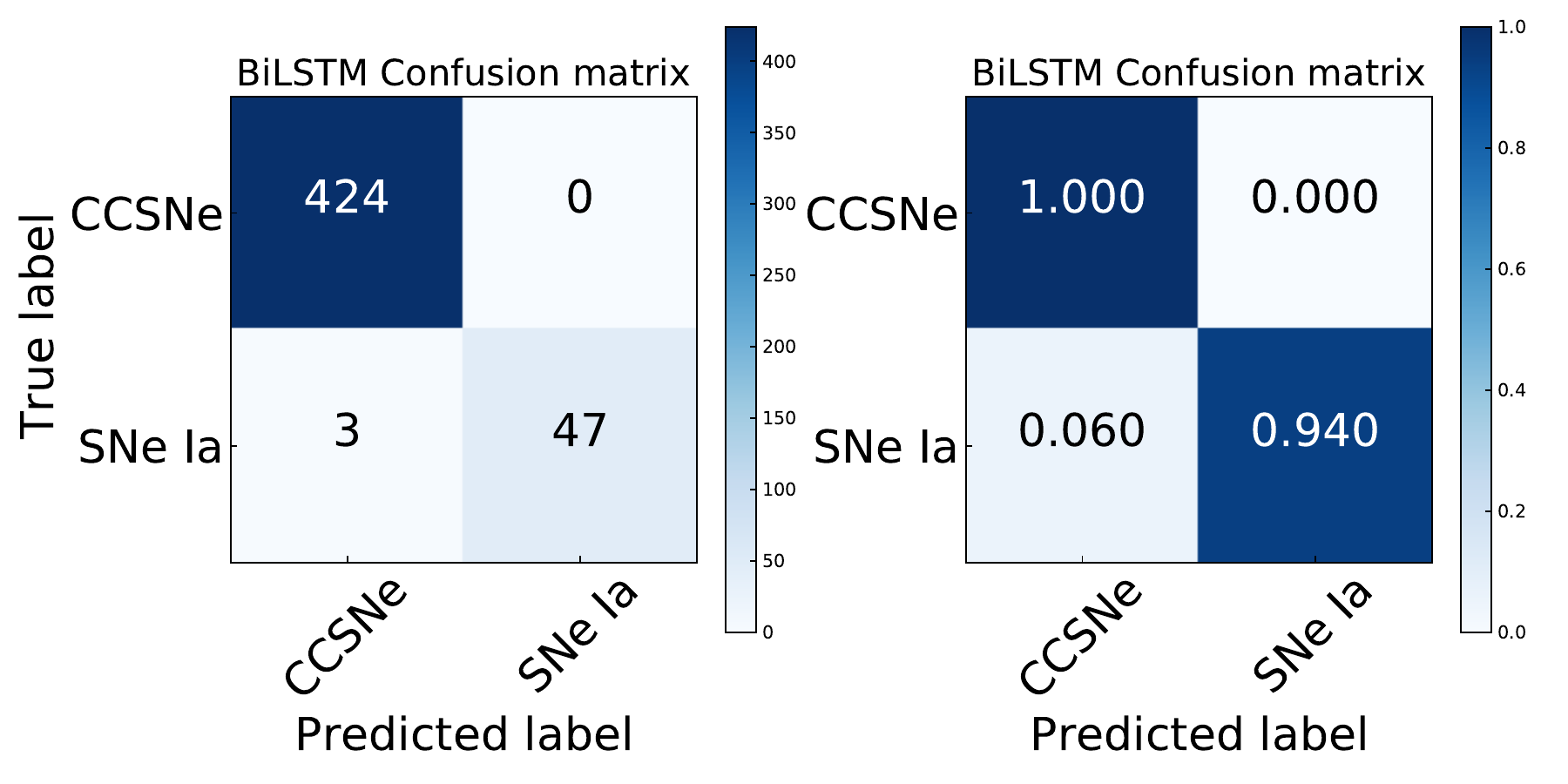}
   {\caption*{(2) Confusion matrix with 2 epochs early-time observations}}
  \end{minipage}
  \quad
  \begin{minipage}[t]{0.495\textwidth}
  \centering
   \includegraphics[width=\textwidth]{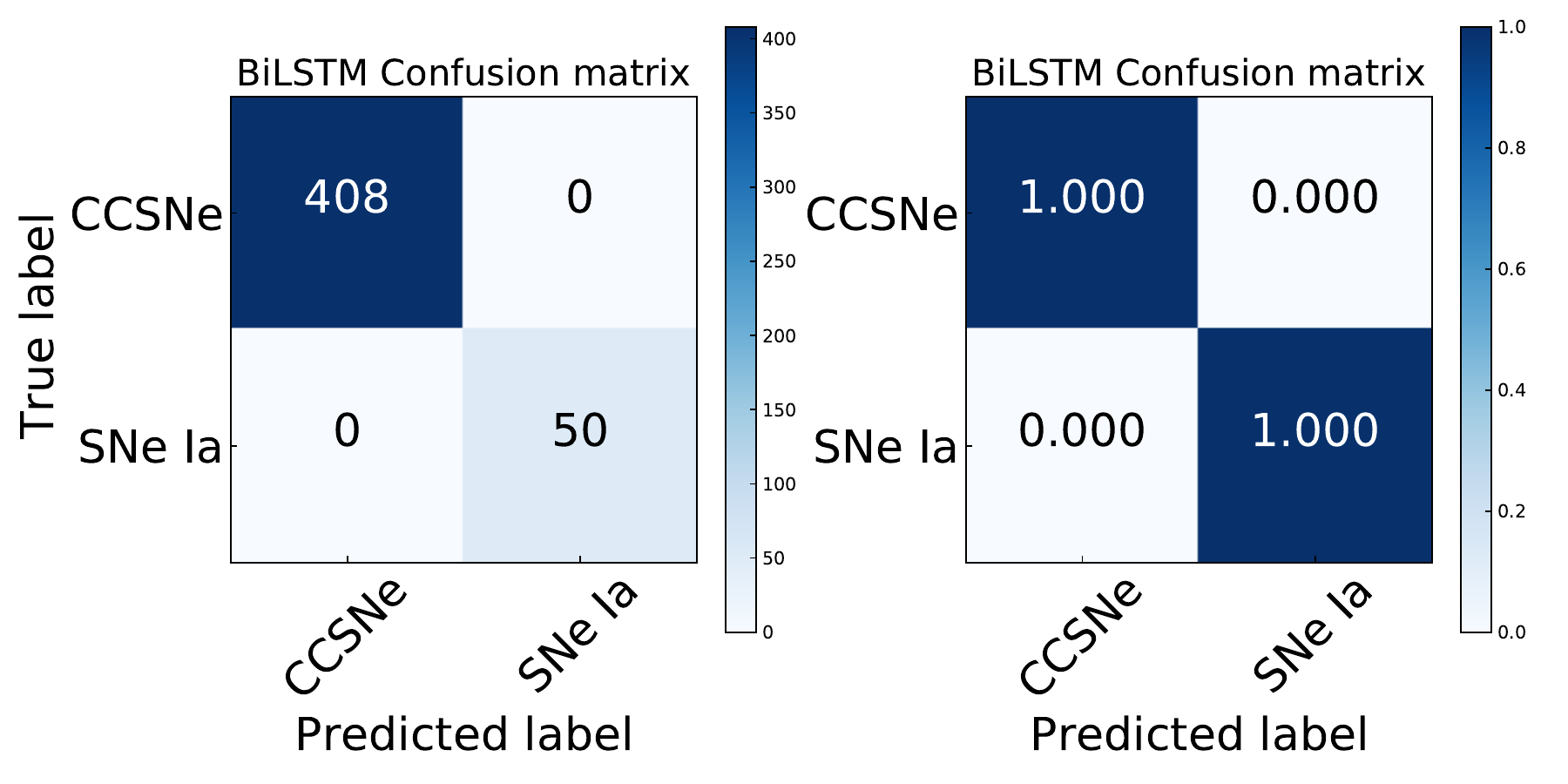}
   {\caption*{(3) Confusion matrix with 3 epochs early-time observations}}
  \end{minipage}
  \begin{minipage}[t]{0.495\textwidth}
  \centering
   \includegraphics[width=\textwidth]{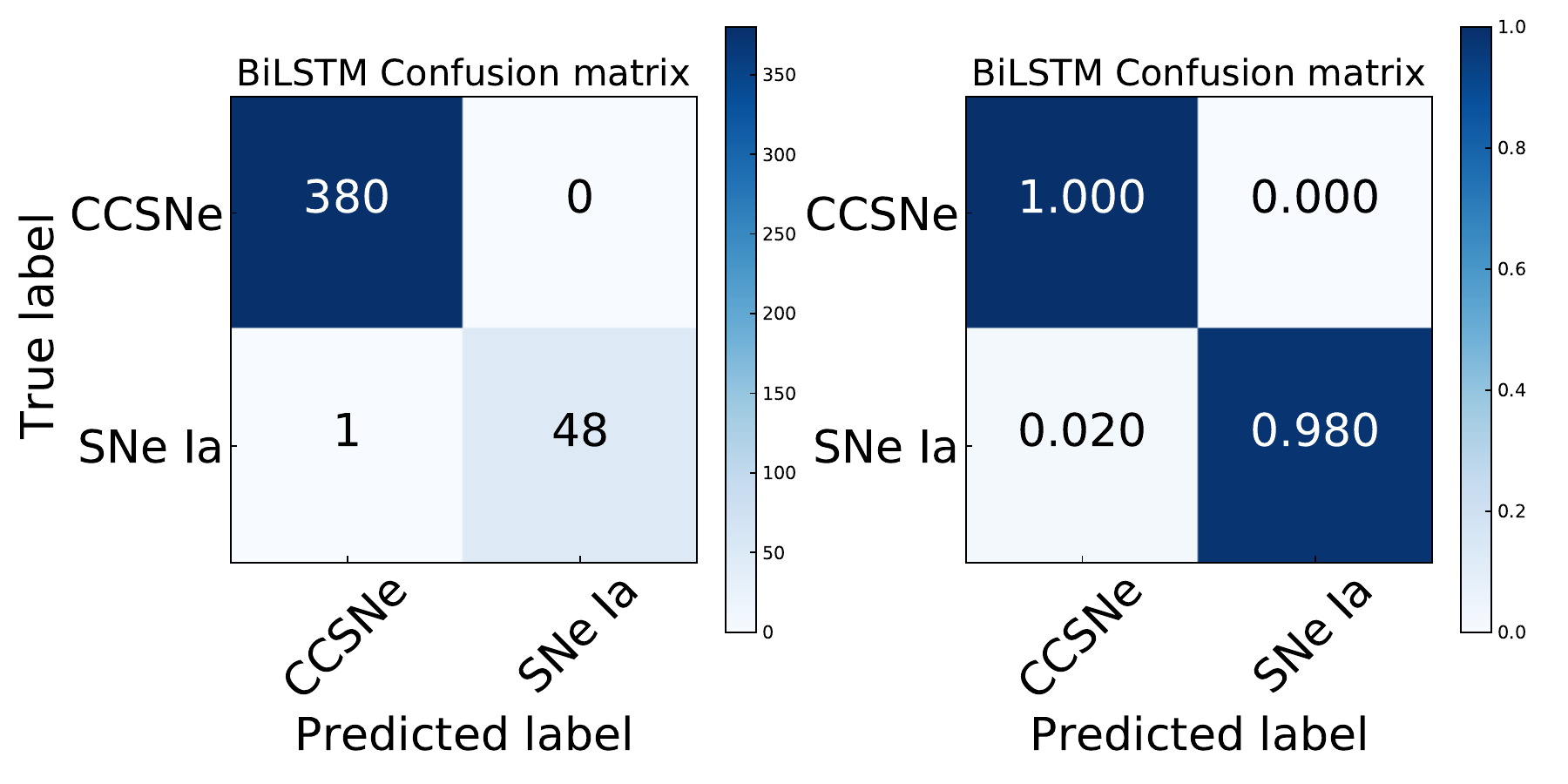}
   {\caption*{(4) Confusion matrix with 4 epochs early-time observations}}
  \end{minipage}
  \quad
  \begin{minipage}[t]{0.495\textwidth}
  \centering
   \includegraphics[width=\textwidth]{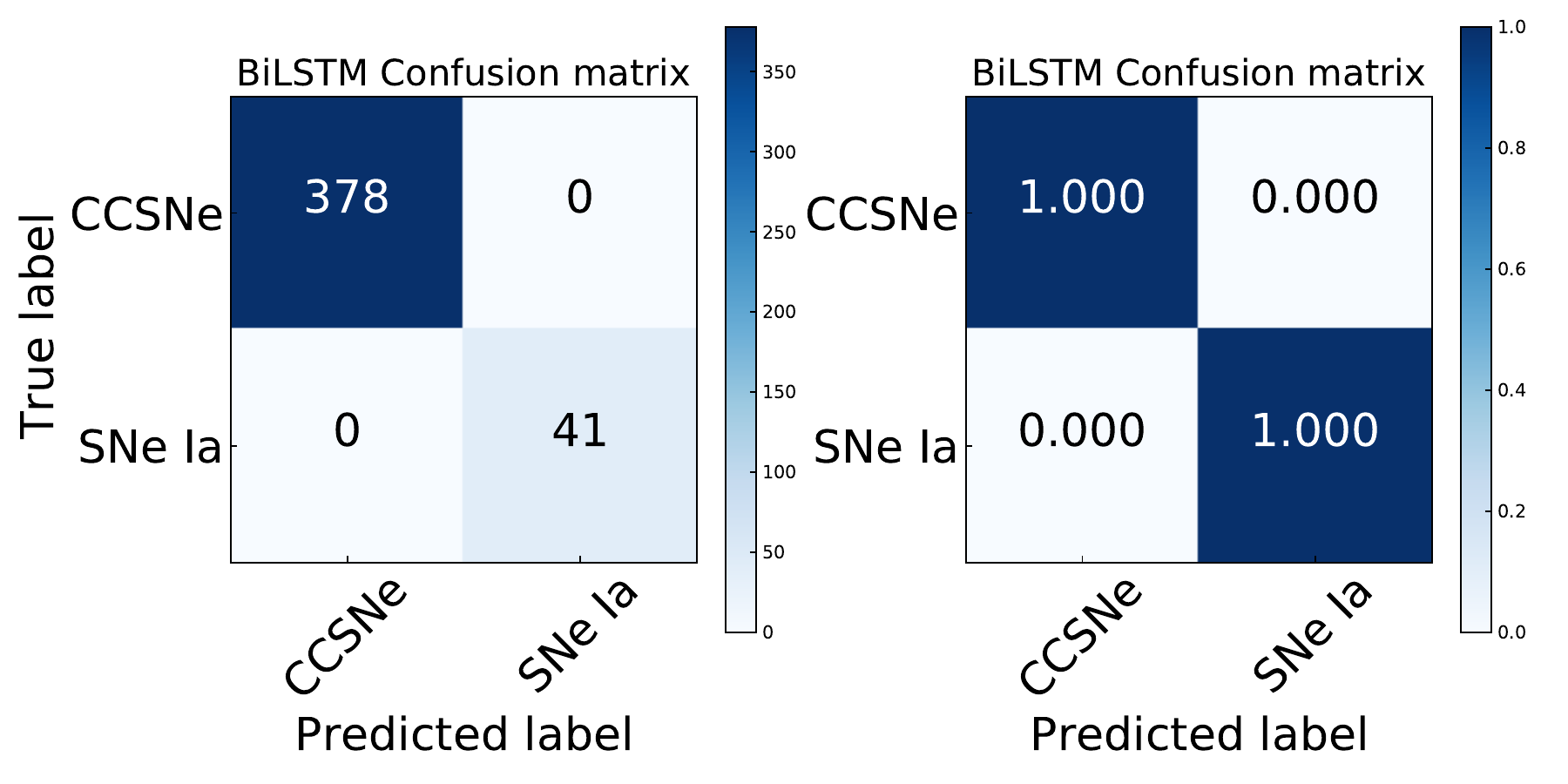}
   {\caption*{(5) Confusion matrix with 5 epochs early-time observations}}
  \end{minipage}
  \begin{minipage}[t]{0.495\textwidth}
  \centering
   \includegraphics[width=\textwidth]{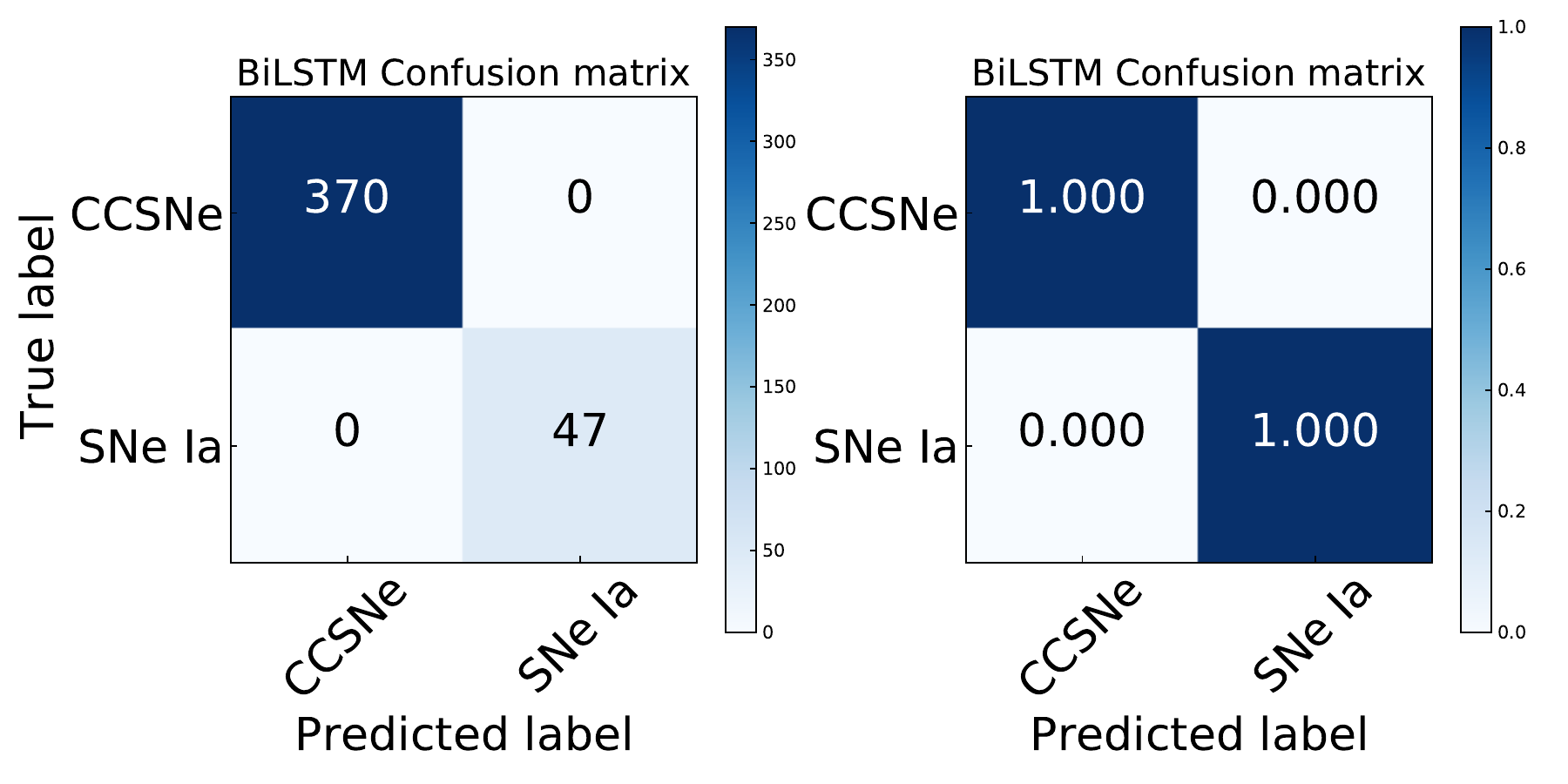}
   {\caption*{(6) Confusion matrix with 6 epochs early-time observations}}
  \end{minipage}
  \quad
  \begin{minipage}[t]{0.495\textwidth}
  \centering
   \includegraphics[width=\textwidth]{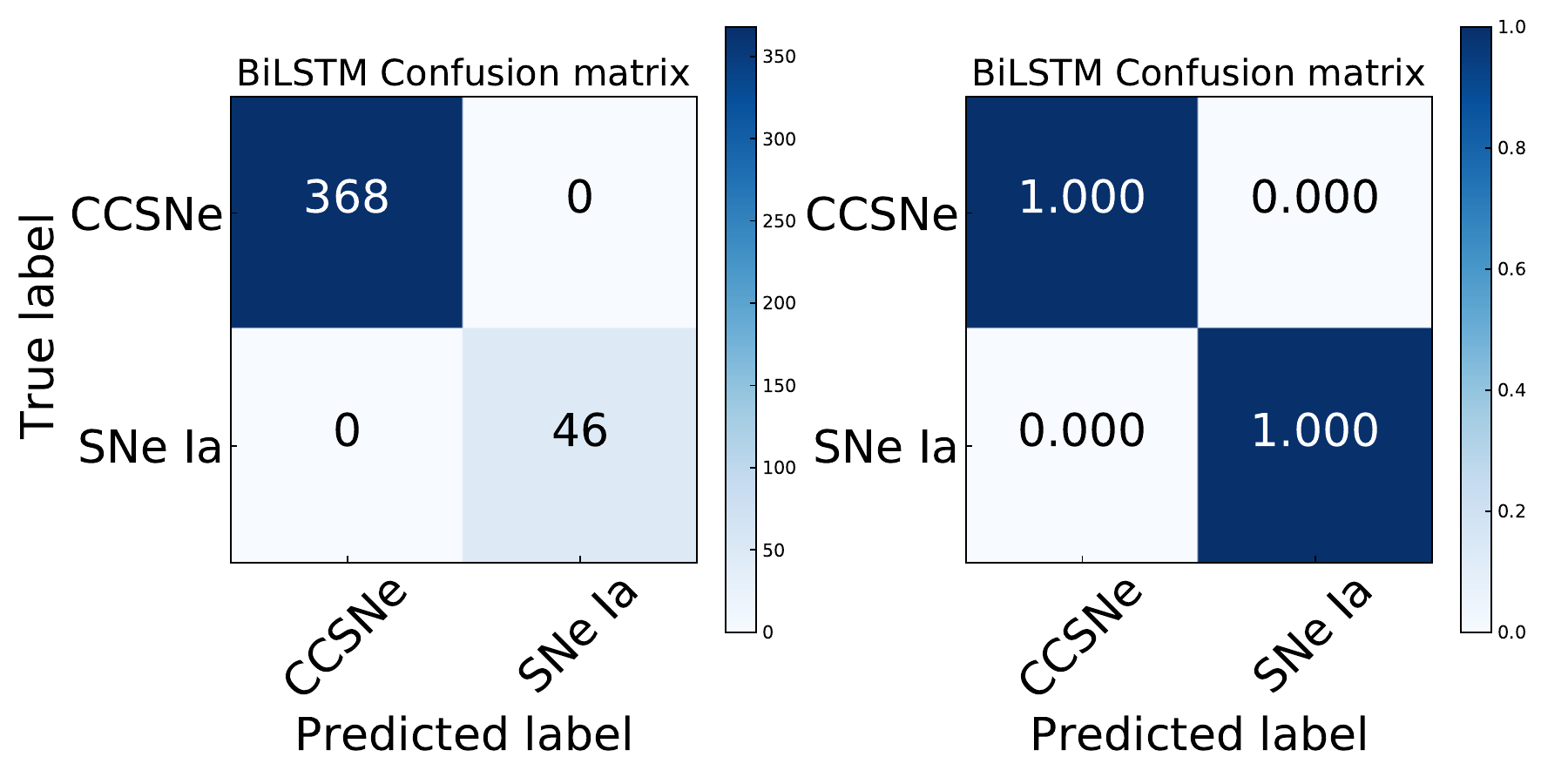}
   {\caption*{(7) Confusion matrix with 7 epochs early-time observations}}
  \end{minipage}
    \begin{minipage}[t]{0.495\textwidth}
  \centering
   \includegraphics[width=\textwidth]{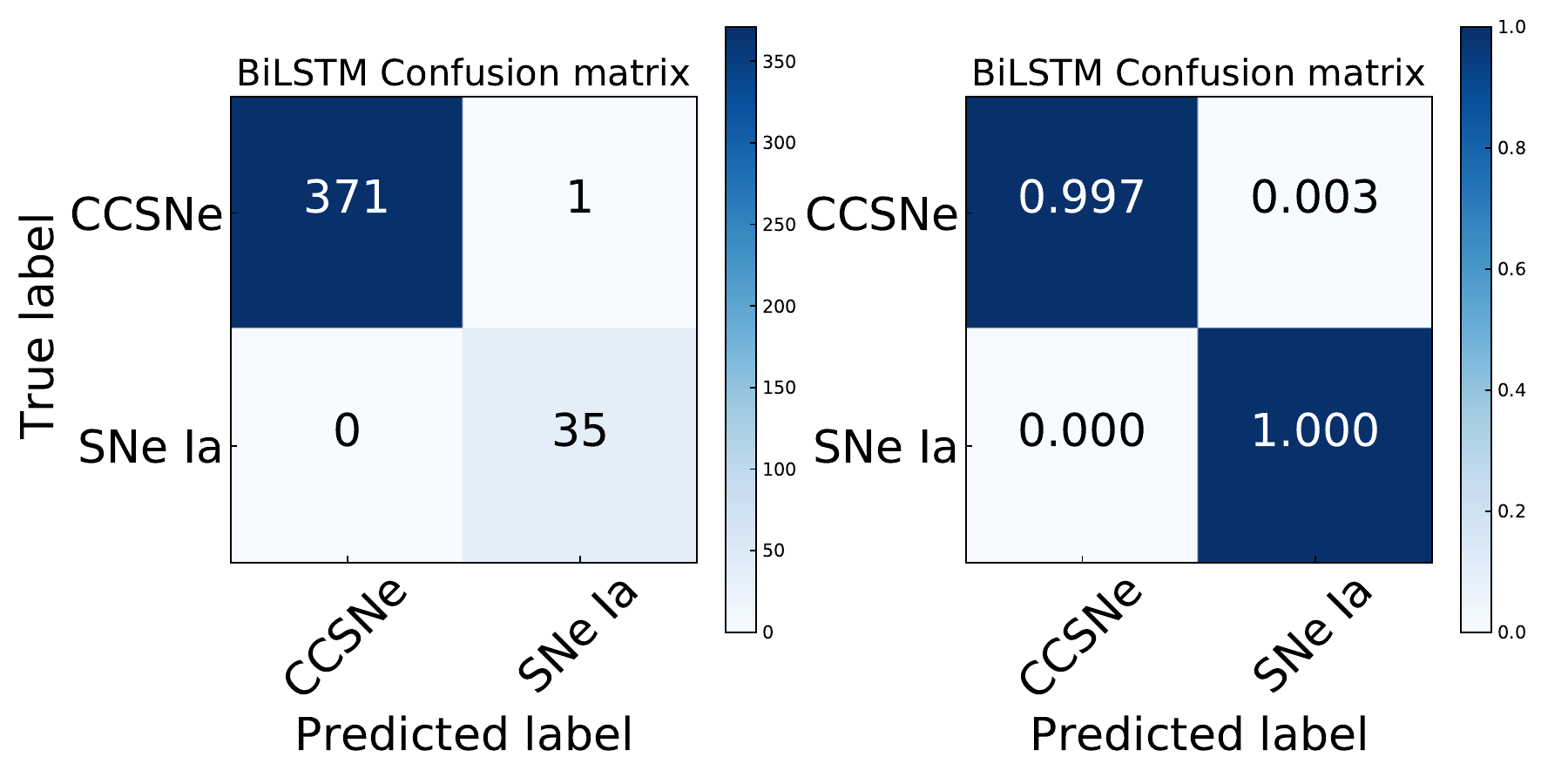}
   {\caption*{(8) Confusion matrix with 8 epochs early-time observations}}
  \end{minipage}
  \quad
    \begin{minipage}[t]{0.495\linewidth}
  \centering
   \includegraphics[width=\textwidth]{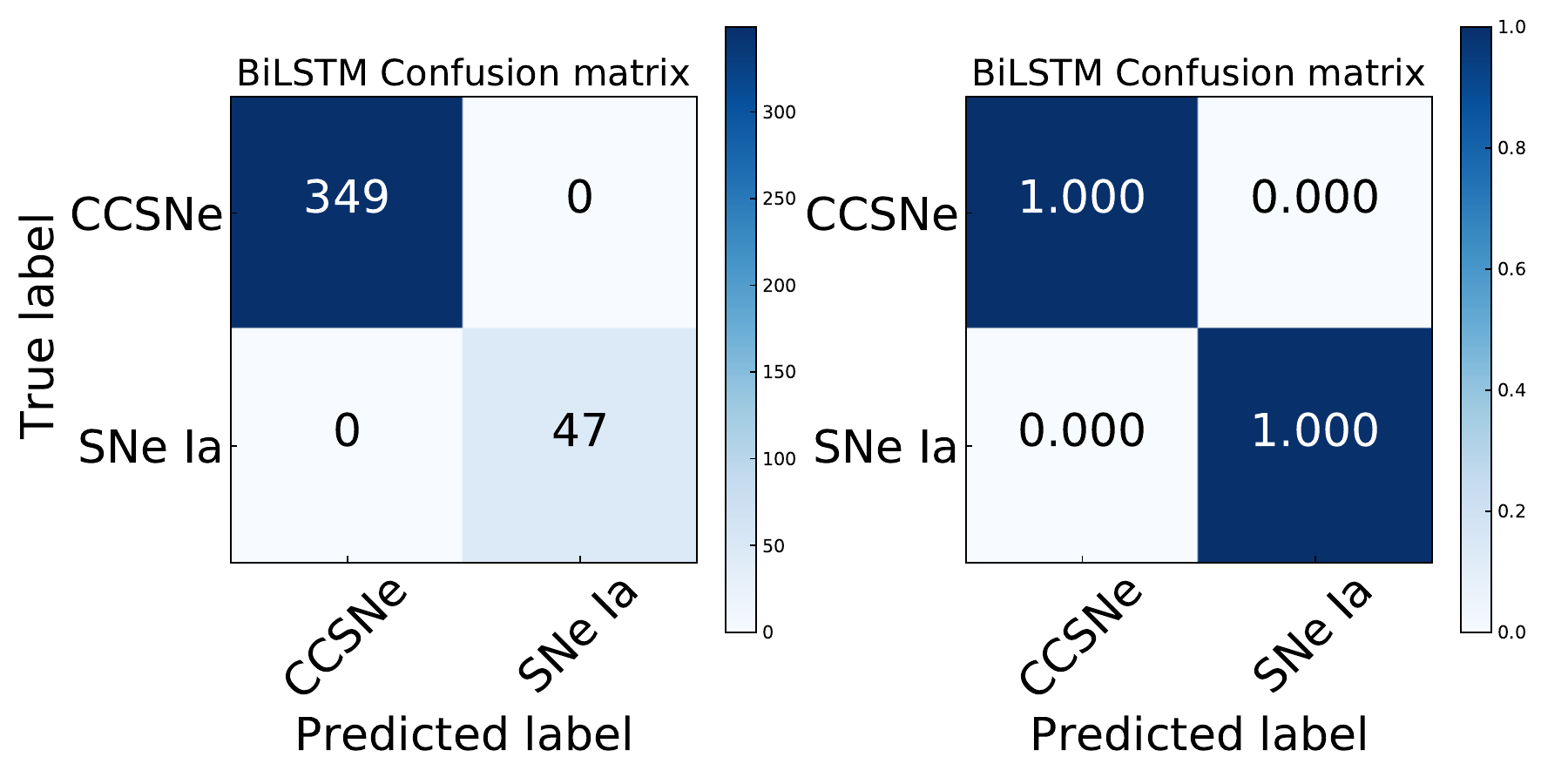}
   {\caption*{(9) Confusion matrix with 9 epochs early-time observations}}
  \end{minipage}
  \begin{minipage}[t]{0.495\textwidth}
  \centering
   \includegraphics[width=\textwidth]{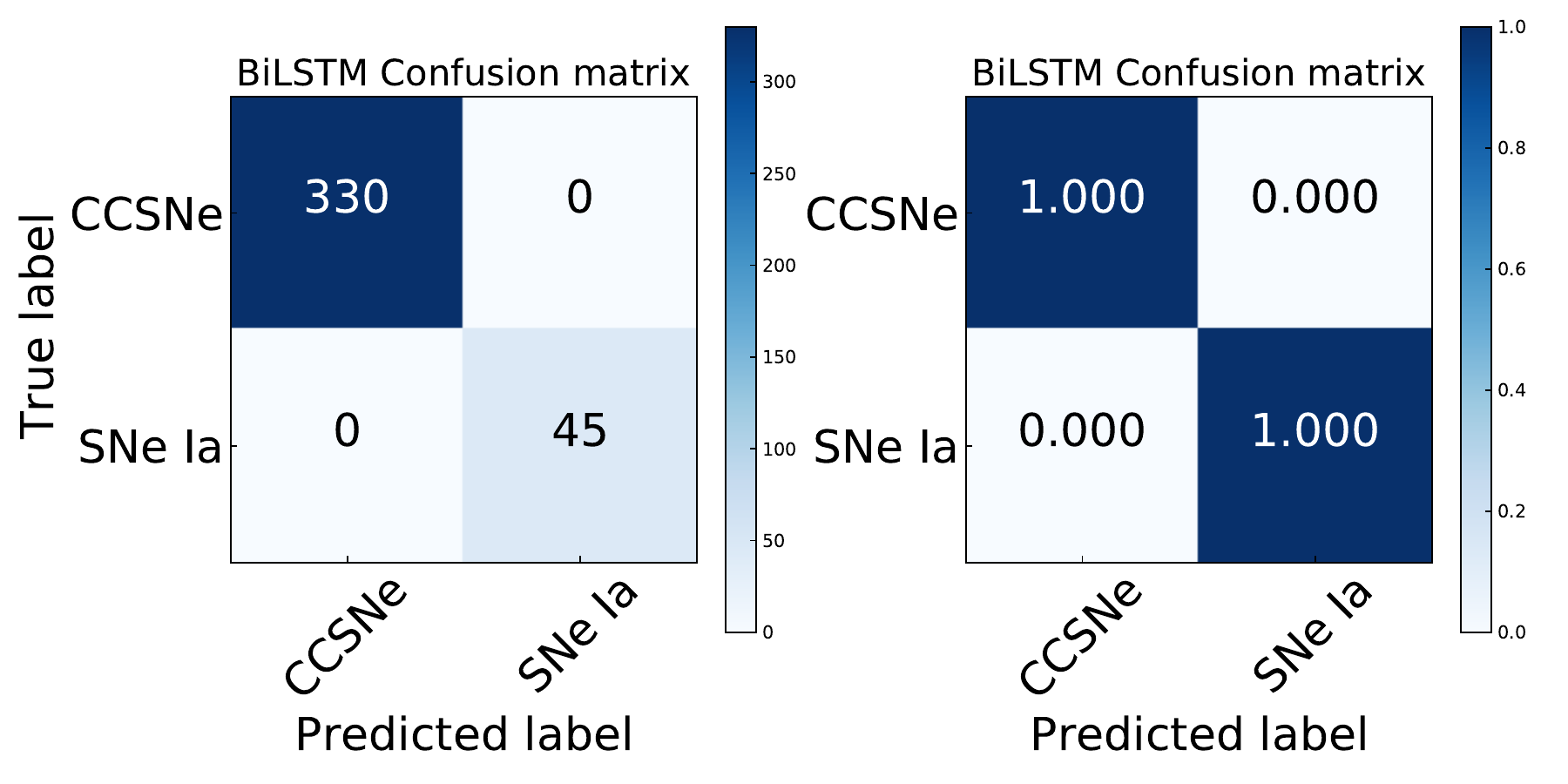}
   {\caption*{(10) Confusion matrix with 10 epochs early-time observations}}
  \end{minipage}
\caption{Confusion matrices, from left to right and top to bottom, each subplot represents the confusion matrix of the corresponding samples (left) and normalized confusion matrix (right) for the pre-maximum epoch of observations, respectively.} 
\label{fig:cm_distprepeak}
\end{figure}

\begin{figure}[h]
  \begin{minipage}[t]{0.495\textwidth}
  \centering
  \includegraphics[width=\textwidth]{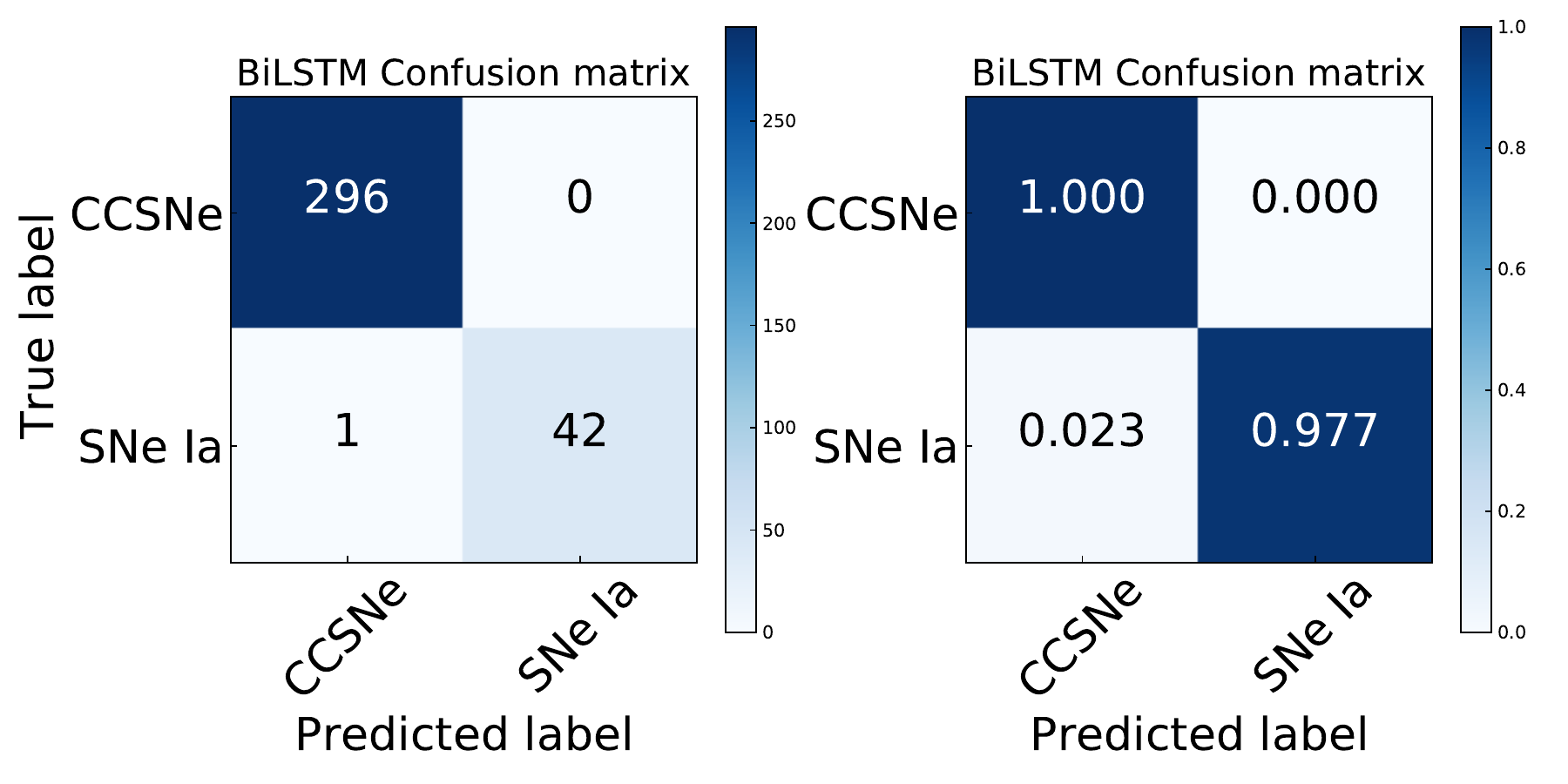}
  {\caption*{(11) Confusion matrix with 11 epochs early-time observations}}
  \end{minipage}
  \begin{minipage}[t]{0.495\textwidth}
  \centering
   \includegraphics[width=\textwidth]{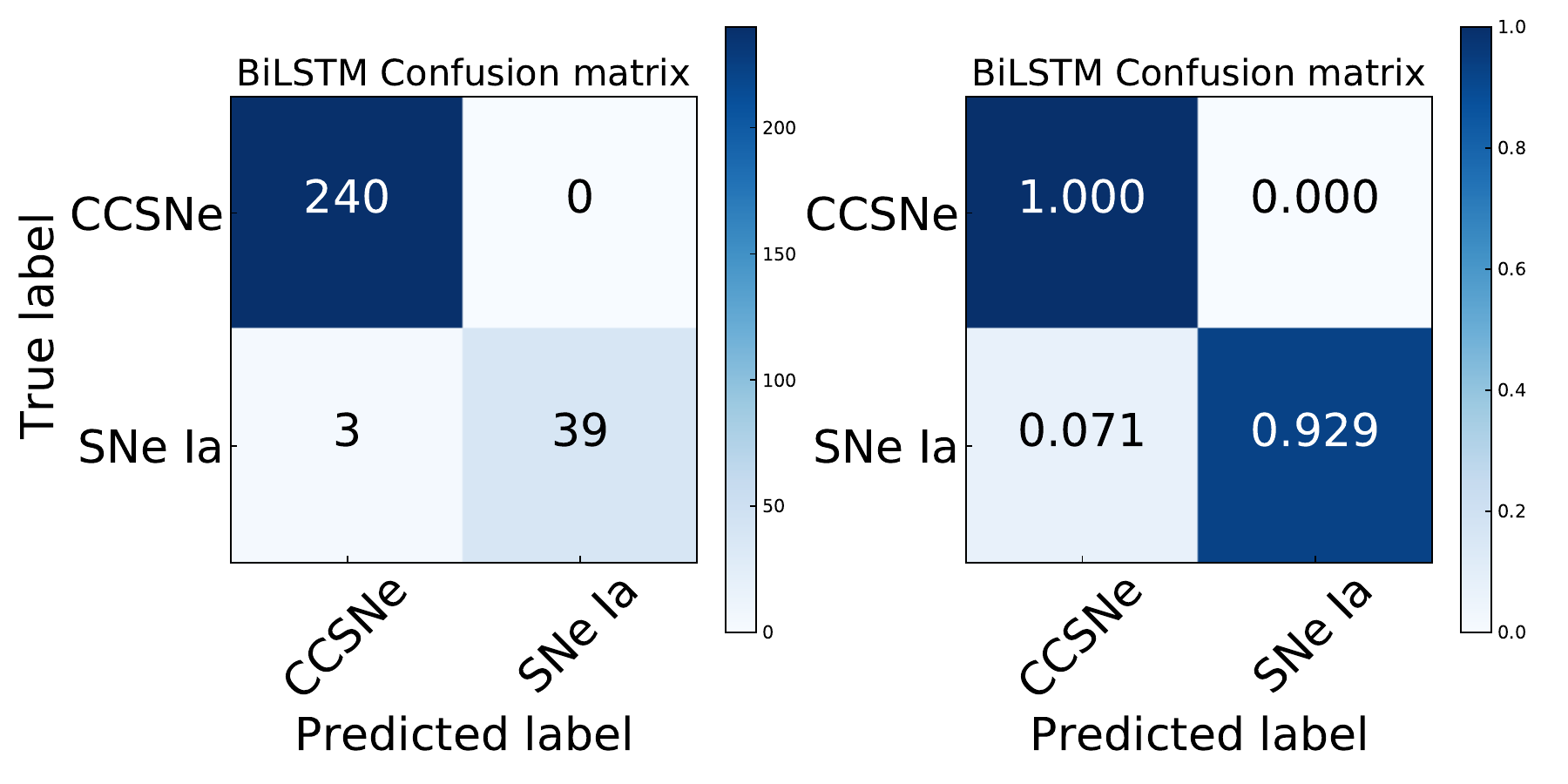}
   {\caption*{(12) Confusion matrix with with 12 epochs early-time observations}}
  \end{minipage}
  \quad
  \begin{minipage}[t]{0.495\textwidth}
  \centering
   \includegraphics[width=\textwidth]{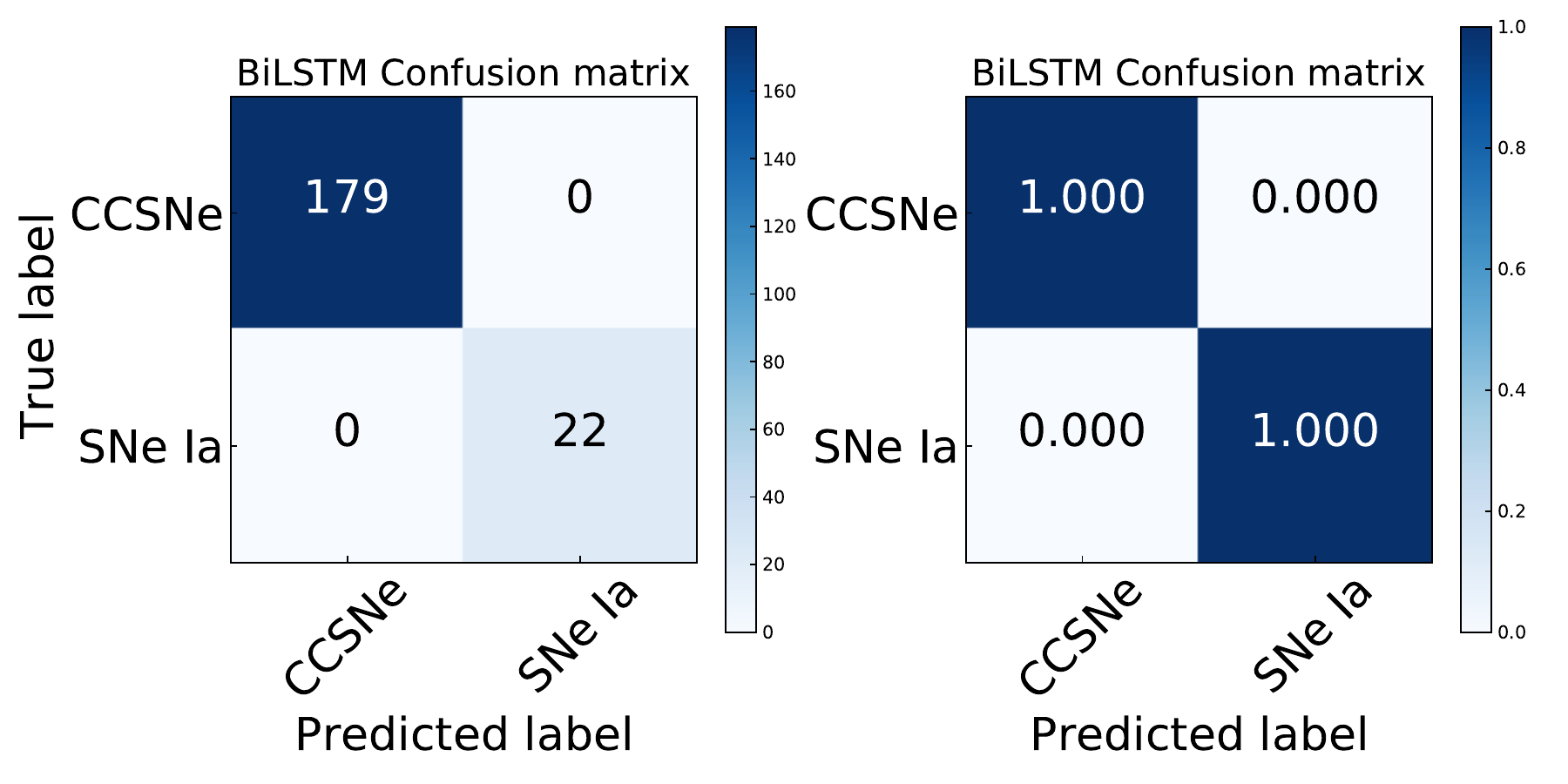}
   {\caption*{(13) Confusion matrix with 13 epochs early-time observations}}
  \end{minipage}
  \begin{minipage}[t]{0.495\textwidth}
  \centering
   \includegraphics[width=\textwidth]{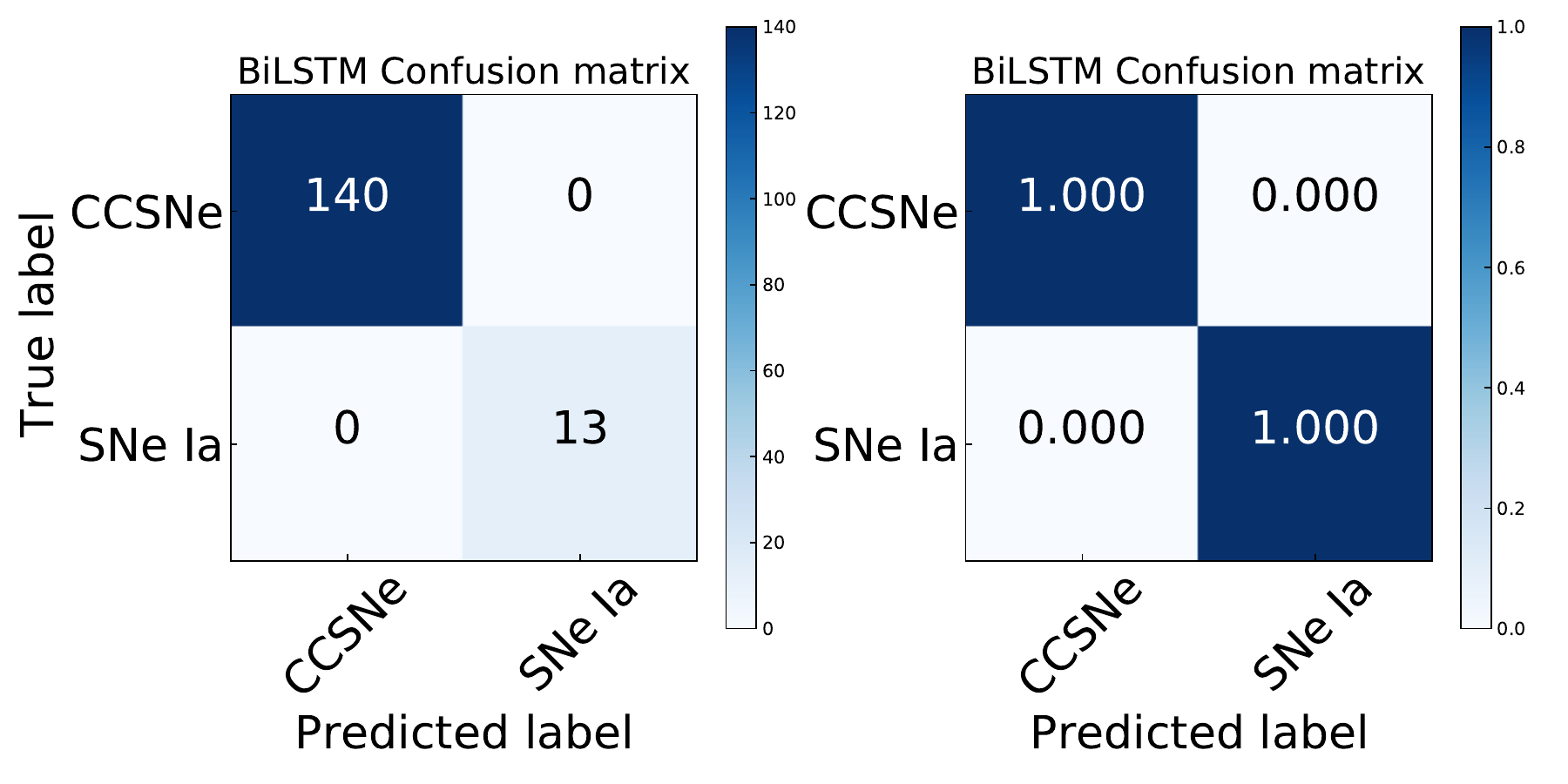}
   {\caption*{(14) Confusion matrix with 14 epochs early-time observations}}
  \end{minipage}
  \quad
  \begin{minipage}[t]{0.495\textwidth}
  \centering
   \includegraphics[width=\textwidth]{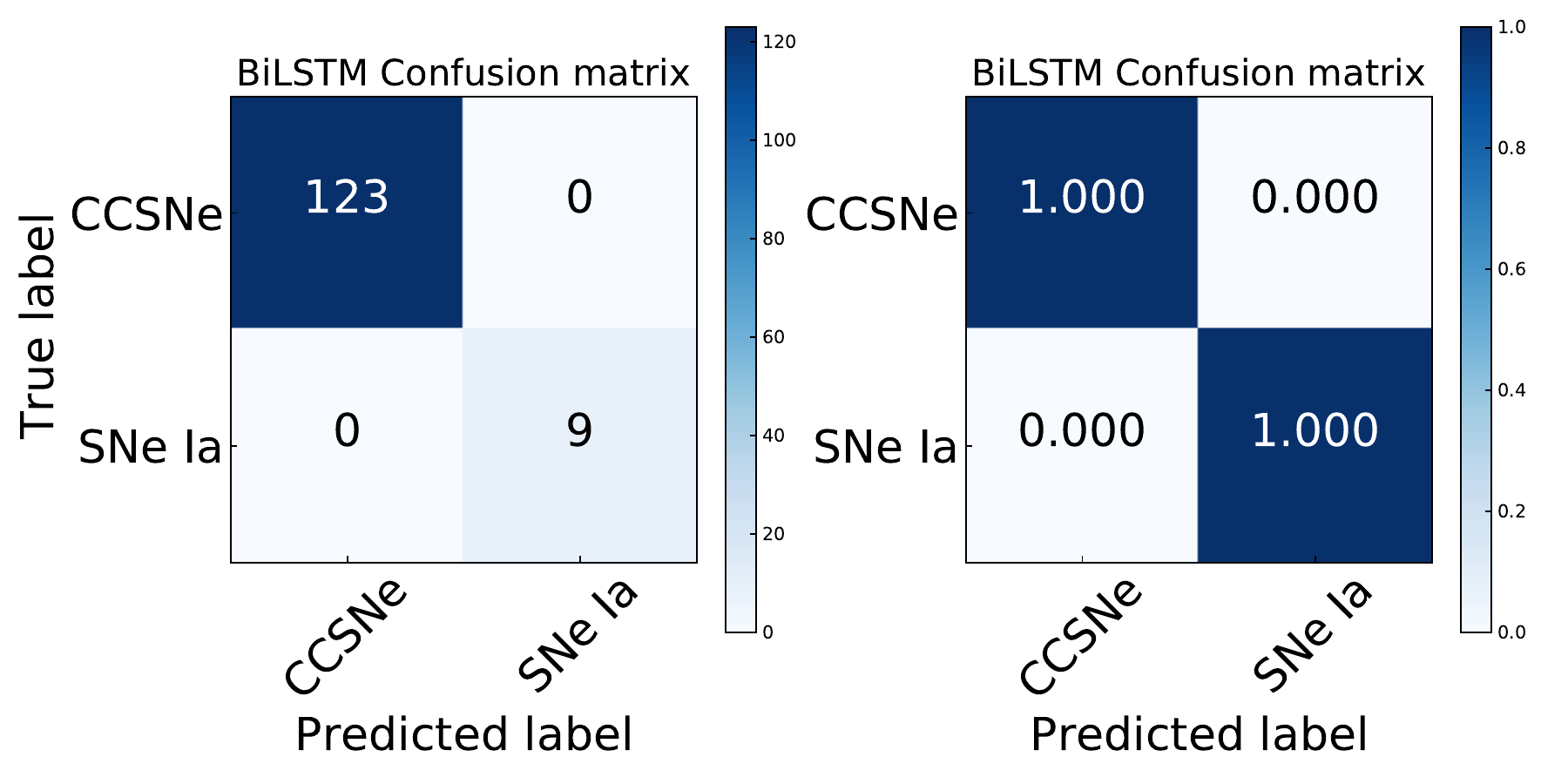}
   {\caption*{(15) Confusion matrix with 15 epochs early-time observations}}
  \end{minipage}
    \begin{minipage}[t]{0.495\textwidth}
  \centering
   \includegraphics[width=\textwidth]{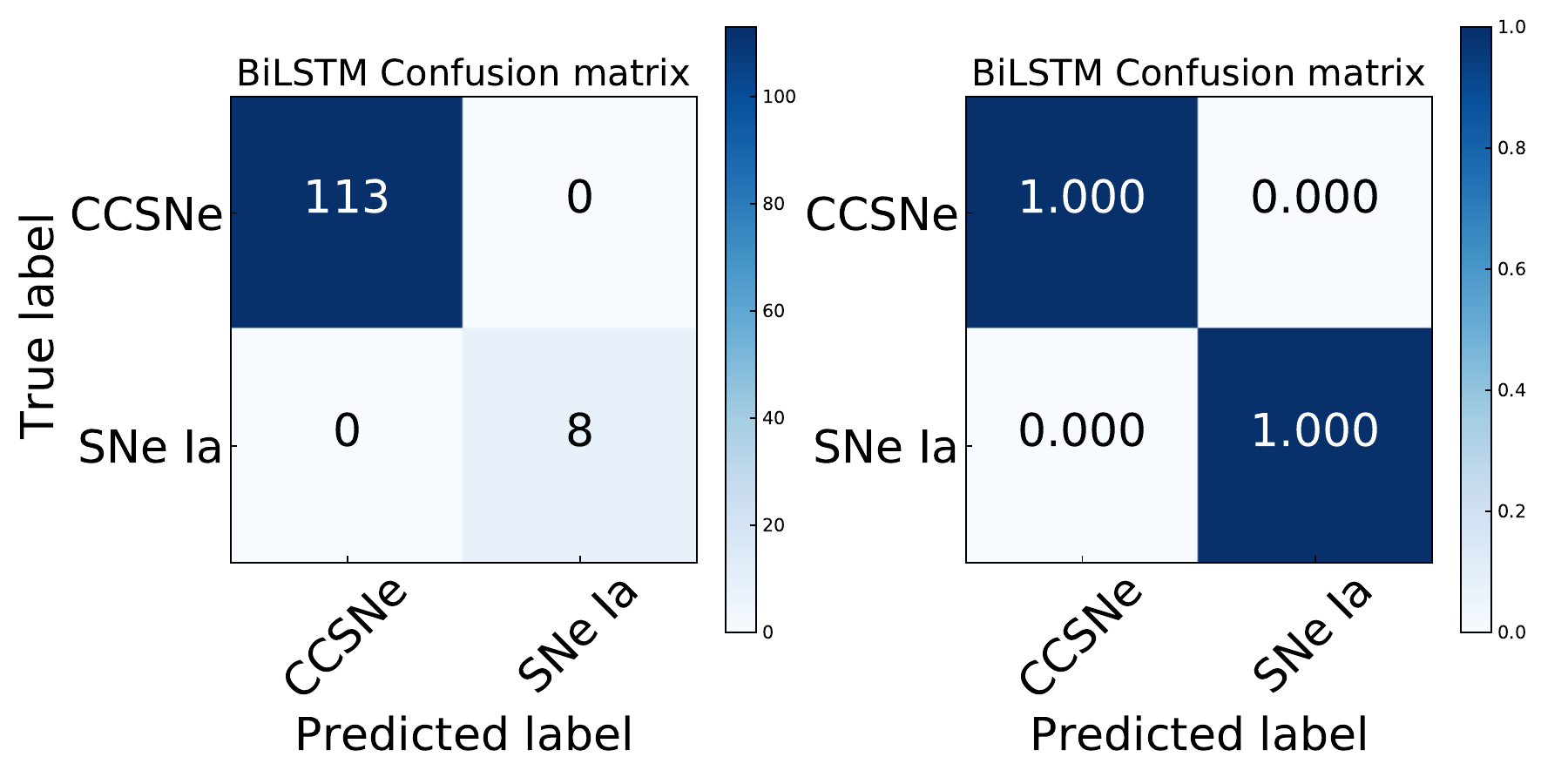}
   {\caption*{(16) Confusion matrix with 16 epochs early-time observations}}
  \end{minipage}
  \quad
  \begin{minipage}[t]{0.495\textwidth}
  \centering
   \includegraphics[width=\textwidth]{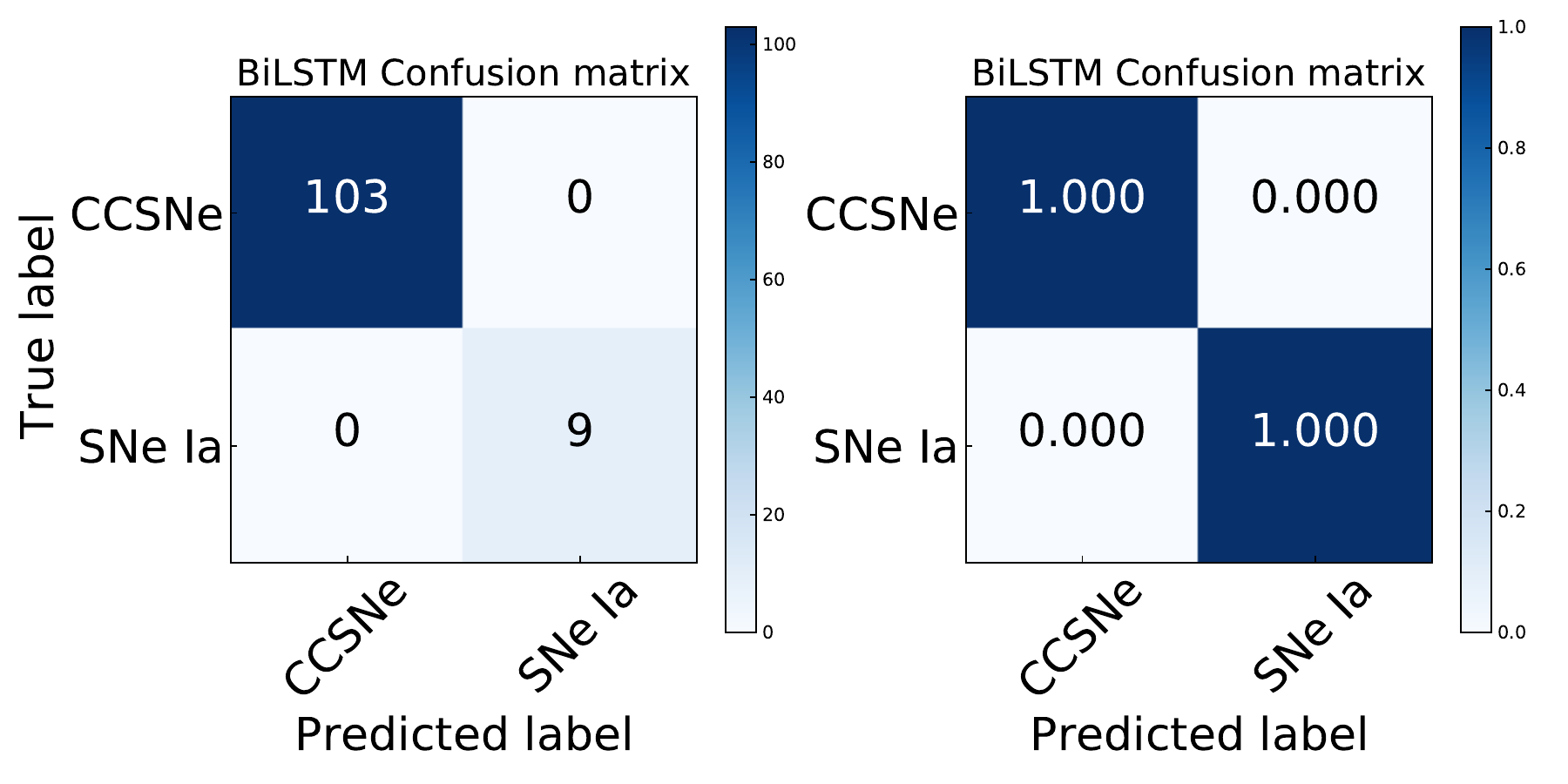}
   {\caption*{(17) Confusion matrix with 17 epochs early-time observations}}
  \end{minipage}
\caption{Continued~\ref{fig:cm_distprepeak}, from left to right and top to bottom, each subplot represents the confusion matrix of the corresponding samples (left) and normalized confusion matrix (right) for the pre-maximum epoch of observations, respectively.}
\label{fig:cm_distprepeak2}
\end{figure}

\clearpage

\bibliographystyle{raa}
\bibliography{ms2024-0117}

\label{lastpage}

\end{document}